\documentclass[12pt,a4paper, amsmath,amssymb]{article}
\usepackage[margin=0.8in]{geometry}
\usepackage{amssymb}
\usepackage{amsmath}
\usepackage{amsfonts}
\usepackage{bm}
\usepackage[utf8]{inputenc}
\usepackage{graphicx}
\usepackage{float}
\usepackage[colorlinks,linkcolor = blue,
urlcolor  = blue,
citecolor = blue,
anchorcolor = blue]{hyperref}
\usepackage{dcolumn}
\usepackage{graphicx}
\usepackage{color}
\usepackage{amsmath}
\usepackage{amssymb}
\usepackage{amsthm}
\usepackage{latexsym}
\usepackage{bm}
\usepackage{multirow}
\usepackage{cancel}
\usepackage{float}
\usepackage{hyperref}
\usepackage{tabularx}
\providecommand{\openone}{\leavevmode\hbox{\small1\kern-3.8pt\normalsize1}}

\newcommand{\slx}{s_L^u}
\newcommand{\slu}{s_L^u}
\newcommand{\sld}{s_L^d}
\newcommand{\clx}{c_L^u}
\newcommand{\clu}{c_L^u}
\newcommand{\cld}{c_L^d}
\newcommand{\srx}{s_R^u}
\newcommand{\sru}{s_R^u}
\newcommand{\srd}{s_R^d}
\newcommand{\crx}{c_R^u}
\newcommand{\cru}{c_R^u}
\newcommand{\crd}{c_R^d}

\newcommand{\sqt}{\sqrt{2}}

\newcommand{\cbasq}{c^2_{\beta-\alpha}}
\newcommand{\sbasq}{s^2_{\beta-\alpha}}

\usepackage{xcolor}
\usepackage{makecell}
\usepackage{ctable} 
\usepackage{adjustbox}
\def\thefootnote{\fnsymbol{footnote}}
\usepackage{hyperref}
\usepackage{cite}

\usepackage{soul}
\definecolor{green}{rgb}{0.0, 0.56, 0.0}
\begin{document}
\begin{center}
	%
	{\Large \textbf{\upshape   The oblique parameters in the 2HDM with Vector-Like Quarks: Confronting  $M_W$ CDF-II Anomaly \\\vspace{0.2cm}
		}}
	\thispagestyle{empty}
	%
	\vspace{1cm}

	{\sc
		H. Abouabid$^1$\footnote{\url{hamza.abouabid@gmail.com
			}},
		A. Arhrib$^1$\footnote{\url{aarhrib@gmail.com
			}},
		R. Benbrik$^2$\footnote{\url{r.benbrik@uca.ac.ma}},
		M. Boukidi$^2$\footnote{\url{mohammed.boukidi@ced.uca.ma}},
		J. El Falaki$^3$\footnote{\url{jaouad.elfalaki@gmail.com
			}}
		\\
	}
	\vspace{1cm}
	{\sl
		$^1$ Université Abdelmalek Essaadi, FSTT, B. 416, Tangier, Morocco.\\
		$^2$Polydisciplinary Faculty, Laboratory of Fundamental and Applied Physics, Cadi Ayyad University, Sidi Bouzid, B.P. 4162, Safi, Morocco.\\
		$^3$ LPTHE, Physics Department, Faculty of Sciences, Ibnou Zohr University, P.O.B. 8106 Agadir, Morocco.\\
		\vspace{0.1cm}
	}
\end{center}
\vspace*{0.1cm}
\begin{abstract}  The CDF collaboration has released a new measurement of the $W$ boson
	mass using their complete data set with 8.8 fb$^{-1}$ in $p\bar{p}$  collisions.
	This result deviates from the Standard Model prediction by around 7$\sigma$.
	We explain how the two Higgs doublet model (2HDM) with vector-like quarks is affected by the recently discovered W boson mass. In our study, we include both theoretical constraints such as perturbative unitarity and vacuum stability as well as a number of experimental constraints.  We also look into how the effective mixing angle, measured by the SLD collaboration in addition to the CDF W-boson mass, is used to determine the $S$ and $T$ parameters.
	In the alignment limit, we investigate the case where the lighter CP-even neutral Higgs boson of the 2HDM is the one found at the LHC and demonstrate how the parameter space of the 2HDM type II in the presence of vector-like quarks is constrained. It is found that in most cases, there is a cancellation between the 2HDM and vector-like quarks contributions, which enlarges the parameter space of both models.

\end{abstract}

\def\thefootnote{\arabic{footnote}}
\setcounter{page}{0}
\setcounter{footnote}{0}

\newpage

\section{Introduction}
Since the advent of the precise measurement programs at SLD, LEP, and Tevatron, ElectroWeak Precision Observables (EWPOs), such as the W boson mass, the effective mixing angle  $\sin^ 2 \theta_{eff}$, and the Z boson width, among others, have been used to precisely assess the viability of the Standard Model (SM) of particle physics. It turns out that some Beyond SM (BSM) extensions can be severely constrained by these EWPOs.
Using their whole data set of  8.8 fb$^{-1}$ in $p\bar{p}$  collisions at CDF detector, the CDF collaboration recently released a new measurement of the $W$ boson mass (Fermilab Tevatron) \cite{CDF:2022hxs}:

\begin{equation}
	M^{\mathrm{CDF}}_W= 80.4335 \pm 0.0094~\mathrm{GeV}
\end{equation}

This measurement is more precise than all prior measures put together, and it is highly accurate.
In addition, the new CDF value agrees with some previous W mass measurements, but there are also some discrepancies
\cite{ParticleDataGroup:2020ssz}.	Based on complex SM radiative correction calculations that relate the
W mass to other SM particles such as the Higgs boson and the top masses,
the SM prediction for the W mass at the per mille level is given by \cite{ParticleDataGroup:2020ssz}:
\begin{equation}
	M_W^{\mathrm{SM}} = 80.357 \pm 0.006 \mathrm{GeV}
\end{equation}
It is clear that the above CDF measurement depicts a remarkable 7$\sigma$ disagreement with the SM prediction. Several analyses have been presented in this direction, including those involving the 2HDM \cite{ Lu:2022bgw, Fan:2022dck, Song:2022xts, Bahl:2022xzi, Babu:2022pdn, Biekotter:2022abc, Han:2022juu, Heo:2022dey, Ahn:2022xax, Benbrik:2022dja, Abouabid:2022lpg, Arcadi:2022dmt, Ghorbani:2022vtv, Lee:2022gyf, Kim:2022xuo, Hessenberger:2022tcx, Atkinson:2022qnl}, the triplet Higgs model \cite{Du:2022brr, Ghoshal:2022vzo, Kanemura:2022ahw, Addazi:2022fbj, Cheng:2022hbo, Bahl:2022gqg}. { Notably, the LHC data has suggested a disfavor towards Higgs triplet models as a solution to this discrepancy\cite{Butterworth:2022dkt}.}   

In the past, during the LEP  era, it was well known that the global fit of SM predictions to LEP data was used to predict the existence of a heavy top quark and a relatively light Higgs boson well before their discovery at the Tevatron and LHC, respectively. Although the CDF measurement needs to be confirmed, it is quite likely that the difference between the experimental value and the expected SM value is due to a non-decoupled new interactions or new particles.
If this is the case and if the scale of the new interaction is not too high, there is a chance that these new particles
will show up in future experiments.

The SM of particle physics has been highly successful in explaining all observed phenomena at LEP and the LHC. However, there are several indications that the SM is only an effective theory of a more fundamental and complete theory that addresses its weaknesses, such as the dark matter puzzle, the need for additional sources of CP violation to explain the baryon asymmetry in the universe, and the issues with neutrino mass and mixing. Most BSM models predict additional vector-like fermions (quarks and leptons) and/or additional Higgs bosons in their spectrum \cite{Aguilar-Saavedra:2013qpa, Arhrib:2016rlj}. It is now well established that SM-like chiral fermions\footnote{{SM-like chiral fermions would have the same properties and interactions as the fermions in the SM, but would be different in terms of their masses and other properties. In the SM, the left-handed and right-handed components of the fermions are treated differently, with only the left-handed components interacting via the weak force.} }(a fourth generation of quarks) have been excluded by high precision measurements of Higgs production rates at the LHC
\cite{Anastasiou:2011qw,Anastasiou:2016cez}.
In the present study, we focus on new states of matter that are spin 1/2 particles and transform as triplets under color. However, unlike SM quarks, their left- and right-handed couplings have the same Electro-Weak (EW) quantum numbers\footnote{This a distinguishing characteristic of VLQs since their left and right-handed components transform identically under the SM electroweak gauge group $SU(2)_L\otimes U(1)_Y$, enabling the inclusion of mass terms in the Lagrangian without violating gauge invariance transformations.}, called Vector-Like Quarks (VLQs) \textcolor{black}{\cite{Benbrik:2015fyz,Arhrib:2016rlj,Aguilar-Saavedra:2013qpa,Angelescu:2015uiz,Aguilar-Saavedra:2009xmz,DeSimone:2012fs,Kanemura:2015mxa,Lavoura:1992np,Chen:2017hak,He:2022zjz,Cao:2022mif,Carvalho:2018jkq,Moretti:2016gkr,Prager:2017hnt,Prager:2017owg,Moretti:2017qby,Deandrea:2017rqp,Aguilar-Saavedra:2017giu}}. This class of models is well motivated since they show up in many composite Higgs models
\cite{Anastasiou:2009rv,Contino:2010rs,Panico:2015jxa,Agashe:2005dk,Gillioz:2012se,Benbrik:2022kpo}
as well as in various  little Higgs models \cite{Arkani-Hamed:2002ikv,Hubisz:2004ft,Han:2003wu}. Furthermore, VLQs are naturally present in various other extensions of the SM, including supersymmetric models \cite{Martin:2009bg}, models incorporating extra dimensions \cite{Kong:2010qd}, and they also emerge as extra particles within the enlarged representations of grand unified theories \cite{Kang:2007ib,Kilic:2010fs}. The existence of VLQs is also strongly motivated by their ability to address the mass instability of the Higgs boson caused by significant radiative corrections at high energy scales \cite{DeSimone:2012fs}. Indeed, a vector-like top quark partner has the potential to fulfill a similar function as the top quark's superpartner in supersymmetric models.\\ 

We will also consider an extension of the SM that includes both VLQs and an additional Higgs doublet \textcolor{black}{(as stated in reference \cite{Arhrib:2016rlj})}.\\
Several searches for VLQs have been carried out by the ATLAS and CMS collaborations, both from single production via an electroweak interaction and from pair production through the strong interaction. No significant deviation from the expected background of the SM has been observed, and exclusion limits have been set on the masses of the VLQs. {Both the ATLAS and CMS experiments have searched $T$ and $B$ production at $\sqrt{s} = 8$ TeV and $\sqrt{13}$ TeV, spanning practically all production and decay channels}. {With an integrated luminosity of 138 $fb^{-1}$,} CMS presented a search \cite{CMS:2022fck} for $T\bar{T}$ and $B\bar{B}$ pair production of VLQs, excluding masses up to 1.54 TeV (1.56 TeV) for vector-like T and B quarks, respectively, at a 95\% confidence level, depending on their branching ratio scenarios. For vector-like T quarks, masses below 1.48 TeV are excluded in any branching fraction scenario, i.e. decay to the final states tZ, bW, and th, where h is the SM Higgs boson. Analogous exclusion limits by ATLAS \cite{Vale:2018bpf} at $95\%$ confidence level have been set on the VLQs masses up to 1030 (1210) GeV for singlet (doublet) vector-like T quark and 1010 (1140) GeV for singlet (doublet) vector-like B quark. It is important to note that all these search limits are given with the assumption that VLQs decay only to third generation SM particles, and they can be lowered if the VLQs have a non-negligible coupling with the first and second generation SM quarks. Besides, if we consider 2HDM+VLQs, in this case, we will have the possibility for other new decay modes of VLQs to BSM Higgs bosons, such as the decay of vector-like T quark to $tH, tA$, and $bH^{\pm}$, and hence the previous exclusion limits can be further weakened down \cite{Arhrib:2016rlj} where $A$, $H$ and $H^\pm$ are extra Higgs bosons predicted by the 2HDM. {The High-Luminosity Large Hadron Collider (HL-LHC) will offer an unprecedented opportunity to explore extended Higgs sectors, such as the 2HDM, by revealing additional Higgs bosons or exotic Higgs decay modes. Reference \cite{Cepeda:2019klc} provides projected HL-LHC limits for searches of the 2HDM scalars. Similarly, VLQs may manifest themselves through distinctive signatures in collision events at the HL-LHC, for example according to \cite{Liu:2016jho}, a $5\sigma$ discovery sensitivity for the individual production of an up-type VLQ with a mass approximately at 1 TeV, even with relatively low couplings to SM quarks q, reaching as low as $V_{Tq} \approx 0.2$. Additionally, for a pair-produced down-type VLQ, the $5\sigma$ discovery reach is possible for masses around 730 GeV \cite{Paul:2020mul}.}

The oblique parameters $S$, $T$ and $U$ represent important tools at the TeV scale that indicate the level of violation of the custodial isospin symmetry. They contribute significantly to electroweak radiative corrections. It is very well known that in the SM, $S=T=U = 0$ at a reference point where the top quark and Higgs boson masses are fixed. Beyond the SM, these parameters are sensitive to new physics and hence measure their effects. The exact calculations of S, T and U oblique parameters are tedious and prone to error, with the help of public codes such as \texttt{FeynArts}, \texttt{FormCalc}, and \texttt{LoopTools}, an automated code has been introduced to improve the reliability of the calculations. In the current work, we present to the best of our knowledge, the first exact calculations of electroweak oblique parameters $S$ and $T$ in the 2HDM type II in the presence of singlets, doublets, and triplets vector-like quarks scenarios in the context of CP conserving case. Our results are given using the well-known scalar Passarino-Veltman functions \cite{Passarino:1978jh}. We examine in detail the conditions that may affect the oblique parameters in our model, then study the impact of the new CDF result on the extra new quarks. The 2HDM+VLQs is a good candidate for analyzing the oblique parameters formalism because the model has new quarks and their couplings to gauge bosons are model-dependent. { We highlight that the inclusion of VLQs within the 2HDM framework enables mass degeneracy among the charged Higgs boson $H^\pm$ and the two heavy neutral Higgs bosons $A$ and $H$. This mass degeneracy is achievable through the cancellation of oblique parameter contributions from both the 2HDM and VLQ states, a scenario that the standard 2HDM alone typically disfavors \cite{Lu:2022bgw}. Moreover, we demonstrate that assuming $U=0$, the $T$ parameter, or $T_{\mathrm{2HDM+VLQ}}$, is found to be positive \cite{Asadi:2022xiy} in all 2HDM+VLQ scenarios.}\\
This paper is organized as follows: In Section~\ref{modelsdesc}, we give a brief review of VLQs and
2HDM in terms of particle content and their mass. We also present how the W boson mass can be parameterized in terms of the oblique parameters $S$, $T$ and $U$. In Section~\ref{numanalysis}, we
show our numerical analysis within the 2HDM type II augmented by VLQs. We conclude in Section~\ref{concl}. Finally, in the appendix we present the contributions of the vector-like quarks to $S$, $T$ and $U$ parameters and we give the relevant  couplings  to our study and explain the CDF-II W-boson mass anomaly within the SM extended with VLQs.

\section{Models description}
\label{modelsdesc}
\subsection{VLQs}
In this section, we introduce our parametrization and notation for VLQs \footnote{In this work we review both the SM and the 2HDM with VLQs. More detailed description can be found in Ref. \cite{Aguilar-Saavedra:2013qpa}.}. As low energy physics measurements severely limit the mixing of VLQs with the first and second generations, we only consider the scenario where the VLQs couple to the third-generation SM quarks. There are seven distinct possibilities of Vector-like $SU(2)_L$ multiplets with renormalizable couplings that are allowed to mix through Yukawa couplings with the SM quarks, i.e we have two electroweak singlets, three electroweak doublets and two electroweak triplets as indicated below:
\begin{align}
	 & T_{L,R}^0 \,, \quad B_{L,R}^0                                          &  & \text{(singlets)} \,, \notag \\
	 & (T^0\,B^0)_{L,R} \,, \quad (X^0\,T^0)_{L,R} \,, \quad (B^0\,Y^0)_{L,R} &  & \text{(doublets)} \,, \notag \\
	 & (X^0\,T^0\,B^0)_{L,R} \,, \quad (T^0\,B^0\,Y^0)_{L,R}                  &  & \text{(triplets)} \,.
\end{align}
Where the heavy quarks have the following electric charges: $Q_T=\frac{2}{3}, Q_B=-\frac{1}{3}, Q_X=\frac{5}{3}$, and $Q_Y=-\frac{4}{3}$.
In the Higgs basis, the Yukawa Lagrangian may be expressed  as:
\begin{equation}
	-\mathcal{L} \,\, \supset  \,\, y^u \bar{Q}^0_L \tilde{H}_2 U^0_R +  y^d \bar{Q}^0_L H_1 D^0_R + M^0_u \bar{U}^0_L U^0_R  + M^0_d \bar{D}^0_L D^0_R + \rm {h.c}.
\end{equation}
where, $U_R\equiv(u_R, c_R, t_R, T_R)$ and $D_R\equiv(d_R, s_R, b_R, B_R)$ while $y^{u,d}$ are 3$\times$4 Yukawa matrices.
In the scenario where the new heavy quarks couple to third generation, the physical mass eigenstates fields $(t,T)$ and $(b,B)$ can be obtained  by rotating the gauge eigenstates through four unitary mixing matrices $U_{L,R}^u$ and $ U_{L,R}^d$, which  are parametrized as follows:
\begin{equation}
	\left(\! \begin{array}{c} t_{L,R} \\ T_{L,R} \end{array} \!\right) =
	U_{L,R}^u \left(\! \begin{array}{c} t^0_{L,R} \\ T^0_{L,R} \end{array} \!\right)
	= \left(\! \begin{array}{cc} \cos \theta_{L,R}^u & -\sin \theta_{L,R}^u e^{i \phi_u} \\ \sin \theta_{L,R}^u e^{-i \phi_u} & \cos \theta_{L,R}^u \end{array}
	\!\right)
	\left(\! \begin{array}{c} t^0_{L,R} \\ T^0_{L,R} \end{array} \!\right) \,.
	\label{ec:mixu}
\end{equation}
\begin{equation}
	\left(\! \begin{array}{c} b_{L,R} \\ B_{L,R} \end{array} \!\right)
	= U_{L,R}^d \left(\! \begin{array}{c} b^0_{L,R} \\ B^0_{L,R} \end{array} \!\right)
	= \left(\! \begin{array}{cc} \cos \theta_{L,R}^d & -\sin \theta_{L,R}^d e^{i \phi_d} \\ \sin \theta_{L,R}^d e^{-i \phi_d} & \cos \theta_{L,R}^d \end{array}
	\!\right)
	\left(\! \begin{array}{c} b^0_{L,R} \\ B^0_{L,R} \end{array} \!\right) \,.
	\label{ec:mixd}
\end{equation}
In the weak eigenstate basis, the mass terms for the top and bottom quarks and the VLQs can be extracted from
\begin{eqnarray}
	\mathcal{L}_\text{mass} & = & - \left(\! \begin{array}{cc} \bar t_L^0 & \bar T_L^0 \end{array} \!\right)
	\left(\! \begin{array}{cc} y_{33}^u \frac{v}{\sqrt 2} & y_{34}^u \frac{v}{\sqrt 2} \\
             y_{43}^u \frac{v}{\sqrt 2}      & M^0\end{array} \!\right)
	\left(\! \begin{array}{c} t^0_R \\ T^0_R \end{array}
	\!\right) \notag \\
	& & - \left(\! \begin{array}{cc} \bar b_L^0 & \bar B_L^0 \end{array} \!\right)
	\left(\! \begin{array}{cc} y_{33}^d \frac{v}{\sqrt 2} & y_{34}^d \frac{v}{\sqrt 2} \\
             y_{43}^d \frac{v}{\sqrt 2}      & M^0\end{array} \!\right)
	\left(\! \begin{array}{c} b^0_R \\ B^0_R \end{array}
	\!\right) +\text{h.c.},
	\label{ec:Lmass}
\end{eqnarray}
Where $y_{ij}$ are the Yukawa couplings and $M^0$ is a bare mass term. For singlets and triplets, $y_{43} = 0$, while for doublets, $y_{34} = 0$. The unitary matrices will be specified in the case where the mass matrices are diagonal in the mass eigenstate basis.
\begin{equation}
	U_L \, \mathcal{M} \, (U_R)^\dagger = \mathcal{M}^q_\text{diag} \,,
	\label{ec:diag}
\end{equation}
with  $\mathcal{M}$ are the mass matrices in Eq.~(\ref{ec:Lmass}) and $\mathcal{M}^q_\text{diag}$ are the diagonalized matrices given by.\\
\begin{equation}
	\mathcal{M}^t_\text{diag}=\left(\! \begin{array}{cc} m_t & 0 \\ 0 & m_T \end{array}
	\!\right) \quad \text{and}\quad \mathcal{M}^b_\text{diag}=\left(\! \begin{array}{cc} m_b & 0 \\0 & m_B \end{array}\!\right)
\end{equation}
Note that the two mixing angles $\theta_L$ and $\theta_R$ are not independent parameters. By using Eq.~(\ref{ec:diag}), and depending on the representation of the VLQs, one can find that:
\begin{eqnarray}
	\tan \theta_R^q & = & \frac{m_q}{m_Q} \tan \theta_L^q \quad \text{(singlets, triplets)} \,. \notag \\
	\tan \theta_L^q & = & \frac{m_q}{m_Q} \tan \theta_R^q \quad \text{(doublets)} \,.
	\label{ec:rel-angle1}
\end{eqnarray}
with $(q,m_q,m_Q) = (u,m_t,m_T)$ and $(d,m_b,m_B)$.\\
Additionally, there are relationships between the mass eigenstates and angles that depend on the representation of the VLQs \cite{Aguilar-Saavedra:2013qpa}
\begin{eqnarray}
	{Doublets:}
	&\mathrm{TB}: &m_T^2 (c_R^{u})^{2}+m_t^2 (s_R^{u})^{2}=m_B^2 (c_R^{d})^{2}+m_b^2 (s_R^{d})^{2}.
	\nonumber \\
	&\mathrm{XT}: &m_X^2=m_T^2 (c_R^u)^{2}+m_t^2 (s_R^{u})^{2}.\nonumber \\
	&\mathrm{BY}:&m_Y^2=m_B^2 (c_R^d)^{2}+m_b^2 (s_R^d)^{2}.
\end{eqnarray}
\begin{eqnarray}
	{Triplets:}& \mathrm{XTB}: & m_X^2=m_T^2 (c_L^u)^{2}+m_t^2 (s_L^u)^{2}.\nonumber \\
	&& \phantom{m_X^2} =m_B^2 (c_L^d)^{2} + m_b^2(s_L^d)^{2}.\end{eqnarray}
From the above equations, it is easy to derive the following relationships:
\begin{eqnarray}
	&&{m_B^2} =\frac{1}{2}\sin^2\left(2\theta^u_L\right)\frac{(m_T^2-m_t^2)^2}{m_X^2-m_b^2}+m_X^2.\nonumber \\
	&&\sin(2\theta_L^d)= \sqrt{2}\frac{m_T^2-m_t^2}{(m_B^2-m_b^2)}\sin(2\theta_L^u).\end{eqnarray}
Similarly for the TBY triplet, we have:
\begin{eqnarray}
	& m_Y^2=m_B ^2 (c_L^d)^2+m_b^2 (s_L^d)^{2}.\nonumber \\
	&\phantom{m_Y^2}=m_T^2 (c_L^u)^{2}+m_t^2 (s_L^u)^{2}. \end{eqnarray}
We may therefore deduce the following relations:
\begin{eqnarray}
	&&{m_B^2}=\frac{1}{8}\sin^2\left(2\theta^u_L\right)\frac{(m_T^2-m_t^2)^2}{m_Y^2-m_b^2}+m_Y^2.\nonumber\\
	&&\sin(2\theta_L^d)= \frac{m_T^2-m_t^2}{ \sqrt{2} (m_B^2-m_b^2)}\sin(2\theta_L^u).
\end{eqnarray}
where the mixing between the top-quark and down-quark vector-like fermions  is described by
$c_{L,R}^{u,d}=\cos\,\theta_{L,R}^{u,d}$ and $s_{L,R}^{u,d}=\sin\, \theta_{L,R}^{u,d}$.\\
It can be easily inferred that all representations of the VLQs can be characterized by only two parameters, except for the TB top partner doublet model which requires three parameters. In our numerical analysis, the following input parameters will be used:

\begin{table}[H]
	\setlength{\tabcolsep}{8pt}
	\renewcommand{\arraystretch}{1.75}
	\begin{adjustbox}{max width=\textwidth}
		\begin{tabular}{l|ccccccccccc}
			\hline\hline
			\multirow{2}{*}{Representations} & \multicolumn{2}{c}{Singlets} &              &  & \multicolumn{3}{c}{Doublets} &                   &             & \multicolumn{2}{c}{Triplets}                                       \\\cline{2-12}
			                                 & T                        & B        &  &                              & TB         & XT    & BY                     &  &  & XTB   & TBY   \\\hline
			Parameters                       & $m_T, s_L^u$                 & $m_B, s_L^d$ &  &                              & $m_T,s_R^u,s_R^d$ & $m_T,s_R^u$ & $m_B,s_R^d$                   &  &  & $ m_T, s_L^u$ & $ m_T, s_L^u$ \\\hline\hline
		\end{tabular}
	\end{adjustbox}
\end{table}
\subsection{2HDM}
One of the simplest extensions of the SM's Higgs sector is the two Higgs doublet model (2HDM) which consist
of adding a new Higgs doublet to the existing one in the SM.
The most general scalar potential which is invariant under $SU_L(2)\times U_Y(1)$
and  satisfy a discrete $Z_2$ symmetry $\phi_{1} \rightarrow +\phi_{1}$ and $\phi_2 \rightarrow -\phi_2$,
can be written as \cite{Branco:2011iw}
\begin{eqnarray}
	V(\phi_1,\phi_2) &=& m_{11}^2(\phi_1^\dagger\phi_1) +
	m_{22}^2(\phi_2^\dagger\phi_2) -
	[ m_{12}^2(\phi_1^\dagger\phi_2)+\text{h.c.}] \nonumber\\
	&+& \frac12\lambda_1(\phi_1^\dagger\phi_1)^2 +
	\frac12\lambda_2(\phi_2^\dagger\phi_2)^2 +
	\lambda_3(\phi_1^\dagger\phi_1)(\phi_2^\dagger\phi_2)~\nonumber\\
	&+&\lambda_4(\phi_1^\dagger\phi_2)(\phi_2^\dagger\phi_1)
	+\frac12\left[\lambda_5(\phi_1^\dagger\phi_2)^2 +\rm{h.c.}\right],
	\label{CTHDMpot}
\end{eqnarray}
where $\lambda_{1,2,3,4,5}$ as well as $m_{11,22,12}^2$  are chosen to be real  ensuring  that the potential is CP conserving.
If both Higgses $\phi_1$ and $\phi_2$  interact with all SM fermions, like in the SM,
we end up with Flavour Changing Neutral Currents (FCNCs) at the tree-level in the Yukawa sector.
For this purpose, a discrete $Z_2$ symmetry has been introduced to prevent the occurrence of large tree-level FCNCs \cite{Glashow:1976nt,Branco:2011iw}. Moreover, based on the $Z_2$ symmetry assignment, there are four possible types of Yukawa sector. In this study, we concentrate on 2HDM Type II, where one doublet couples to down-type quarks and leptons, while the other couples to up-type quarks.

After EWSB takes place, we are left with five physical Higgs bosons: two CP-even $h$ and $H$ with $m_h < m_H$, one CP-odd $A$, and a pair of charged Higgs $H^\pm$. In our study, we assume that the light CP-even Higgs $h$ will be identified as the Higgs boson  observed at the LHC, and we assume $m_h \approx 125$ GeV. The other parameters of the model are:
\begin{itemize}
	\item $\tan\beta=v_2/v_1$ which is the ratio of the vevs.
	\item $\sin(\beta-\alpha)$ , where $\alpha$ is the mixing angle between the CP-even components of the  Higgs fields
	\item $m_{12}^2$: $Z_2$ soft breaking term
	\item $m_{H, A, H^\pm}$: the physical masses of the extra Higgs bosons.
\end{itemize}

\subsection{$W$ mass and ($S$, $T$, $U$) parameters}
The $W$-boson mass $M_W$  is one of the significant EWPOs due to its precise experimental measurements. It  can be calculated as a function of the oblique parameters $S$, $T$ and $U$ \cite{Peskin:1991sw,Peskin:1990zt}. The oblique parameters in 2HDM differ from those in the SM because of the additional contributions from the Higgs boson loops and the modified couplings of the SM-type Higgs bosons.

The W-boson mass can be calculated from the oblique parameters, which can be expressed as follows \cite{Grimus:2008nb}:
\begin{eqnarray}\label{eq4}
	\left.M_W^2=M_W^2\right\rvert^{SM}{}\left[1+ \frac{\alpha(M_Z) }{\left(c_W^2-s_W^2\right)}\left(-\frac{1}{2}S+c_W^2T+\frac{c_W^2-s_W^2}{4s_W^2}U\right)\right].
\end{eqnarray}
Where $c_W=\cos \theta_W$ and $s_w=\sin \theta_W$,  $\theta_W$ is Weinberg angle.
When compared to the contributions from $S$ and $T$, the contribution from the oblique parameter U is expected to be negligible in many new physics models. Hence, we present the results for $S$ and $T$ assuming $U = 0$.\\
Additionally, it is possible to assess the impact of changes in the $W$-boson mass on the effective weak mixing angle $\sin^2\theta_{\mathrm{eff}}$, which is calculated as described in \cite{Biekotter:2022abc}:

\begin{eqnarray}
\sin^2\theta_{eff} = \left.\sin^2\theta_{\mathrm{eff}}\right\rvert^{SM}{} -\alpha\frac{c_W^2s_W^2}{c_W^2-s_W^2}(T - \frac{s_W^2}{c_W^2}S + \frac{1}{2}U)
\end{eqnarray}

with the SM values $\left.M_W\right\rvert^{SM}{}=80.357$ GeV and $\left.\sin^2\theta_{\mathrm{eff}}\right\rvert^{SM}{}= 0.231532$.\\
In the framework of 2HDM with VLQs, the oblique parameters $S$ and $T$ have two separate contributions: one from the VLQs and the other from the Higgs bosons in the 2HDM:
\begin{eqnarray}
	&S_{SM+VLQs}= S_{SM}+S_{VLQs}, \quad \ T_{SM + VLQs}= T_{SM}+T_{VLQs} \nonumber \\
	& S_{2HDM+VLQs}= S_{2HDM}+S_{VLQs}, \quad \ T_{2HDM+VLQs}= T_{2HDM}+T_{VLQs}
\end{eqnarray}
The detailed analytical formulations for $S_{2HDM,VLQs}$ and $T_{2HDM,VLQs}$ in terms of the Passarino-Veltman functions can be found in Appendix \ref{app}. With the help of LoopTools~\cite{Hahn:1999mt,Hahn:2010zi}, the numerical evaluations of the scalar functions are carried out.

We compared our analytical expression for the contributions of 2HDM to $S$ and $T$ to those in  \cite{Kanemura:2015mxa} and
\cite{Eriksson:2009ws}  and found good agreement.
In the case of singlet and doublet representations, the VLQs contributions
can be computed with the use of Ref.~\cite{Lavoura:1992np}. This has been used
for a number of VLQ models, as mentioned in Ref.~\cite{Chen:2017hak}.
We used dimensional regularization to compute the $S$ and $T$ parameters for each of the different VLQs under examination using  \texttt{FeynArts} and \texttt{FormCalc} packages \cite{Hahn:2000kx,Hahn:2001rv}. We made sure that our calculations for $S$ and $T$ were UV finite and independent of the renormalization scale, both analytically and numerically. In the case of the $T$ parameter, we compared our findings to those in Refs.\cite{Lavoura:1992np,Chen:2017hak} and found good agreement for the singlet and doublet cases, but our results diverged from those in Ref. \cite{Chen:2017hak} for the triplets XTB and TBY. However, our findings for the $S$ parameter also differed from those in Ref. \cite{Chen:2017hak}. In reality, Ref.\cite{Chen:2017hak} merely applied the findings from Ref.\cite{Lavoura:1992np} for singlet and doublet situations to the triplet case. The author of Ref.\cite{He:2022zjz} used the method described in Ref.~\cite{Lavoura:1992np} to derive the correct equation for $S$ and $T$ in the case of an XTB triplet model. We calculated both $S$ and $T$ parameters as functions of the standard Passarino-Veltman functions  for the XTB and TBY triplet models and took into account the gauge bosons external momentum. This is a major point of contention with Ref.\cite{Chen:2017hak}, which adopts a crude approximation that ignores the gauge bosons' external momentum for the $S$ parameter. We further cross-checked our results with new computations in Refs. \cite{Cao:2022mif,He:2022zjz} and found good agreement for the $T$ parameter. However, we should note that for the $S$ parameter, we used the full analytic calculation and did not omit any external momentum, as is typically the case in Ref \cite{Cao:2022mif}.


\section{Numerical Results}
\label{numanalysis}
Before presenting our results, we will describe how the theoretical requirements and experimental measurements were used to constrain the parameter space of the 2HDM + VLQs models.\\
From a theoretical perspective, the perturbativity, vacuum stability, and unitarity constraints are enforced as the following:
\begin{itemize}
	\item \textbf{Unitarity:} A variety of scattering processes  are required to be unitary ~\cite{Kanemura:1993hm,Akeroyd:2000wc,Arhrib:2000is}.

	\item \textbf{Perturbativity:} The quartic couplings of the scalar potential should obey the following conditions:$|\lambda_i|<8\pi$ for each $i=1,..5$ ~\cite{Branco:2011iw}.

	\item \textbf{Vacuum stability:} The scalar potential must be bounded from below and positive in any direction of the fields $\Phi_i$. As a consequence, the parameter space must satisfy the following conditions~\cite{Barroso:2013awa,Deshpande:1977rw}:
	      \begin{align}
		      \lambda_1 > 0,\quad\lambda_2>0, \quad\lambda_3>-\sqrt{\lambda_1\lambda_2} ,\nonumber \\ \lambda_3+\lambda_4-|\lambda_5|>-\sqrt{\lambda_1\lambda_2}.\hspace{0.5cm}
	      \end{align}
	      The experimental constraints used in our analysis include:

	\item {\bf SM-like Higgs boson discovery}: Compatibility of the SM-like scalar with
	      the observed Higgs boson. We require that the relevant quantities, calculated with \texttt{HiggsSignals-2.6.1} \cite{Bechtle:2008jh,Bechtle:2011sb,Bechtle:2013wla,Bechtle:2014ewa,Bechtle:2015pma,Bechtle:2020pkv,Bechtle:2013xfa,Bechtle:2020uwn}, satisfy measurements  at 95\% C.L.{ For further details on the measurements implemented in this version, we refer to Tabs.~\ref{tab:ATLAS} and \ref{tab:CMS}.\\\vspace{2cm}

	      \begin{table}[H]
	      	\centering
	      	\footnotesize
	      	\renewcommand{\arraystretch}{1.05}
	      	\setlength{\tabcolsep}{5pt}
	      	\vspace{-1cm}
	      	\begin{tabularx}{\textwidth}{X c c  }
	      		\toprule\toprule
	      		Channel                                                                             & Luminosity [$\text{fb}^{-1}$]   & Ref.                                                 \\
	      		\midrule\midrule
	      		VBF, $H\to b\bar{b}$                                                                & $30.6$                        & \cite{ATLAS:2018jvf}  \\
	      		$t\bar{t}H$, $H\to b\bar{b}$                                             & $36.1$                        & \cite{ATLAS:2017fak}                                     \\
	      		
	      		$t\bar{t}H$                                               & $79.9$                         & \cite{ATLAS:2019nvo} \\
	      		
	      		$gg\to H$, $H\to W^+W^-$                                                            & $36.1$                         & \cite{ATLAS:2018xbv} \\
	      		VBF, $H\to W^+W^-$                                                                  & $36.1$                       & \cite{ATLAS:2018xbv} \\
	      		VBF, $H\to ZZ$                                                      & $139.0$                            & \cite{ATLAS:2020rej}  \\
	      		
	      		$gg\to H$, $H\to ZZ$                                               & $139.0$                         & \cite{ATLAS:2020rej}  \\
	      		
	      		$t\bar{t}H$, $H\to ZZ$                                                              & $139.0$                      & \cite{ATLAS:2019jst} \\
	      		$gg\to H$, $H\to \gamma\gamma$                                              & $139.0$                        & \cite{ATLAS:2019jst} \\
	      		
	      		$t\bar{t}H$, $H\to \gamma\gamma$                                                    & $139.0$                       & \cite{ATLAS:2019jst} \\
	      		VBF, $H\to \tau^+\tau^-$                                                            & $36.1$                      & \cite{ATLAS:2018ynr} \\
	      		$gg\to H$, $H\to \tau^+\tau^-$                                                      & $36.1$                        & \cite{ATLAS:2018ynr} \\
	      		$W/ZH$, $H\to W^+W^-$                                                                 & $36.1$                        & \cite{ATLAS:2019vrd} \\
	      		
	      		$W/ZH$, $H\to b\bar{b}$                       & $139.0$                            & \cite{ATLAS:2020fcp}   \\
	      		\bottomrule\bottomrule
	      	\end{tabularx}
	      	\caption{ATLAS Limits included in \texttt{HiggsSignals-5}}%
	      	\label{tab:ATLAS}
	      \end{table}
	      \vspace{2cm}
	      \begin{table}[H]
	      	\centering
	      	\footnotesize
	      	\renewcommand{\arraystretch}{1.05}
	      	\setlength{\tabcolsep}{5pt}
	      	\vspace{-1cm}
	      	\begin{tabularx}{\textwidth}{X c c }
	      		\toprule\toprule
	      		Channel                                                     & Luminosity [$\text{fb}^{-1}$] & Ref.                                               \\
	      		\midrule\midrule
	      		$pp\to H$, $H\to \mu^+\mu^-$                                & $35.9$                            & \cite{CMS:2018nak}\\
	      		$WH$, $H\to b\bar{b}$                                       & $35.9$                            & \cite{CMS:2017odg}\\
	      		$pp\to H$, $H\to b\bar{b}$                        & $35.9$                           & \cite{CMS:2017bcq}\\
	      		$t\bar{t}H$, $H\to b\bar{b}$                       & $35.9 \oplus 41.5$            & \cite{CMS:2018hnq,CMS:2019lcn}\\
	      		
	      		$t\bar{t}H$, $H\to b\bar{b}$                   & $41.5$                        & \cite{CMS:2018hnq}\\
	      		$t\bar{t}H$                 & $35.9 \oplus 41.5$            & \cite{CMS:2018fdh,CMS:2018dmv}\\

	      		$gg\to H$, $H\to W^+W^-$                            & $137.0$                      & \cite{CMS:2020dvg} \\
	      		
	      		VBF, $H\to ZZ$                                              & $137.1$                        & \cite{CMS:2019chr}\\
	      		$gg/b\bar{b}\to H$, $H\to ZZ$                               & $137.1$                        & \cite{CMS:2019chr}\\
	      		$VH$, $H\to ZZ$                                             & $137.1$                      & \cite{CMS:2019chr}\\
	      		$t\bar{t}H,tH$, $H\to ZZ$                                   & $137.1$                        & \cite{CMS:2019chr}\\
	      		$gg\to H$, $H\to \gamma\gamma$                      & $77.4$                       & \cite{CMS:2019xnv} \\
	      		
	      		VBF, $H\to \gamma\gamma$                                    & $77.4$                       &\cite{CMS:2019xnv} \\
	      		$t\bar{t}H$, $H\to \gamma\gamma$                            & $137.0$                         & \cite{CMS:2020cga} \\
	      		VBF, $H\to \tau^+\tau^-$                                    & $77.4$                         &\cite{CMS:2019pyn} \\
	      		$gg\to H$, $H\to \tau^+\tau^-$                      & $77.4$                         &\cite{CMS:2019pyn} \\
	      		
	      		\bottomrule\bottomrule
	      	\end{tabularx}
	      	\caption{CMS Limits included in \texttt{HiggsSignals-2} .}
	      	\label{tab:CMS}
	      \end{table}}

	\item {\bf BSM Higgs boson exclusions}: Exclusion limits at 95\% C.L. from direct Higgs boson searches at LEP, Tevatron and LHC are enforced via \texttt{HiggsBounds-5.10.1}.  ~\cite{Bechtle:2008jh,Bechtle:2011sb,Bechtle:2013wla,Bechtle:2014ewa,Bechtle:2015pma,Bechtle:2020pkv,Bechtle:2013xfa,Bechtle:2020uwn} 
{The principal channels of investigation relevant to the Type-II 2HDM comprise:
	
		\begin{itemize}
			\item $p p \rightarrow A \rightarrow Z Z \rightarrow l^+ l^+ l^- l^-,\;l^+l^-qq,\;l^+l^- \nu \bar{\nu} $ \cite{CMS:2018amk}
			\item $b b \rightarrow A \rightarrow Z h \rightarrow l^+ l^- b b $ \cite{CMS:2019qcx}
			\item $p p \rightarrow H/A \rightarrow \tau^+ \tau^- $ \cite{ATLAS:2020zms}
			\item $p p \rightarrow H \rightarrow Z Z \rightarrow l^+l^-l^+l^-, l^+l^-qq,l^+l^-\nu\bar{\nu} $\cite{CMS:2018amk}
			\item $p p \rightarrow H \rightarrow V V $\cite{ATLAS:2018sbw}
			\item $p p \rightarrow A \rightarrow H Z \rightarrow b\bar{b}l^+l^- $\cite{ATLAS:2018oht}
			\item $p p \rightarrow H \rightarrow A Z \rightarrow b\bar{b}l^+l^- $\cite{ATLAS:2018oht}
			\item $ g g \rightarrow A \rightarrow h Z \rightarrow b\bar{b}l^+l^- $\cite{ATLAS:2020pgp}
					\item $p p \rightarrow tbH^+ \rightarrow tbtb$ \cite{ATLAS:2021upq}
		\end{itemize}}

\end{itemize}
We then performed a systematic scan over the following physical parameters using the public code \texttt{2HDMC-1.8.0}\cite{Eriksson:2009ws}\footnote{We have modified the code as to implement the analytical expressions of
	$S_{VLQs}$ and $T_{VLQs}$ given in the Appendix \ref{app}}.

\begin{itemize}
	\item The SM parameters : $m_t$ = 173.40 GeV, $m_b$ =4.2 GeV, $m_Z$=91.1876 GeV,
	      $m_h$=125 GeV and $\alpha$=1/137.036.
	\item The 2HDM parameters : $m_h$ = 125 GeV, $m_{H,A} \in [300,900]$ GeV,  $m_{H^\pm} \in [600,900]$ GeV, $\tan\beta \in [1,15]$ and $\sin(\beta-\alpha)$=1. Where we have assumed that the charged Higgs boson is heavier than 600 GeV in order to comply with $B_S\to X_s \gamma$ measurement.
	\item The VLQs masses : $m_{T}\in [700,2000]$ GeV and $m_{B}\in [700,2000]$ GeV.
	\item The VLQs mixing angles are chosen in the range [-0.5,0.5].
\end{itemize}
It should be noted that, in order to select points within the 2$\sigma$
ranges of the new CDF measurement, we only consider points that satisfy $\chi^2_{M_W^\mathrm{CDF-II}}\leq 4$ where:
\begin{eqnarray}
	\chi^2_{M_W^\mathrm{CDF-II}}=\frac{\left(M_W^\mathrm{2HDM+VLQ}-M_W^\mathrm{CDF-II}\right)^2}{\left(\Delta M_W^\mathrm{CDF-II}\right)^2}
\end{eqnarray}

\subsection{2HDM + VLQs}
In this subsection, we will illustrate results for the seven VLQ models studied in this analysis. We start with the singlets,
and then move to the doublets and finish with the triplets.
\subsubsection{2HDM+T }
\begin{figure}[t!]
	\centering

	\includegraphics[width=0.45\textwidth]{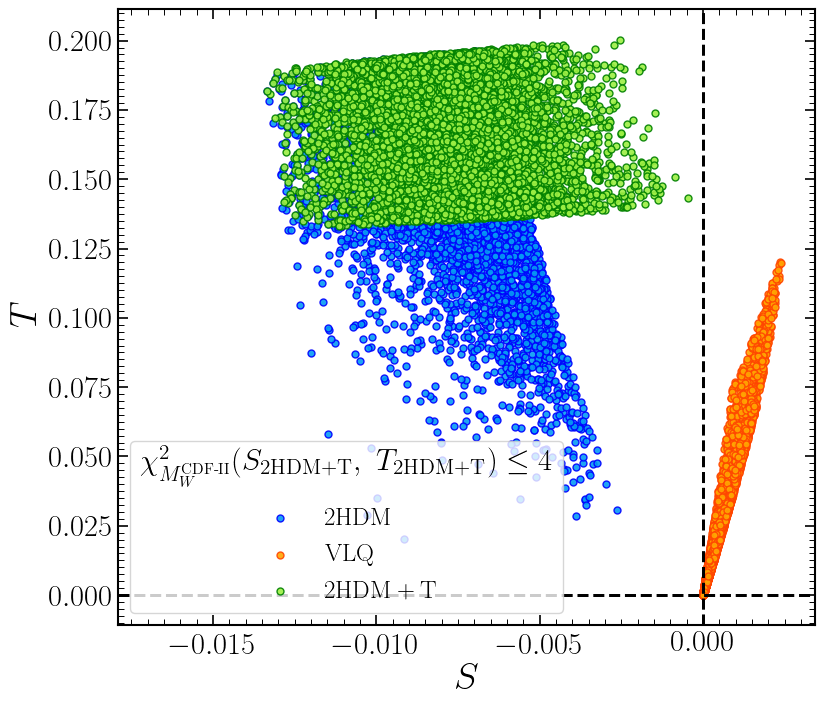}
	\includegraphics[width=0.49\textwidth]{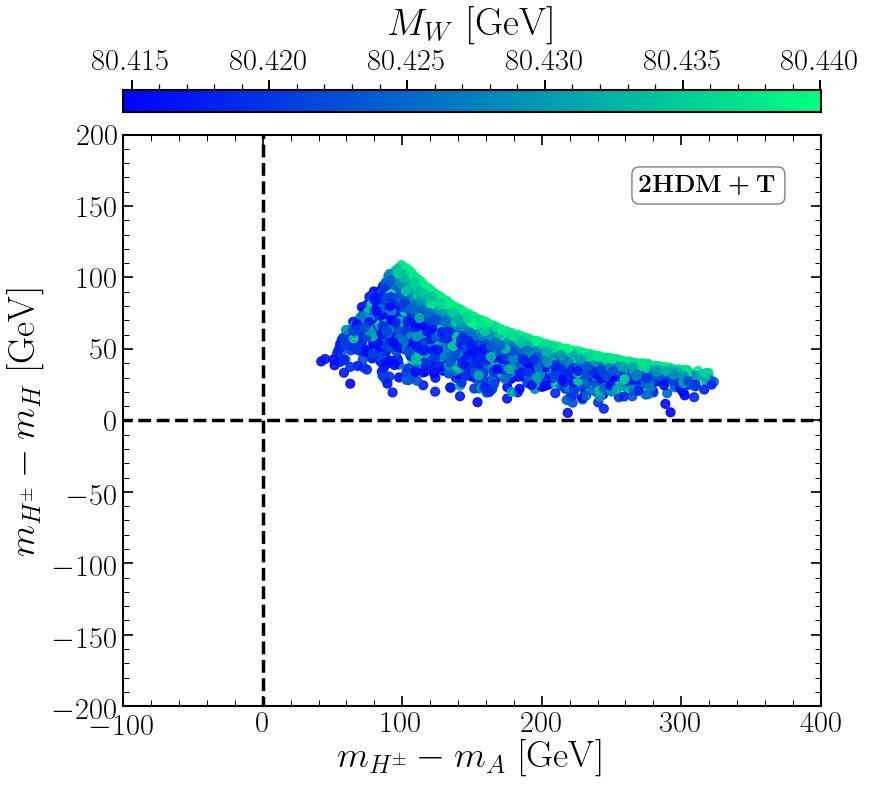}
	\caption{Left: Scan results of the 2HDM+T singlet in the $(S,~T)$ plane. The blue points indicate the 2HDM, while the VLQ are indicated by orange points. The combination of the 2HDM and VLQ is illustrated by green points. Right: the same data points are displayed in the $(m_{H^\pm}-m_A , m_{H^\pm}-m_H)$ plane, with the color code representing the value of $M_W$. }\label{figg1}
\end{figure}
In Fig.~\ref{figg1} (left panel), we present the results of our analysis in the $(S,T)$ plane, with each contribution displayed separately. The results from the two Higgs doublet model (2HDM) are represented in blue, the VLQ in orange, and the combination of the 2HDM and VLQ  are shown in  green. The plot illustrates that the $T$ parameter is consistently positive for both the 2HDM and VLQ. On the other hand, the $S$ parameter is positive in the VLQ, and negative in the 2HDM. Furthermore, it is noteworthy that the cancellation between the 2HDM and VLQ  leads to the $S$ parameter consistently assuming negative values in this representation.
In the right panel of Fig.~\ref{figg1}, we show the allowed parameter space in the $(m_{H^\pm}-m_A , m_{H^\pm}-m_H)$ plane within the 2HDM+T singlet  top partner model. The color coding indicates the value of the W boson mass. One can observe that the CDF $M_W$ measurement value can only be reached  in the case where $m_{H^\pm}>m_H,m_A$. Like in the 2HDM, the degenerate case $m_{H^\pm}=m_A=m_H$ is strongly disfavored here by the CDF-II data {as predicted in Refs. \cite{ Lu:2022bgw, Fan:2022dck, Song:2022xts, Bahl:2022xzi, Babu:2022pdn, Biekotter:2022abc, Han:2022juu, Heo:2022dey, Ahn:2022xax, Benbrik:2022dja, Abouabid:2022lpg, Arcadi:2022dmt, Ghorbani:2022vtv, Lee:2022gyf, Kim:2022xuo, Hessenberger:2022tcx, Atkinson:2022qnl}}. Moreover, it can also be seen that  the mass splitting between the charged Higgs boson and the CP-even Higgs $H$ can go up to 100 GeV, while the mass splitting between $H^\pm$ and the CP-odd Higgs boson $A$ can be at most of the order of 40 GeV to 320 GeV.\\
\begin{figure}[H]
	\centering
	\begin{minipage}{0.48\textwidth}
		\centering
		\includegraphics[width=\textwidth]{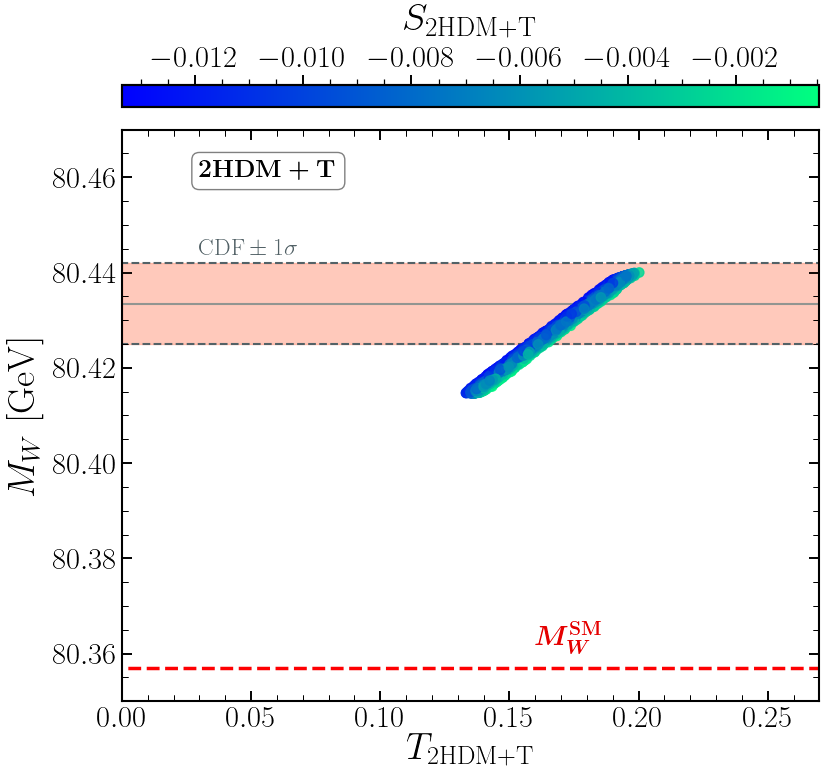}
	\end{minipage}\hspace{0.5cm}
	\begin{minipage}{0.48\textwidth}
		\centering
		\includegraphics[width=\textwidth]{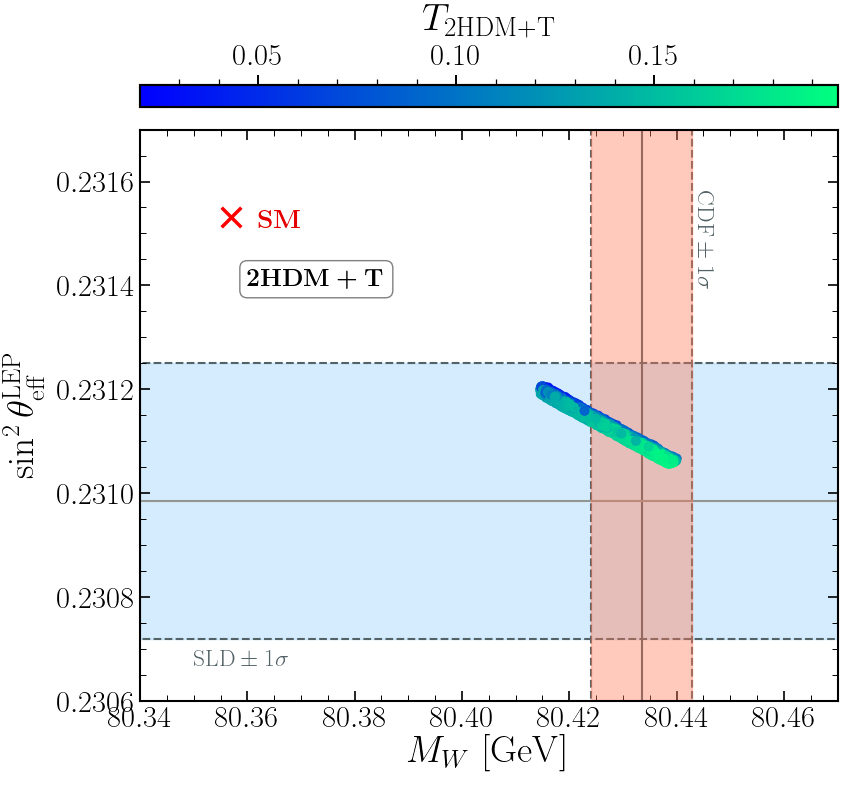}
	\end{minipage}
	\caption{Scan result of the 2HDM+T singlet in the $(T,M_W)$ plane left and $(M_W , \sin^2 \theta^{\mathrm{LEP}}_{\mathrm{eff}})$ right. The light red band indicates the $M_W$ value with the associated 1$\sigma$ uncertainty measured recently by the CDF collaboration, while the light green band  shows the result of $\sin^2\theta_{\mathrm{eff}}$ and its associated 1$\sigma$ uncertainty measured  by the SLD collaboration.}\label{fig1}
\end{figure}
In the left panel of Fig.~\ref{fig1}, we display the allowed parameters space of our scan with the predicted values for $M_W$ as a function of the $T$ parameter within the 2HDM+T singlet top partner model. The color coding of the points represents the values of the $S$ parameter. The red-dashed line shows the SM prediction for the W boson mass, and the light red region indicates the new CDF measurement at a level of $\pm 1\sigma$. It is evident from this plot that there is a positive correlation between $M_W^{2HDM+T}$ and the $T$ parameter as demanded by Eq.~(\ref{eq4}). The values of the $S$ parameter in the 2HDM+T singlet top partner model preferred by the CDF-II measurement of $M_W$ are in the range between -0.012 and -0.002. In the right panel of Fig.~\ref{fig1}, we display the parameter points of the 2HDM+T singlet in the $(M_W , \sin^2 \theta^{\mathrm{LEP}}_{\mathrm{eff}})$ plane. The color coding shows the size of $T$ parameter. The light green region indicates the measurements of $\sin^2\theta_{\mathrm{eff}}$ within $\pm 1\sigma$ confidence level reported by the SLD
collaboration, while the light red band shows the
W boson mass value at $\pm 1\sigma$ recently measured by the CDF. At first sight, one can see that  there is good agreement between the preferred parameter space of the CDF $M_W$ measurement and the value measured by the SLD experiment. Furthermore, It can be observed that there is an obvious strong negative correlation between $\sin^2 \theta^{\mathrm{LEP}}_{\mathrm{eff}}$ and $M_W$ since in the $T$ parameter dominant case, a positive contribution to the W boson mass prediction corresponds to a negative contribution to the effective weak mixing angle. We find that the parameter $T$ within the 2HDM+T singlet top partner model takes values in the range between 0.05 and 0.25.

\subsubsection{2HDM+B}
\begin{figure}[H]
	\centering
	\begin{minipage}{0.48\textwidth}
		\centering
		\includegraphics[width=\textwidth]{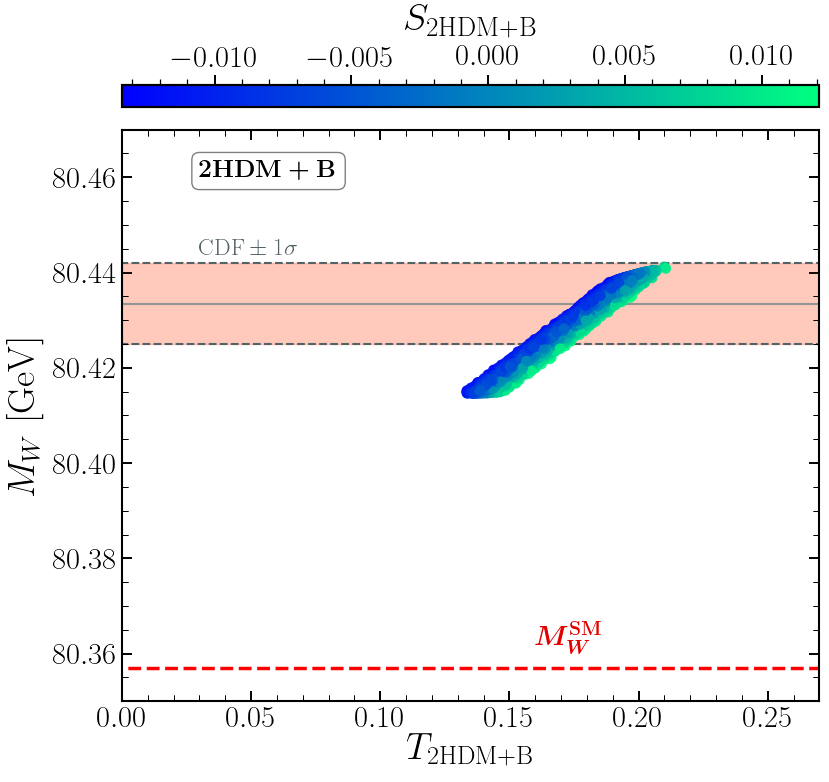}
	\end{minipage}\hspace{0.5cm}
	\begin{minipage}{0.48\textwidth}
		\centering
		\includegraphics[width=\textwidth]{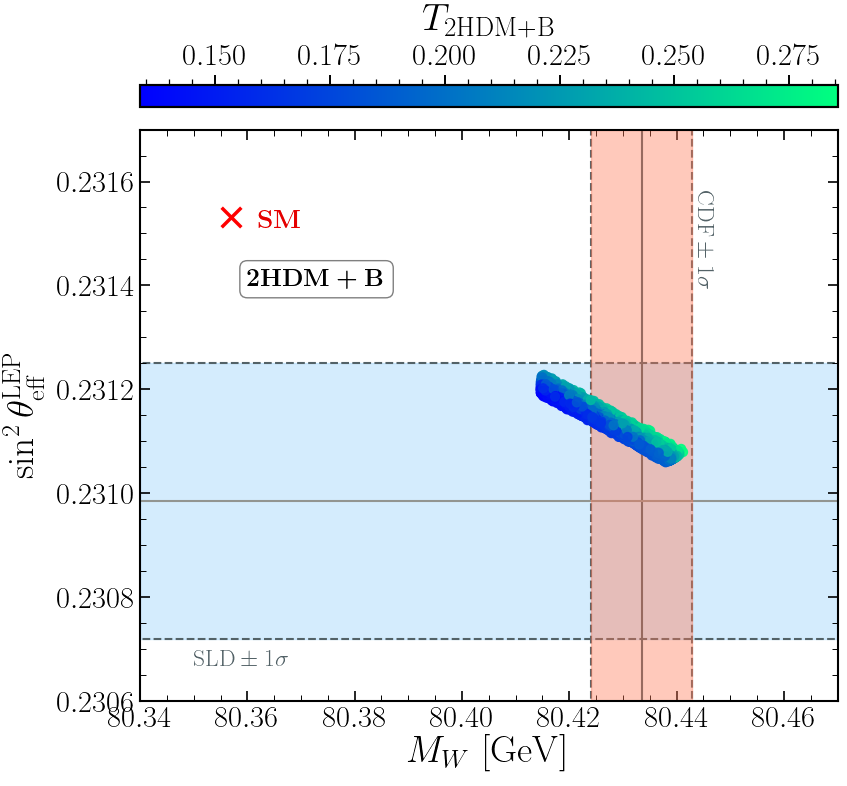}
	\end{minipage}
	\caption{Scan result of the 2HDM+B singlet   in the $(T,M_W)$ plane left and $(M_W , \sin^2 \theta^{\mathrm{LEP}}_{\mathrm{eff}})$ right. The light red band indicates the $M_W$ value with the associated 1$\sigma$ uncertainty measured recently by the CDF collaboration, while the light green band  shows the result of $\sin^2\theta_{\mathrm{eff}}$ and its associated 1$\sigma$ uncertainty measured  by the SLD collaboration.}\label{fig3}
\end{figure}
We depict the allowed parameter points in the left panel of Fig.~\ref{fig3}. The predicted values for the W boson mass, $M_W$, are displayed as a function of the $T$ parameter within the 2HDM+B singlet bottom partner model. The color coding of the points represents the values of the $S$ parameter and indicates that values of S between $-0.01$ and 0.01 are preferred by the CDF's measurement of $M_W$. In the right panel, we show the predictions for $M_W$ and $\sin^2 \theta^{\mathrm{LEP}}_{\mathrm{eff}}$, with the color code indicating the values of the $T$ parameter. We can see that the points preferred by the CDF's measurement of $M_W$ are consistent with the values of $\sin^2 \theta^{\mathrm{LEP}}_{\mathrm{eff}}$ measured by the SLD experiment. Moreover, the $T$ parameter in the 2HDM+B singlet bottom partner model is found to be in the range between 0.14 and 0.26.
\begin{figure}[H]
	\centering
	\includegraphics[width=0.47\textwidth]{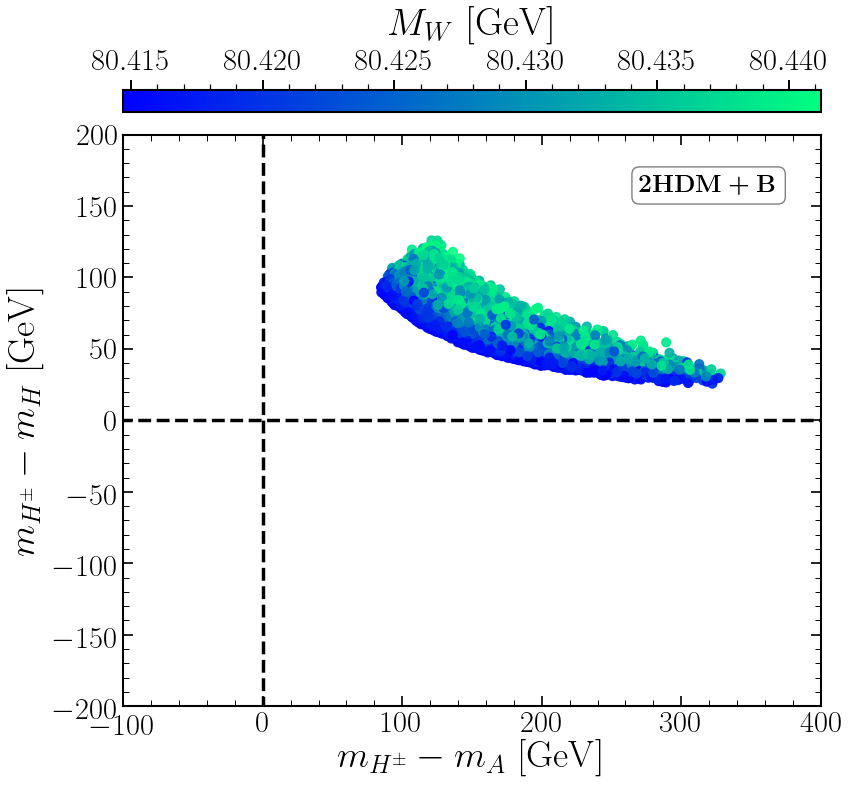}
	\caption{Scan results of the  2HDM + B singlet in the $(m_{H^\pm}-m_A , m_{H^\pm}-m_H)$ plane , with the colour code indicating  $M_W$ GeV.}\label{fig4}
\end{figure}
In Fig.~\ref{fig4}, we plot the allowed parameter space in the $(m_{H^\pm}-m_A , m_{H^\pm}-m_H)$ plane in the 2HDM+B singlet bottom partner model. The color code represents the value of the W boson mass. We can see that there is only one mass hierarchy and, similarly to the behavior shown in the 2HDM+T singlet top partner model, the CDF $M_W$ measurement value can be explained only when $m_{H^\pm}>m_H,m_A$. Furthermore, it can be observed that the mass splitting between the charged Higgs boson and the CP-even Higgs $H$ can be as high as 140 GeV, while the mass difference between $H^\pm$ and the CP-odd Higgs boson $A$ can reach 320 GeV.

\subsubsection{2HDM + TB}
\begin{figure}[H]
	\centering

	\includegraphics[width=0.45\textwidth]{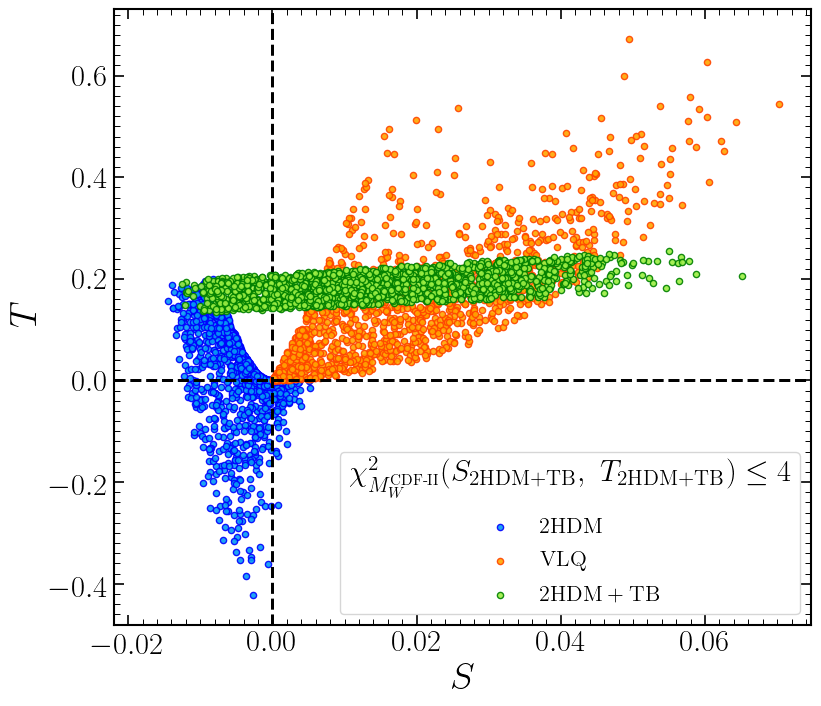}
	\includegraphics[width=0.49\textwidth]{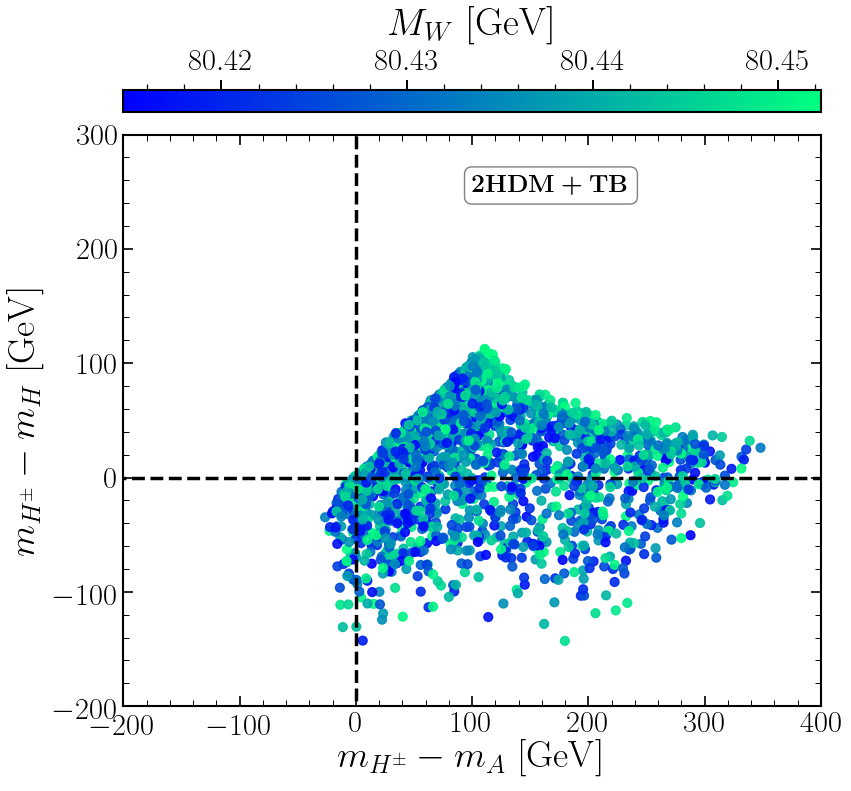}

	\caption{Left: Scan results of the 2HDM+TB doublet in the $(S,T)$ plane. The blue points indicate the 2HDM, while the VLQ are indicated by orange points. The combination of the 2HDM and VLQ is illustrated by green points. Right: the same data points are displayed in the $(m_{H^\pm}-m_A , m_{H^\pm}-m_H)$ plane, with the color code representing the value of $M_W$. }\label{figg4}
\end{figure}

In Fig.~\ref{figg4} (left panel), we present the results of our analysis of the 2HDM+TB doublet scenario in the $(S, T)$ plane. The contributions of the 2HDM and the VLQs are depicted separately in blue and orange, respectively, while the combination of both models is shown in green. The plot indicates that the $T$ parameter has both positive and negative values in the 2HDM, while it is solely positive in the doublet VLQs. Similarly, the $S$ parameter displays similar behavior, with positive values in the doublet VLQs and both positive and negative values in the 2HDM. Moreover, the cancellation between the 2HDM and VLQs in this scenario results in a consistent positive value for the $T$ parameter and a combination of positive and negative values for the $S$ parameter.

In the right panel of Fig.~\ref{figg4}, we present the allowed parameter space in the $(m_{H^\pm}-m_H)$ versus $(m_{H^\pm}-m_A)$  plane for the 2HDM+TB doublet. The color code represents the value of the W boson mass. It has been previously demonstrated in 2HDM studies \cite{Lu:2022bgw, Fan:2022dck, Song:2022xts, Bahl:2022xzi, Babu:2022pdn, Biekotter:2022abc, Han:2022juu, Heo:2022dey, Ahn:2022xax, Benbrik:2022dja, Abouabid:2022lpg, Arcadi:2022dmt, Ghorbani:2022vtv, Lee:2022gyf, Kim:2022xuo, Hessenberger:2022tcx, Atkinson:2022qnl}  that the charged Higgs boson mass $m_{H^\pm}$ must be non-degenerate with the masses of the two neutral Higgs boson  $m_H$ and $m_A$ in order to explain the current anomaly in the W boson mass. However, the degenerate scenario, where $m_{H^\pm}=m_A=m_H$, is still allowed in the 2HDM+TB top partner doublet model. Additionally, the CDF-II $M_W$ measurement value can also be explained in  the case where $(m_{H^\pm}-m_A)(m_{H^\pm}-m_H)>0$ and also it is interesting to observe that the case where $m_A<m_{H^\pm}<m_H$ is not excluded. Note that, in the 2HDM this mass splitting region is not allowed because the $T$ parameter is negative but here the presence of the top partner doublet $TB$ pushed it to the preferred positive values required by the $M_W$ mass anomaly. Furthermore, we see that the mass splitting between the charged Higgs boson and the CP-odd Higgs boson can reach up to 360 GeV, while the mass difference between the charged Higgs boson and the CP-even Higgs boson $H$ can be at most  100 GeV. The large mass gap between $m_{H^\pm}$ and $m_A$ can guarantee the dominant decay mode of  $H^\pm \rightarrow A W^\pm$  which is just gauge coupling and have no mixing suppression. Future experimental searches for this decay mode would be able to confirm or rule out the explanation of the recent $M_W$ mass anomaly within this model.
\begin{figure}[H]
	\centering
	\begin{minipage}{0.48\textwidth}
		\centering
		\includegraphics[width=\textwidth]{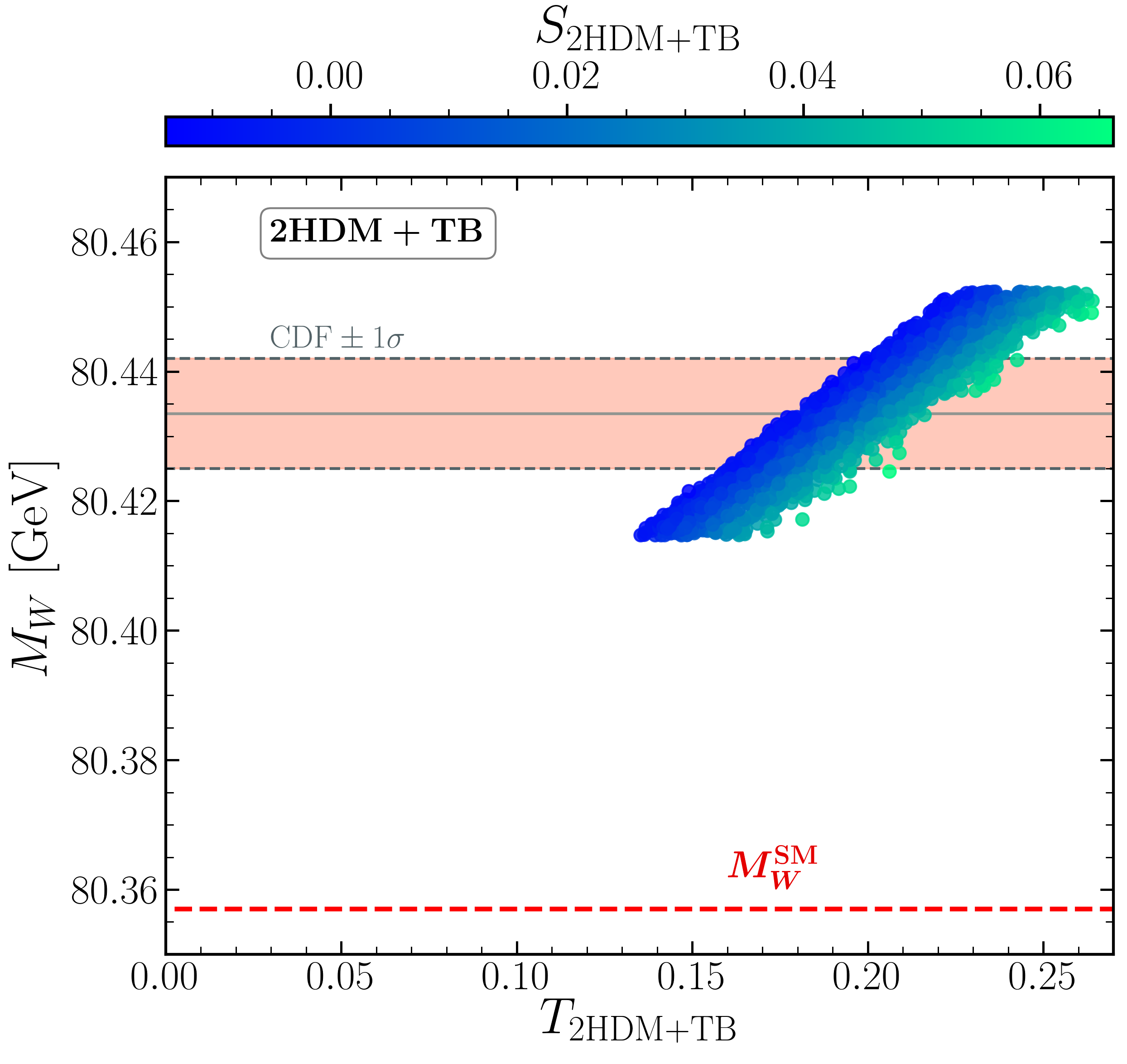}

	\end{minipage}\hspace{0.5cm}
	\begin{minipage}{0.48\textwidth}
		\centering
		\includegraphics[width=\textwidth]{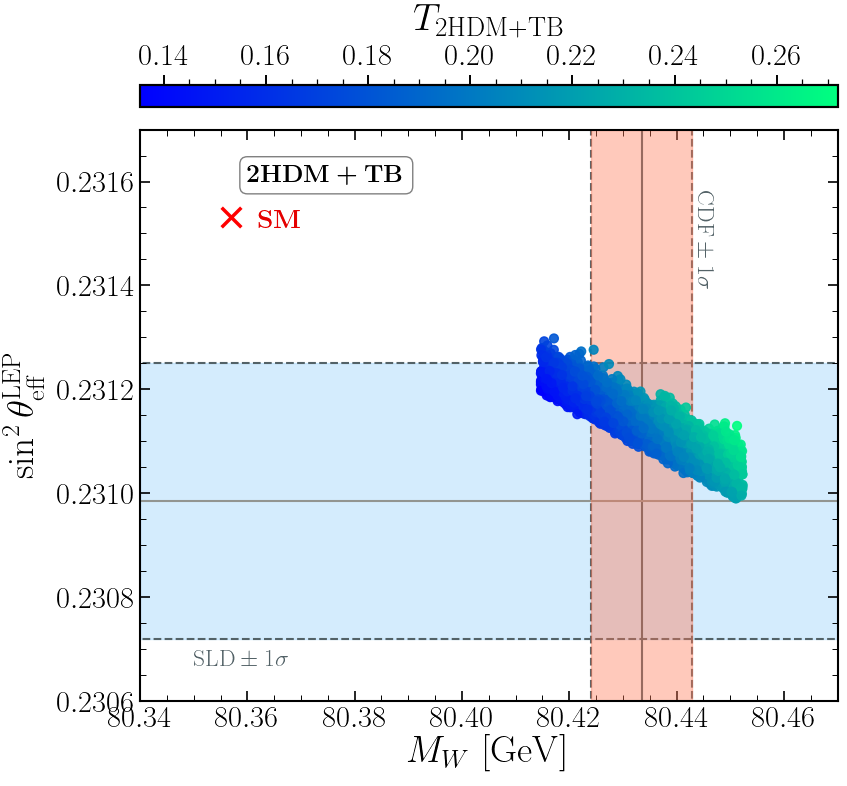}
	\end{minipage}
	\caption{Scan result of the 2HDM+TB doublet  in the $(T,M_W)$ plane left and $(M_W , \sin^2 \theta^{\mathrm{LEP}}_{\mathrm{eff}})$ right. The light red band indicates the $M_W$ value with the associated 1$\sigma$ uncertainty measured recently by the CDF collaboration, while the light green band  shows the result of $\sin^2\theta_{\mathrm{eff}}$ and its associated 1$\sigma$ uncertainty measured  by the SLD collaboration.}\label{figg5}
\end{figure}
We display in the left panel of Fig.~\ref{figg5}  the allowed parameters points in the plane of $M_W$ versus $T$ within the 2HDM+TB top partner doublet model. The points are color-coded to show the values of the $S$ parameter. The red dashed line indicates the SM prediction for the W boson mass and the light red region shows the new CDF-II measurement at $\pm 1\sigma$ CL. It can seen that $M_W^{2HDM+TB}$ is a linear function of the $T$ parameter. Furthermore, the regions consistent with CDF-II measurement of $M_W$ require the values of the $S$ parameter to be in the range between 0 and 0.06. In the right panel of Fig.~\ref{figg5}, we visualize our results within the 2HDM+TB top partner doublet model in the $\sin^2\theta_{\mathrm{eff}}$ versus $M_W$ plane. The color-coding indicates the allowed values of the $T$ parameter. It is interesting to see that full agreement is achieved between the SLD measurement of the weak mixing angle and the favored parameter space required by the CDF-II data. Also, we see that the preferred range for the $T$ parameter is almost entirely positive and varies between 0.14 and 0.26.

\subsubsection{2HDM+BY or 2HDM+XT}
\begin{figure}[H]
	\centering
	\begin{minipage}{0.48\textwidth}
		\centering
		\includegraphics[width=0.94\textwidth]{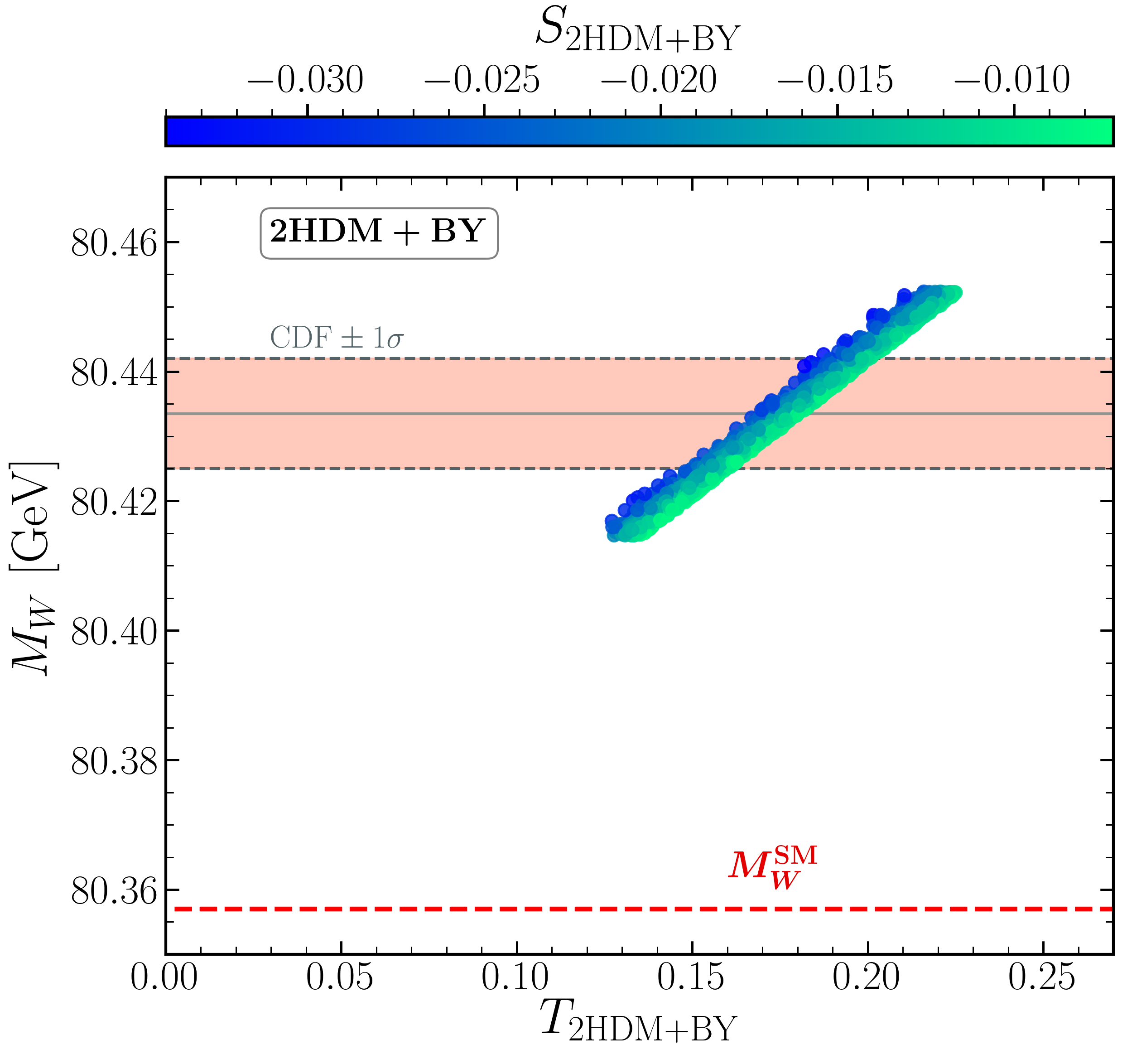}
		
	\end{minipage}\hspace{0.5cm}
	\begin{minipage}{0.48\textwidth}
		\centering
		\includegraphics[width=0.94\textwidth]{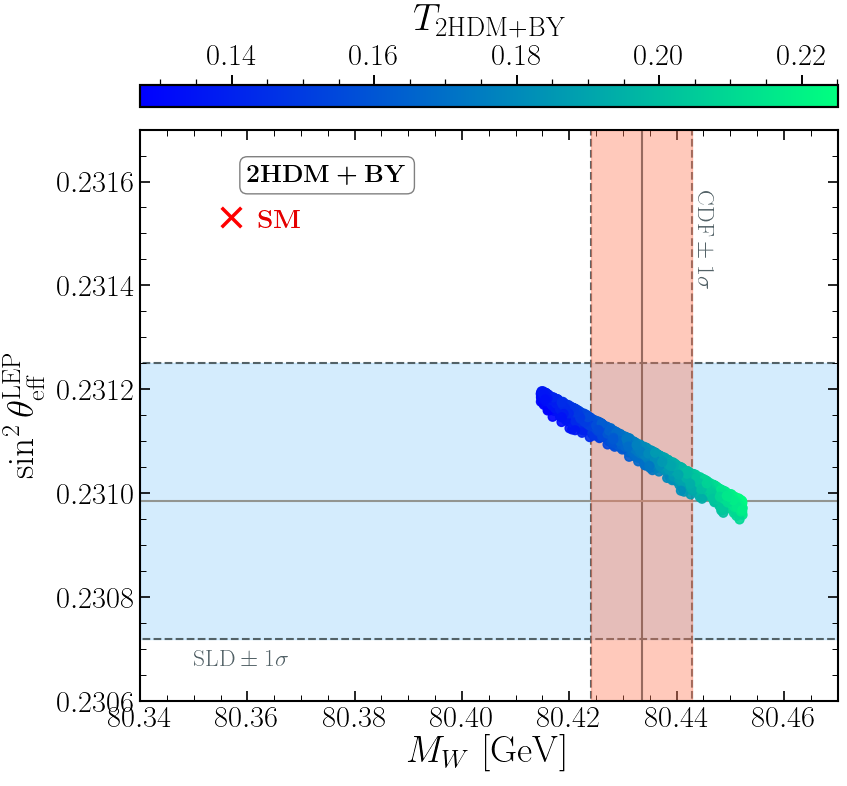}
	\end{minipage}
	\begin{minipage}{0.48\textwidth}
		\centering
		\includegraphics[width=0.94\textwidth]{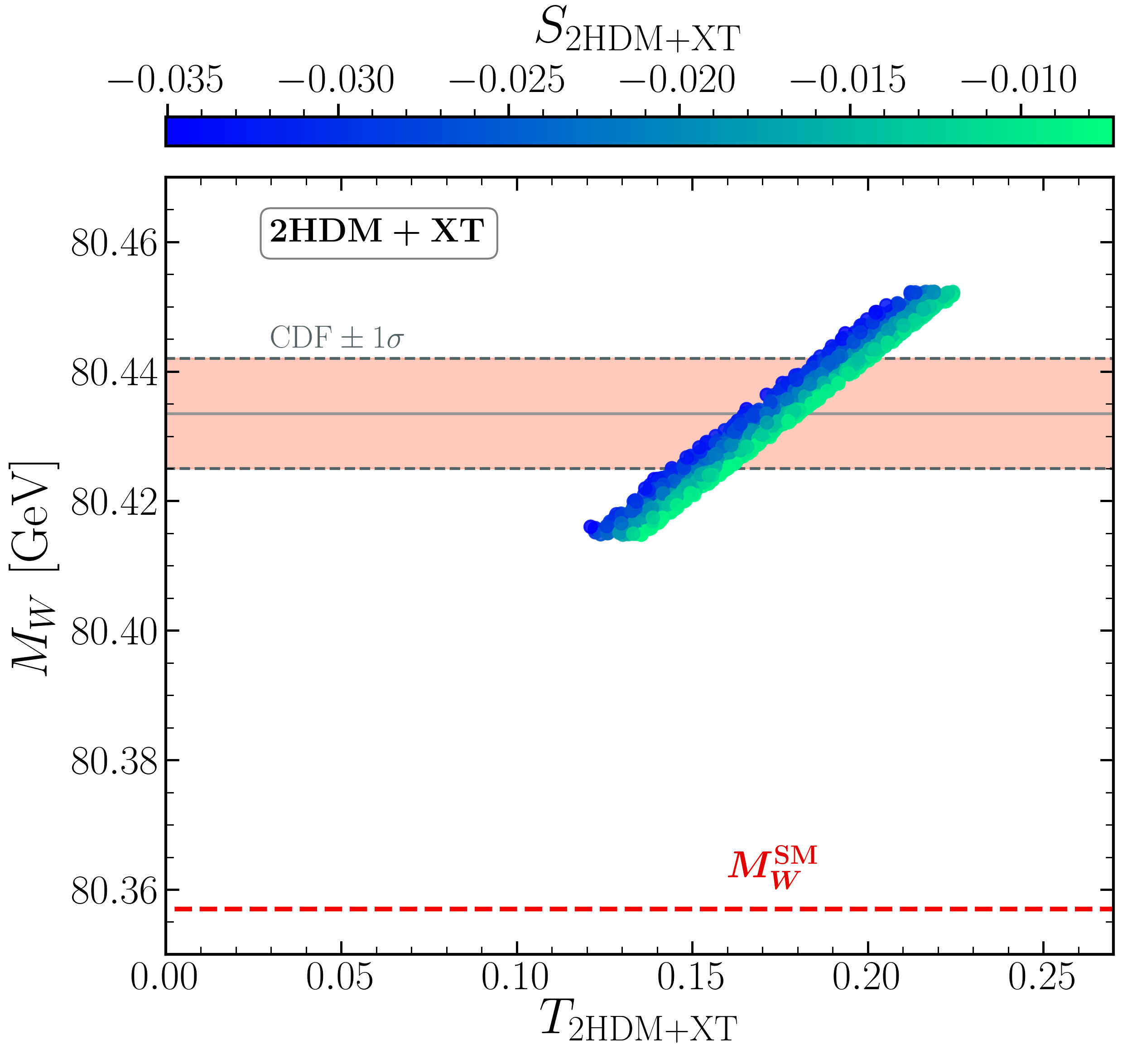}
		
	\end{minipage}\hspace{0.25cm}
	\begin{minipage}{0.48\textwidth}
		\centering
		\includegraphics[width=0.94\textwidth]{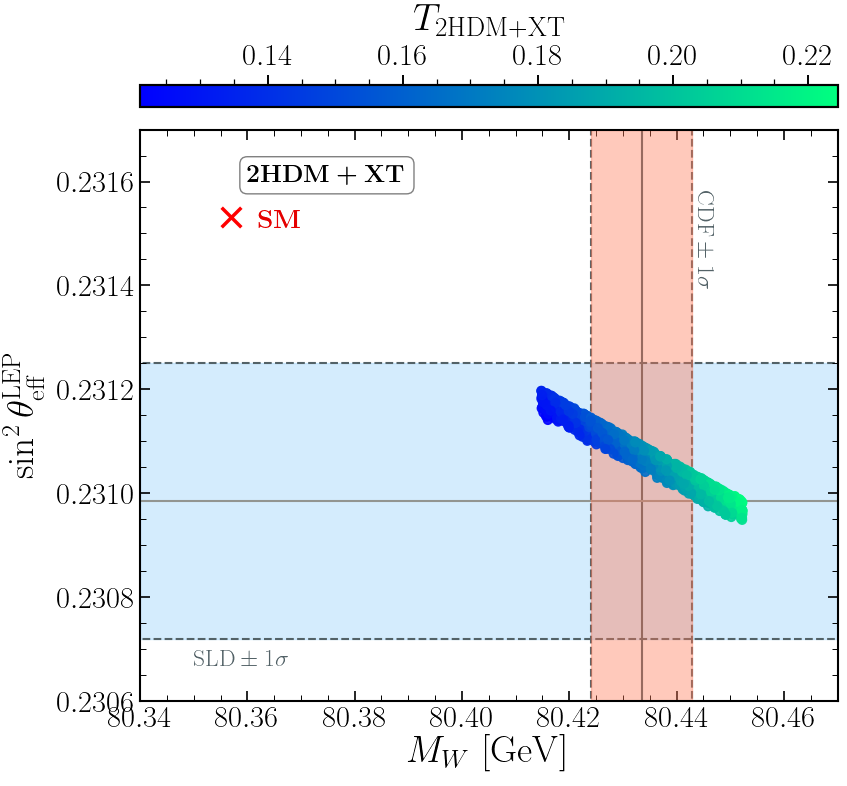}
	\end{minipage}
	\caption{Scan result of the 2HDM+BY (2HDM+XT in the $T$ versus $M_W$ plane in the upper left panel (lower left panel) and in the $M_W$ versus $\sin^2 \theta^{\mathrm{LEP}}_{\mathrm{eff}}$ plane in the upper right panel (lower right panel). The light red band indicates the $M_W$ value with the associated 1$\sigma$ uncertainty measured recently by the CDF collaboration, while the light green band  shows the result of $\sin^2\theta_{\mathrm{eff}}$ and its associated 1$\sigma$ uncertainty measured  by the SLD collaboration.}\label{figg7}
\end{figure}
We display in the upper left and lower left  panels of Fig.~\ref{figg7}  the allowed parameter points in the $M_W$ versus $T$ plane in the frameworks of the 2HDM+BY and  2HDM+XT models, respectively. The color coding of the points shows the allowed values of $S$ parameter. The red-dashed line indicates the SM prediction for the W boson mass and the light red region displays the new CDF-II measurement at the level of $\pm 1\sigma$. One can observe that the regions consistent with CDF-II measurement of $M_W$ require the values of the $S$ parameter to be in the range between -0.030 and -0.01 in the 2HDM+BY model, while the favored values of the $S$ parameter in the 2HMD+XT model are still negative and range between -0.035 and -0.01. In the upper right and lower right panels of Fig.~\ref{figg7} we visualize our results in the $\sin^2\theta_{\mathrm{eff}}$ versus $M_W$ plane within the 2HDM+BY and 2HDM+XT models, respectively. The color code indicates the allowed values taken by the $T$ parameter. It is interesting to see that a full agreement is achieved between the weak mixing angle measured by the SLD collaboration and the favored parameters space required by the CDF-II data. Also, we see that the preferred range for the $T$ parameter for both models needed to explain the anomaly is almost entirely positive and varies between 0.14 and 0.22.

\begin{figure}[H]
	\centering
	\includegraphics[width=0.49\textwidth]{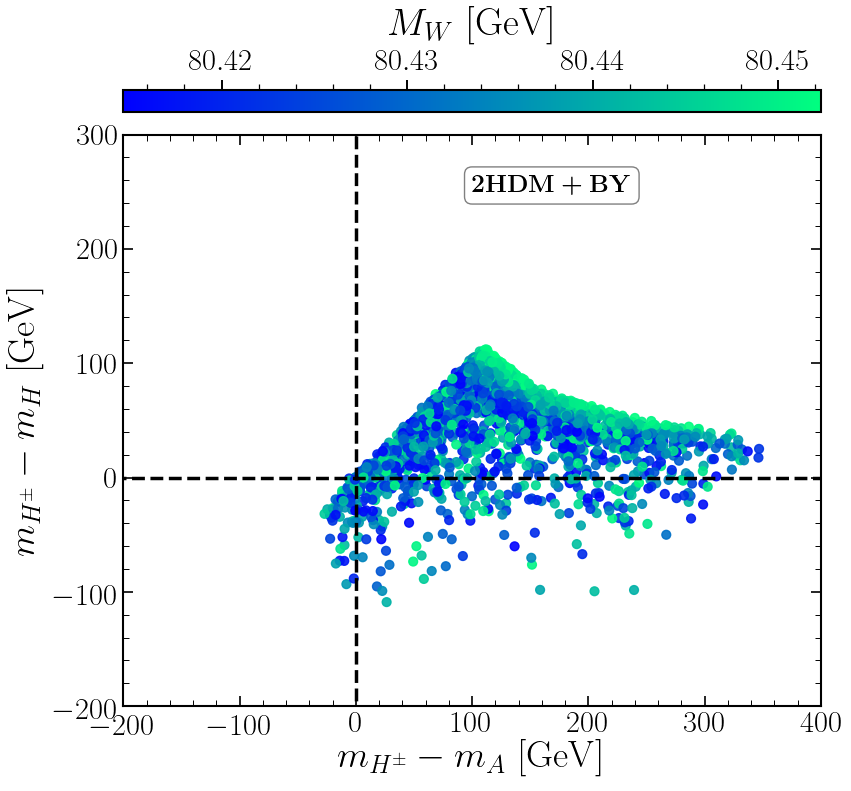}\includegraphics[width=0.49\textwidth]{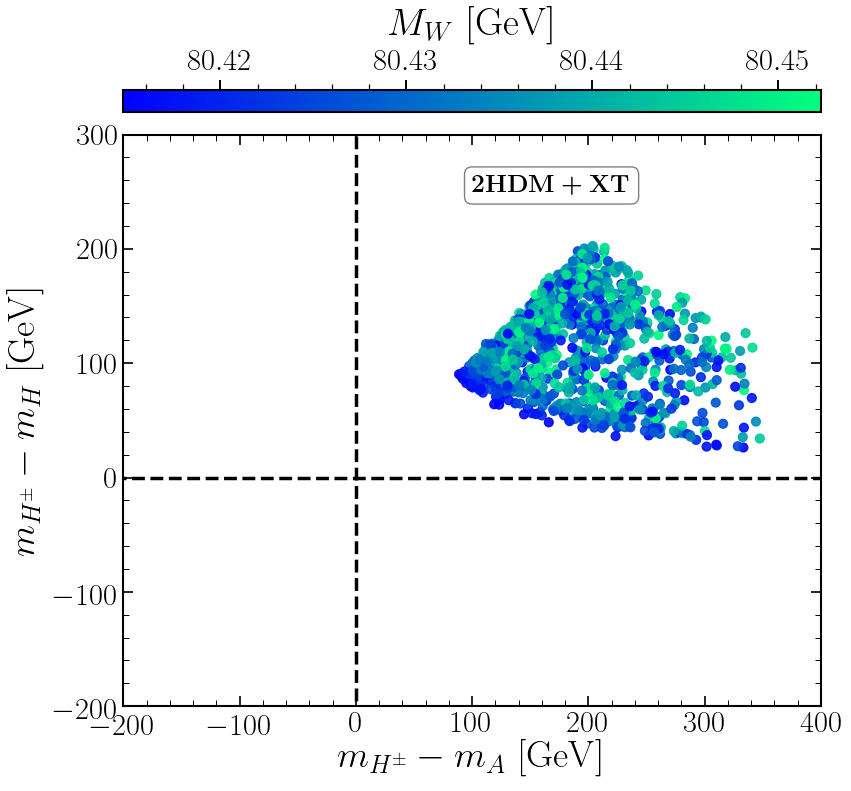}
	\caption{The points from the parameter scan of the  2HDM+BY (2HDM+XT)models in the $m_{H^\pm}-m_A$  versus $m_{H^\pm}-m_H$ plane in the left panel (right panel), with the color code indicating the W-boson mass $M_W$. All points have $\chi^2_{M_W^\mathrm{CDF-II}}\le 4$. }\label{figg8}
\end{figure}
In the left panel of Fig.~\ref{figg8}, we show the allowed parameter points in the $(m_{H^\pm}-m_H)$ versus $(m_{H^\pm}-m_A)$  plane within the 2HDM+BY model. The color code indicates the value of the W boson mass. It is interesting to see that within this model the charged Higgs boson mass $m_{H^\pm}$ can be degenerate with the two neutral BSM Higgs boson masses $m_H$ and $m_A$  to accommodate the CDF-II W-boson mass. Additionally, the recent anomaly can also be explained in the case where $(m_{H^\pm}-m_A)(m_{H^\pm}-m_H)>0$ and again it is interesting to observe that the case where $m_A<m_{H^\pm}<m_H$ is not excluded due to the presence of (B,Y) doublet which gives the favored large positive values of the $T$ parameter required to accommodate the W-mass reported by the CDF-II collaboration. Furthermore, we see that the mass splitting between the charged Higgs boson and CP-odd Higgs boson can go up to 360 GeV whilst the mass splitting between $H^\pm$ and CP-even Higgs boson $H$ can at most be of the order of 100 GeV. In the right panel of Fig.~\ref{figg8} we present the allowed parameter points within the 2HDM+XT doublet model. One can observe that both splittings $m_{H^\pm}-m_H$  and $m_{H^\pm}-m_A$ are required to be simultaneously positive and always non-zero in order to be consistent  with the CDF-II data. The maximum  mass splitting between the charged Higgs boson and the CP-even Higgs $H$ can reach 200 GeV, while the maximum splitting between $H^\pm$ and the CP-odd Higgs boson $A$ can go up to 340 GeV.
\subsubsection{2HDM+TBY or 2HDM+XTB}

\begin{figure}[H]
	\centering
	\begin{minipage}{0.48\textwidth}
		\centering
		\includegraphics[height=7.cm,width=8.0cm]{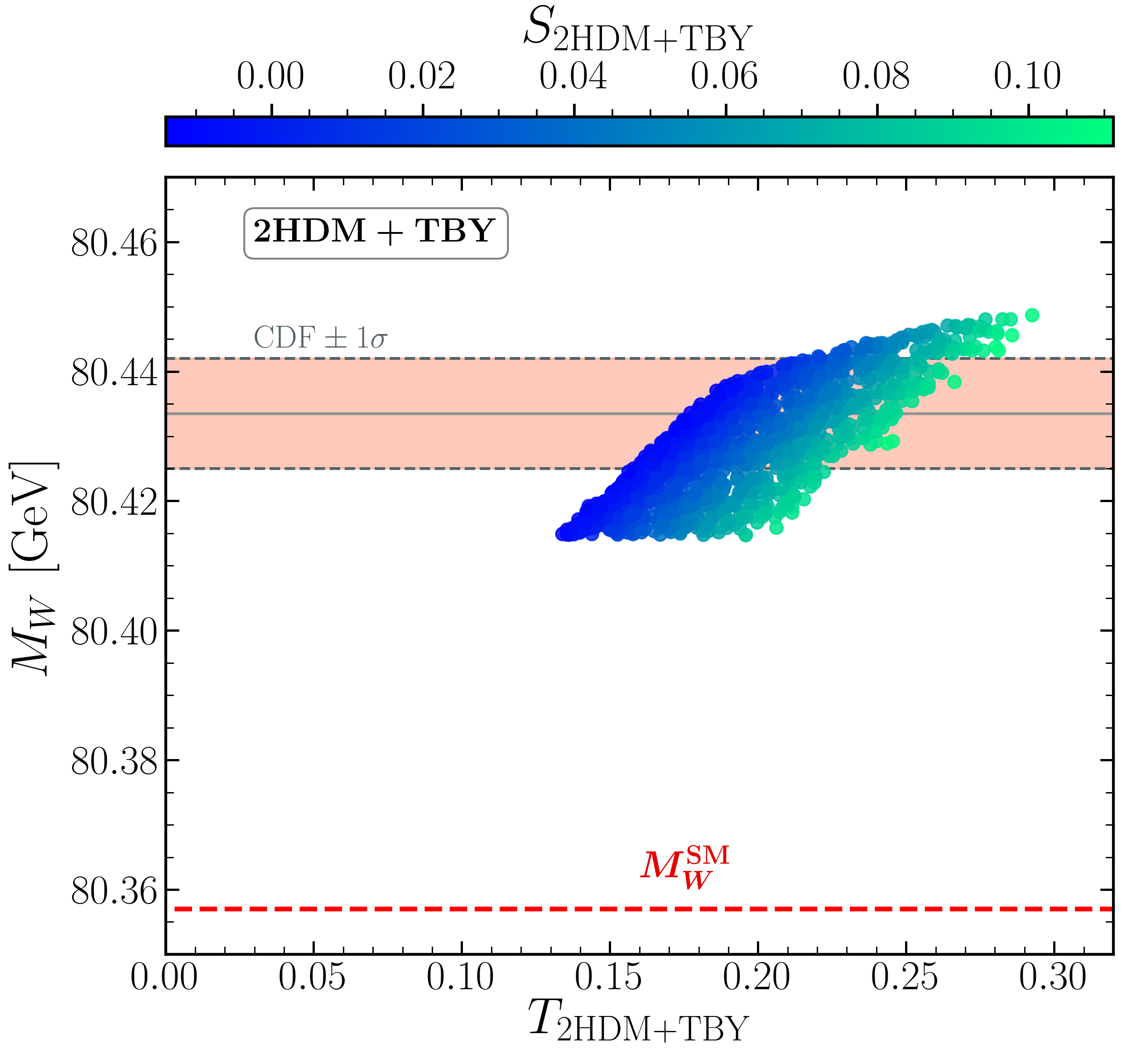}

	\end{minipage}\hspace{0.5cm}
	\begin{minipage}{0.48\textwidth}
		\centering
		\includegraphics[height=7.cm,width=8.0cm]{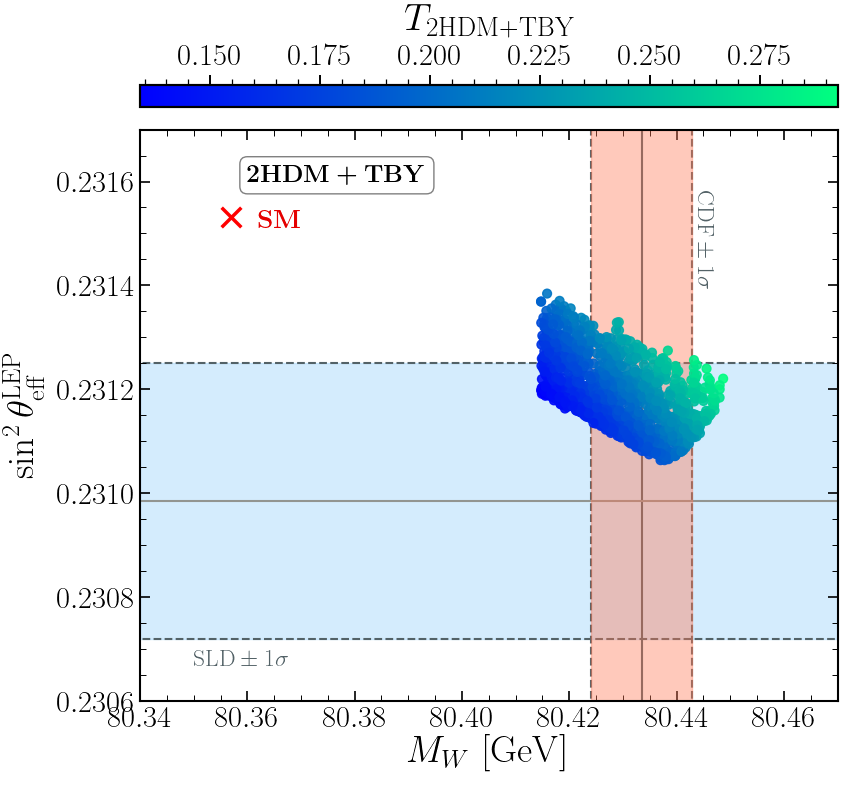}
	\end{minipage}
	\begin{minipage}{0.48\textwidth}
		\centering
		\includegraphics[height=7.cm,width=8.0cm]{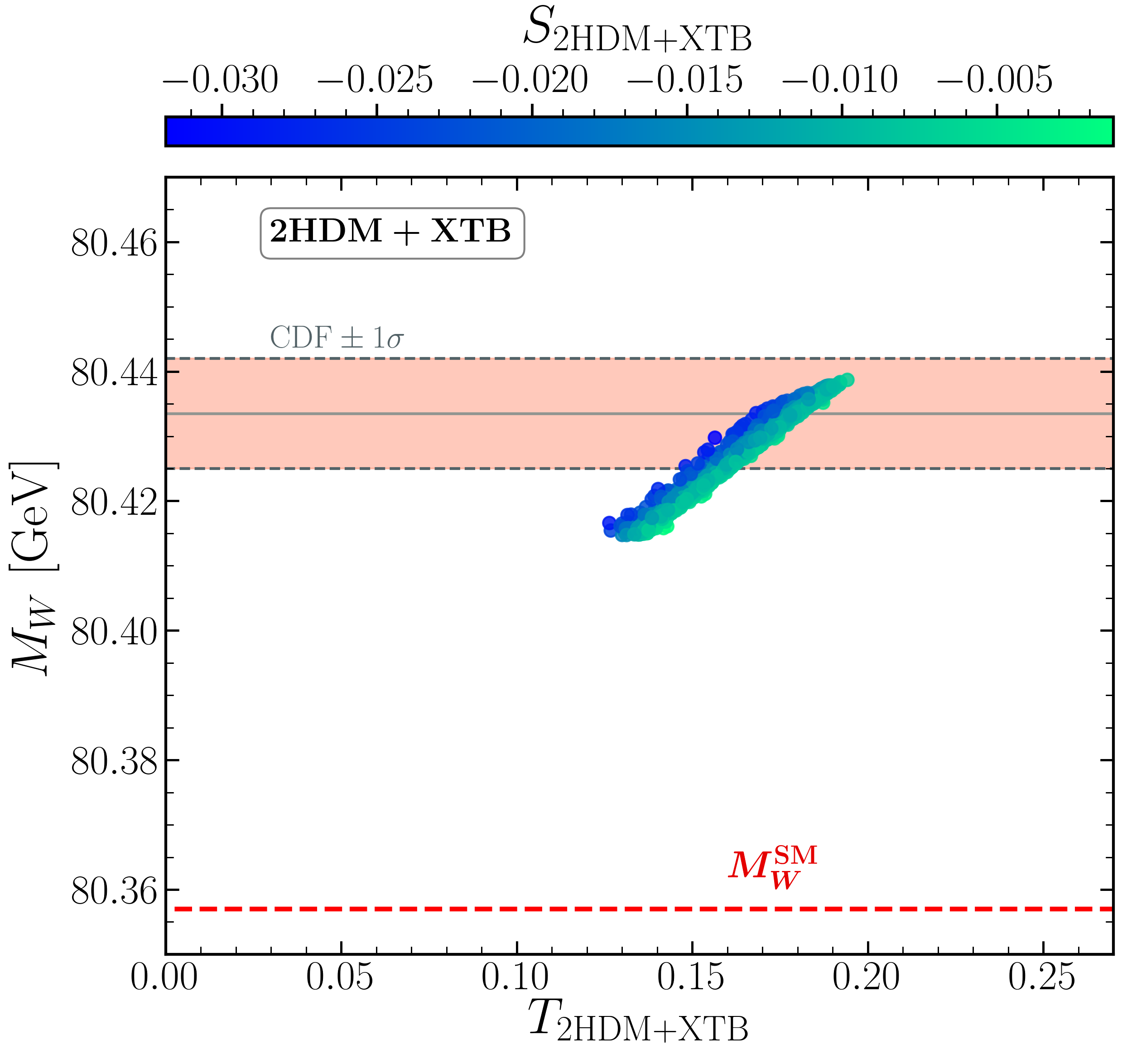}

	\end{minipage}\hspace{0.5cm}
	\begin{minipage}{0.48\textwidth}
		\centering
		\includegraphics[height=7.cm,width=8.0cm]{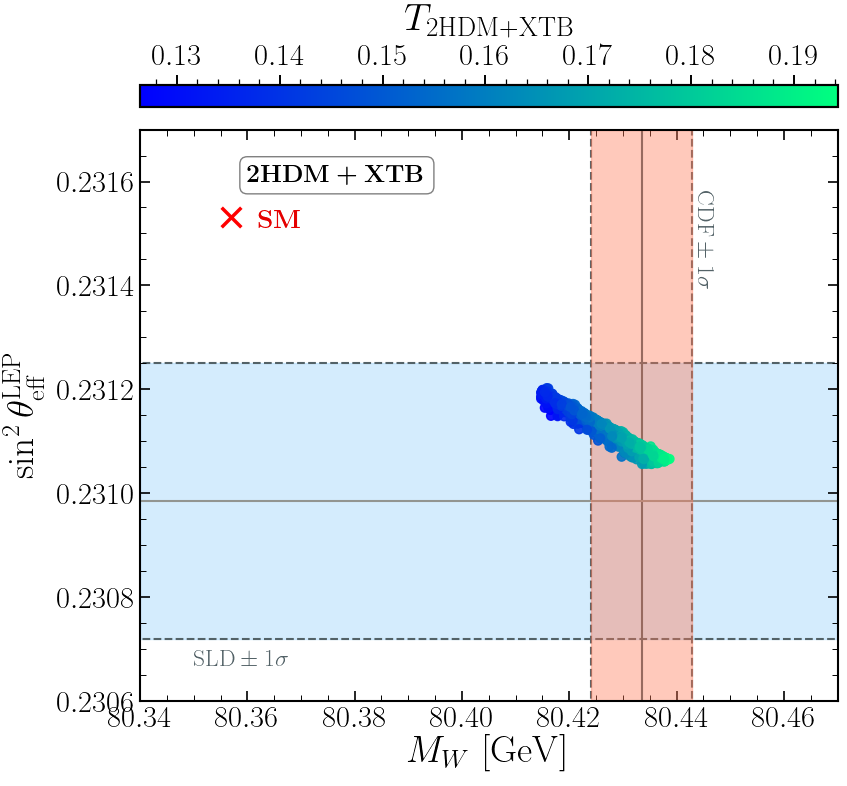}
	\end{minipage}
	\caption{Scan result of the 2HDM+TBY (upper panels)  and  (2HDM+XTB) (lower panels)
		in the $T$ versus $M_W$ plane in the upper left panel (lower left panel) and in the $M_W$ versus $\sin^2 \theta^{\mathrm{LEP}}_{\mathrm{eff}}$ plane in the upper right panel (lower right panel). The light red band indicates the $M_W$ value with the associated 1$\sigma$ uncertainty measured recently by the CDF collaboration, while the light green band  shows the result of $\sin^2\theta_{\mathrm{eff}}$ and its associated 1$\sigma$ uncertainty measured  by the SLD collaboration.}\label{figg9}
\end{figure}
We present in the upper-left and lower-left panels of Fig. \ref{figg9} the allowed points from the parameter scan in the $M_W$ versus $T$ plane in the framework of the 2HDM+TBY and 2HDM+XTB triplet models, respectively. The color coding of the points indicates the allowed values of the $S$ parameter. The red-dashed line shows the SM prediction for the W-boson mass, and the light red region displays the new CDF-II measurement at the $\pm 1\sigma$ level. We can see that the regions consistent with the new CDF measurement require the values of the $S$ parameter to be in the range between -0.014 and 0.11 in the 2HDM+TBY model, while a negative $S$ parameter contribution is required to accommodate the anomaly in the 2HDM+XTB model, with favored values of the $S$ parameter ranging between -0.032 and -0.0012. In the upper-right and lower-right panels of Fig. \ref{figg9}, we visualize, respectively, our results within the 2HDM+TBY and 2HDM+XTB models in the $\sin^2\theta_{\mathrm{eff}}$ versus $M_W$ plane. The color code indicates the allowed values taken by the $T$ parameter. Besides the fact that $M_W$ is anti-correlated with the weak mixing angle $\sin^2\theta^{\mathrm{LEP}}_{\mathrm{eff}}$, it is interesting to note that in both models there is full agreement between the weak mixing angle measured by the SLD collaboration and the points from the parameter space preferred by the CDF-II data. Also, we see that the favored range for the $T$ parameter within the 2HDM+TBY model that is required to explain the anomaly is almost entirely positive and varies between 0.13 and 0.3, while it ranges between 0.13 and 0.19 in the framework of the 2HDM+XTB model.
\begin{figure}[H]
	\centering
	\includegraphics[height=7.2cm,width=8.0cm]{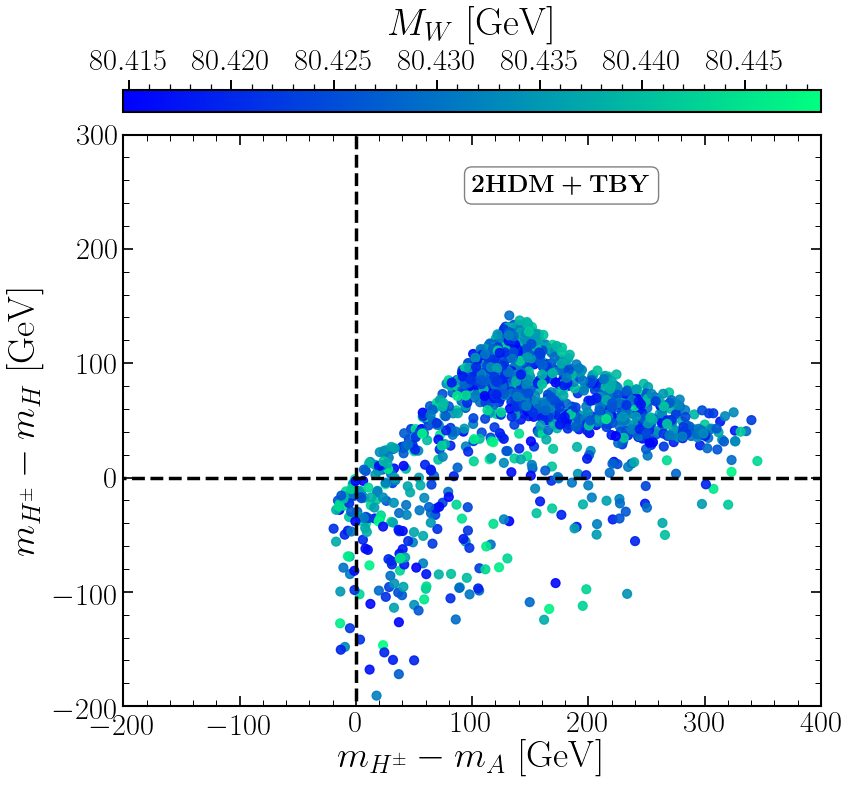}	\includegraphics[height=7.2cm,width=8.0cm]{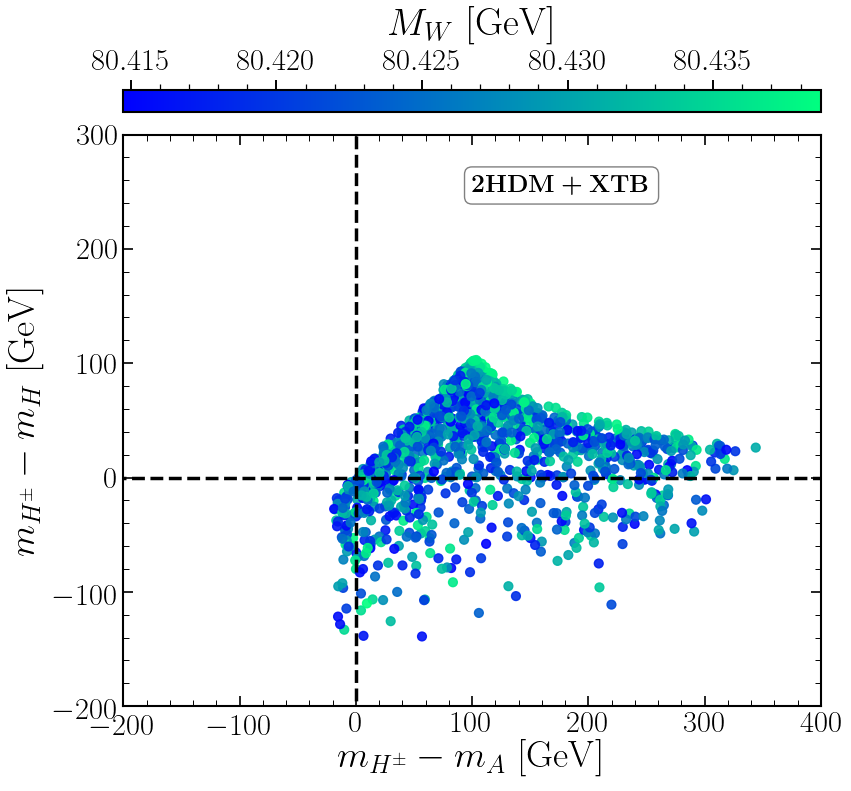}
	\caption{The points from the parameter scan of the  2HDM+TBY (2HDM+XTB) models in the $m_{H^\pm}-m_A$  versus $m_{H^\pm}-m_H$ plane in the left panel (right panel), with the color code indicating the W-boson mass $M_W$.}\label{figg10}
\end{figure}
In the left panel of Fig.~\ref{figg10}, we present the allowed parameter points in the $(m_{H^\pm}-m_H)$ versus $(m_{H^\pm}-m_A)$ plane within the 2HDM+TBY triplet model. The color code indicates the value of the W boson mass. At a glance, it is evident that the CDF value of $M_W$ can be reproduced in the scenario where the three BSM scalars are mass-degenerate, i.e., $m_{H^\pm}=m_A=m_H$. Additionally, the CDF-II $M_W$ measurement can also be explained in the case where $(m_{H^\pm}-m_A)(m_{H^\pm}-m_H)>0$. It is worth noting that the scenario with $m_H<m_{H^\pm}<m_A$ is still allowed and can accommodate the anomaly. Besides, we see that the mass splitting between the charged Higgs boson and the CP-odd Higgs can reach up to 160 GeV, while the mass splitting between $H^\pm$ and the CP-even Higgs $H$ can go up to 350 GeV.\\
In the right panel of Fig.~\ref{figg10}, the allowed parameter space within the 2HDM+XTB model is presented. It can be seen that the degenerate scenario, in which the masses of the charged Higgs boson ($m_{H^\pm}$), the CP-even Higgs ($m_H$), and the CP-odd Higgs ($m_A$) are equal, cannot explain the CDF value of the W boson mass. Furthermore, it is shown that the mass difference between the charged Higgs boson and the CP-even Higgs $H$ is restricted to a range of 150 GeV to 110 GeV, while the mass difference between the charged Higgs boson and the CP-odd Higgs $A$ is limited to a range between -10 GeV and 350 GeV.
\section{Conclusion}
\label{concl}
The very recent measurement result of the W-boson mass $M_W$ reported by the CDF collaboration
deviates from the SM prediction by about 7$\sigma$. If the CDF-II result is correct, it would indicate the need for new physics beyond the Standard Model to explain the observed mass of the W boson that matches the CDF-II data.
In this paper, we consider the implications of the recent W boson mass anomaly in the framework of 2HDM type II augmented by vector-like quarks. We studied all seven distinct possibilities of vector-like $SU(2)_L$ multiplets including singlets, doublets, and triplets.
We computed the electroweak precision observables S and T using the standard scalar Passarino-Veltman functions and provided the generic analytic expressions for $S$ and $T$ and listed the couplings used for  specific models such us the
singlet, doublet and triplet VLQs models. In order to explain the CDF measurement we need a positive $T$  and this is possible
if we consider VLQs alone, which is all time positive. In the case of the 2HDM augmented with VLQs, both $S$ and $T$
contributions are split into two separate contributions, one from VLQs and one from 2HDM which do not mix.
We found that in most cases, there is a cancellation between VLQs and 2HDM contributions, which enlarges  the parameter space of both models. After this observation we performed a $\chi^2$ analysis
and found that  it is possible to solve the tension between the new CDF $M_W$ measurement and the SM prediction.
We also noticed that the scenario with 2HDM alone with a degenerate spectrum that is ruled out by the CDF measurement could survive if we include both 2HDM and VLQs.
We also discussed the consequences of the 2HDM augmented with VLQs on the effective
mixing  angle $\sin^2\theta_{\text{eff}}$.

\section*{Acknowledgments}
We thank Junjie Cao for useful discussions. This work is supported by the Moroccan Ministry of Higher Education and Scientific Research
MESRSFC and CNRST: Projet PPR/2015/6. MB is grateful for the technical support of CNRST/HPC-MARWAN.
\bibliographystyle{JHEP}

\appendix\label{app}
\section{VLQs contributions to the oblique parameters S, T and U}
In this appendix we give the analytical expressions for S, T and U in the 2HDM as well as in the generic expression in the
VLQ models as a function of the well-known scalar Passarino-Veltman functions.

\subsection{Passarino-Veltman functions}
The Passarino-Veltman functions are defined as follow:

\begin{eqnarray}
	A_0(m_1^2) &=& \frac{(2 \pi \mu)^{4-D}}{i \pi^{2}}\int_{}^{}{d^Dq \frac{1}{d_1}},\\
	B_0;B^{\mu};B^{\mu\nu}(p_1^2,m_1^2,m_2^2) &=& \frac{(2 \pi \mu)^{4-D}}{i \pi^{2}}\int_{}^{}{d^Dq \frac{1;q^\mu;q^\mu q^\nu}{d_1d_2}}
\end{eqnarray}

With the denominators $d_i$ are defined by $ d_1= q^2-m_1^2$ and $d_2=(q+p_1)^2-m_2^2$ and $\mu$ is the renormalization scale. $p_1$ is the external momentum and $q$ is the internal momentum which is integrated out.

The functions $B^\mu$ and $B^{\mu \nu} $ can be decomposed as,

\begin{eqnarray}
	B^\mu &=& p_1^\mu B_1,\quad \ B^{\mu \nu } = g^{\mu \nu} B_{00} + p_1^\mu p_1^\nu B_{11}.
\end{eqnarray}

Thus, the functions $B_{00}$ and $B_1$ can be calculated by contracting $B^\mu$ and $B^{\mu\nu}$ with $p_1$ and $g_{\mu \nu}$ respectively,

\begin{eqnarray}
	B_1(p^2, m_1^2, m_2^2) &=& \frac{1}{2}\left[
		A_0(m_1^2) - A_0(m_2^2) - \left(p^2 + m_1^2 - m_2^2\right) B_0(p^2, m_1^2, m_2^2) \right]\\
	B_{00}(p^2, m_1^2, m_2^2) &=& -\frac{
		p^2 - 3 (m_1^2 + m_2^2)}{18} + \frac{m_1^2 B_0(p^2, m_1^2, m_2^2)}{3}\notag \\ && +
	\frac{A_0(m_2^2) + (p^2 + m_1^2 - m_2^2) B_1(p^2, m_1^2, m_2^2)}{6}
\end{eqnarray}

\subsection{ $S$, $T$ and $U$ parameters}

The oblique parameters can be parameterized in terms of the gauge bosons two-point functions as follows:

\begin{eqnarray}
	S &=& \frac{4 s_W^2 c_W^2 }{\alpha} \Re e \Bigg[ \frac{\Pi_{\gamma \gamma}(m_Z^2)}{m_Z^2}-\frac{\Pi_{Z Z}(m_Z^2)-\Pi_{Z Z}(0)}{m_Z^2} -\Big( \frac{c_W^2-s_W^2}{c_W s_W}\Big)\Big( \frac{\Pi_{Z \gamma}(m_Z^2)+\Pi_{Z \gamma}(0)}{m_Z^2}\Big)  \Bigg] \nonumber \\ \label{eq:S_expression} \\
	T &=& \frac{1}{\alpha}\Re e \left[ \frac{\Pi_{Z Z}(0)}{m_Z^2}-\frac{\Pi_{W W}(0)}{m_W^2}-\frac{2 s_W }{c_W} \frac{\Pi_{Z \gamma}(0)}{m_Z^2}\label{eq:T_expression}  \right] \\
	U &=& \frac{4 s_W^2 }{\alpha} \Re e \Bigg[ \frac{\Pi_{W W}(0)-\Pi_{W W}(m_W^2)}{m_W^2} - c_W^2 \frac{\Pi_{ZZ}(0)-\Pi_{ZZ }(m_Z^2)}{m_Z^2} -2 s_W c_W \frac{\Pi_{\gamma Z}(0)-\Pi_{\gamma Z}(m_Z^2)}{m_Z^2}+ \nonumber \\ && s_W^2 \frac{\Pi_{\gamma \gamma}(m_Z^2)}{m_Z^2}  \Bigg] \label{eq:U_expression}
\end{eqnarray}

Where $\alpha$ is the fine-structure constant and $c_w$ ($s_w$) denotes the cosine (sine) of the Weinberg angle.

\subsection{$S$, $T$ and $U$ parameters in 2HDM}

The contributions of the 2HDM to the oblique parameters S, T and U can be given as follows:

\begin{eqnarray}
	\Delta S&= &\frac{c_W^2}{\pi m_W^2}
	\Bigg\{c_{\beta-\alpha}^2 \bigg[m_Z^2 \Big(-B_0 (0,m_{h}^2,m_Z^2)+ B_0 (m_Z^2,m_{h}^2,m_Z^2)+B_0 (0,m_{H}^2,m_Z^2)\nonumber \\ &&  - B_0 (m_Z^2,m_{H}^2,m_Z^2)\Big) +B_{00}(m_Z^2,m_{h}^2,m_{A}^2)-B_{00}(0,m_{h}^2,m_{A}^2)+B_{00}(0,m_{h}^2,m_Z^2) \nonumber \\ && -B_{00}(m_Z^2,m_{h}^2,m_Z^2)-B_{00}(0,m_{H}^2,m_Z^2) +B_{00}(m_Z^2,m_{H}^2,m_Z^2)\bigg] \nonumber \\ && +s_{\beta-\alpha}^2 \bigg[B_{00}(m_Z^2,m_{H}^2,m_{A}^2)-B_{00}(0,m_{H}^2,m_{A}^2)\bigg]-B_{00}(m_Z^2,m_{H^\pm}^2,m_{H^\pm}^2)\nonumber \\ && +(c_{W}^2-s_{W}^2)^2 B_{00}(0,m_{H^\pm}^2,m_{H^\pm}^2)+2 A_0(m_{H^\pm}^2) s_W^2 c_W^2\Bigg\}
\end{eqnarray}

\begin{align}
	\Delta T= & \frac{1}{4 m_{W}^2\pi s_W^2}\Big\{ m_W^2\cbasq B_0( 0, m_{h}^2, m_{W}^2)+ \sbasq B_{00}( 0, m_{h}^2, m_{W}^2)
	\nonumber                                                                                                                 \\ &
	- \cbasq  \left[m_Z^2  B_0( 0, m_{h}^2, m_{Z}^2)-   B_{00}( 0, m_{h}^2, m_{Z}^2)\right] \nonumber                         \\ &
	- \cbasq \left[ m_{W}^2  B_0( 0, m_{H}^2, m_{W}^2) -  B_{00}( 0, m_{H}^2, m_{W}^2) \right]
	+ \cbasq \left[ m_{Z}^2  B_0( 0, m_{H}^2, m_{Z}^2) -  B_{00}( 0, m_{H}^2, m_{Z}^2) \right]\nonumber                       \\ &
	- 2m_{H^\pm}^2s_W^4B_0( 0, m_{H^\pm}^2, m_{H^\pm}^2)+(4s_W^4-1)B_{00}( 0, m_{H^\pm}^2, m_{H^\pm}^2)\nonumber              \\ &
	- \cbasq \left[  B_{00}( 0, m_{h}^2, m_{A}^2)-  B_{00}( 0, m_{h}^2, m_{H^\pm}^2) \right]\nonumber                         \\ &
	-\sbasq \left[ B_{00}( 0, m_{H}^2, m_{A}^2) -  B_{00}( 0, m_{H}^2, m_{H^\pm}^2) \right]
	+ B_{00}( 0, m_{A}^2, m_{H^\pm}^2)  - B_{00}( 0, m_{W}^2, m_{h}^2)\nonumber                                               \\& - B_{00}( 0, m_{W}^2, m_{Z}^2) + B_{00}( 0, m_{Z}^2, m_{W}^2) -2m_{H^\pm}^2s_W^4\Big\}
\end{align}

\begin{eqnarray}
	\Delta U&=& \frac{1}{\pi m_W^2}\Bigg\{c_{\beta-\alpha}\Big[
	m_Z^2  c_W^2 B_0(0,m_{h}^2,m_Z^2)-m_Z^2 c_W^2 B_0(0,m_{h}^2,m_W^2)\nonumber \\ && -m_Z^2 c_W^2 B_0(m_Z^2,m_{h}^2,m_Z^2)
	+m_Z^2 c_W^2 B_0(m_W^2,m_{h}^2,m_W^2)-m_Z^2  c_W^2 B_0(0,m_{H}^2,m_Z^2)\nonumber \\ &&
	+m_Z^2  c_W^2 B_0(0,m_{H}^2,m_W^2)+m_Z^2  c_W^2 B_0(m_Z^2,m_{H}^2,m_Z^2)
	-m_Z^2  c_W^2 B_0(m_W^2,m_{H}^2,m_W^2) \nonumber \\ &&
	- c_W^2 B_{00}(m_Z^2,m_{h}^2,m_{A}^2)+ c_W^2 B_{00}(0,m_{h}^2,m_{A}^2)- c_W^2 B_{00}(0,m_{h}^2,m_Z^2)\nonumber \\ &&
	+ c_W^2 B_{00}(m_Z^2,m_{h}^2,m_Z^2)+ c_W^2 B_{00}(0,m_{H}^2,m_Z^2)- c_W^2 B_{00}(m_Z^2,m_{H}^2,m_Z^2)\nonumber \\ &&
	+B_{00}(m_W^2,m_{h}^2,m_{H^\pm}^2) - B_{00}(0,m_{h}^2,m_{H^\pm}^2)  - B_{00}(0,m_{H}^2,m_W^2)\nonumber \\ && + B_{00}(m_W^2,m_{H}^2,m_W^2)\Big]
	-s_{\beta-\alpha} \Big[ c_W^2 B_{00}(m_Z^2,m_{H}^2,m_{A}^2) + c_W^2 B_{00}(0,m_{H}^2,m_{A}^2)\nonumber \\ &&
	- B_{00}(0,m_{h}^2,m_W^2)
	+ B_{00}(m_W^2,m_{h}^2,m_W^2)+ B_{00}(m_W^2,m_{H}^2,m_{H^\pm}^2) - B_{00}(0,m_{H}^2,m_{H^\pm}^2)\Big]\nonumber \\ &&
	+B_{00}(m_W^2,m_{A}^2,m_{H^\pm}^2)
	-B_{00}(0,m_{A}^2,m_{H^\pm}^2)+B_{00}(0,m_W^2,m_{h}^2)
	-B_{00}(m_W^2,m_W^2,m_{h}^2)\nonumber \\ &&
	-c_W^2 B_{00}(m_Z^2,m_{H^\pm}^2,m_{H^\pm}^2)
	-\frac{1}{2} c_W^2 (-4 c_{2 W}
	+c_{4 W}+1) B_{00}(0,m_{H^\pm}^2,m_{H^\pm}^2)
	-B_{00}(0,m_Z^2,m_W^2)\nonumber \\ &&
	+B_{00}(0,m_W^2,m_Z^2)
	+B_{00}(m_W^2,m_Z^2,m_W^2)
	-B_{00}(m_W^2,m_W^2,m_Z^2)+2 A_0(m_{H^\pm}^2) s_W^4 c_W^2 \Bigg\}
\end{eqnarray}
It is understood that in the above expressions for $\Delta S$, $\Delta T$, and $\Delta U$, the SM bosonic contribution has been subtracted from the 2HDM contribution.

\subsection{$S$, $T$ and $U$ parameters in VLQs}

We use  \texttt{FeynArts/FormCalc} public codes to calculate the analytical expressions of the self-energy of the gauge bosons, presented in equations (\ref{eq:S_expression}) and (\ref{eq:T_expression}), in a scenario where VLQs interact only with third generation SM quarks. By taking into account only the contribution of those heavy quarks (both SM top and bottom as well as the new VLQs)  to the self-energy, as the other contributions are the same as in the SM, the expressions of the electroweak gauge boson self-energies can be written as:
\begin{eqnarray}
	-\Pi_{\gamma Z}(q^2) &=& \sum_{f}^{}{g_{\gamma z}(c_{Zff}^L,c_{Zff}^R,Q_f,m_f^2,q^2)} \\
	-\Pi_{\gamma \gamma}(q^2) &=& \sum_{f}^{}{g_{vv}(Q_f,Q_f,m_f^2,m_f^2,q^2)}\\
	-\Pi_{Z Z}(q^2) &=&  2 \sum_{f_i \neq f_j}^{}{\delta_0{(|Q_{f_i}-Q_{f_j}|)} g_{vv}\left(c_{Zf_if_j}^L,c_{Zf_if_j}^R,m_{f_i}^2,m_{f_j}^2,q^2\right)}\notag\\ && +\sum_{f}^{}{g_{vv}\left(c_{Zff}^L,c_{Zff}^R,m_f^2,m_f^2,q^2\right)} \\
	-\Pi_{W W}(q^2) &=&   \sum_{f_i \neq f_j}^{}{\delta_1{( |Q_{f_i}-Q_{f_j}|)} g_{vv}\left(c_{Wf_if_j}^L,c_{Wf_if_j}^R,m_{f_i}^2,m_{f_j}^2,q^2\right)}
\end{eqnarray}

Where $c_{Vff}$ are the couplings between the gauge boson Z or W and two fermions ($f=t,b,T,X,B,Y$) given in App.~\ref{app:coupling}
, $\delta_d$ is the Dirac function with a non-null value at d, and the functions $g_{ab}$ are defined as,

$g_{vv}$ and $g_{Z\gamma}$ are given by:
\begin{eqnarray}
	g_{vv}\left(x,y,m_1^2,m_2^2,k^2 \right)&=&\frac{N_c}{8 \pi ^2}\Big[(x^2+y^2) \big( A_0(m_2^2) -2 B_{00}(k^2,m_1^2,m_2^2)+ k^2 B_1(k^2,m_1^2,m_2^2)\big)\nonumber \\ && +\big((x^2+y^2)m_1^2-2x y m_1m_2\big)B_0(k^2,m_1^2,m_2^2) \Big]\\
	g_{\gamma z}\left(x,y,Q,m^2,k^2 \right)&=& \frac{-N_c}{8 \pi ^2}\Big[(x + y) Q \big(A_0(m^2) - 2 B_{00}(k^2, m^2, m^2) \notag \\ && + k^2 B_1(k^2, m^2, m^2)\big)\Big]
\end{eqnarray}

Where the $N_c$ is the color factor with $N_c=3$ for quarks. $A_0$, $B_0$, $B_{00}$ and $B_1$ are the usual Passarino-Veltman functions.

The SM contribution to S and T is calculated by considering the exchange of only top and bottom quarks in the self-energies of the gauge bosons, and the SM couplings used are given by Eq.(\Ref{eq:SM_Vff}).

After computing the expressions of the $S$, $T$ and $U$ parameters for both the SM and the SM augmented with VLQs, it is necessary to perform a subtraction to extract the contribution of new physics to the oblique parameters.

\begin{eqnarray}
	\Delta S= S_{SM+VLQs} -S_{SM}, \
	\Delta T= T_{SM+VLQs} -T_{SM}, \
	\Delta U= U_{SM+VLQs} -U_{SM}
\end{eqnarray}

It is important to note that we checked that both S, T and U are UV finite and renormalization scale $\mu$ independent in all the aforementioned  models.

The oblique parameters of the extensions with VLQs have been studied in many works~\cite{Cao:2022mif,Aguilar-Saavedra:2013qpa,Chen:2017hak,Lavoura:1992np}. In our analysis, we have compared our results numerically with some of those papers and found the following conclusions:
\begin{itemize}
	\item Our expressions for the oblique parameter $T$ are in agreement with all the previously mentioned papers in all extensions, except in the SM+XTB case where there is a disagreement with those of the Ref.~\cite{Chen:2017hak}.
	\item The approximation proposed in Eqs.(25),(26),(27), and (28) of~\cite{Lavoura:1992np} can lead to large deviations in the evaluation of the oblique parameter $S$ in the SM +$B$, $XT$, $BY$ extension, but when using the general terms proposed in Eqs. (17), (18), (22), and (23) of~\cite{Lavoura:1992np}, these deviations disappear in the case of large $s^d_R$ in the SM +$TB$ extension.

	\item The expression for the $S$ parameter in the triplet extensions cannot be obtained in the same manner as in~\cite{Chen:2017hak} where they extended the formulations from Ref.\cite{Lavoura:1992np} to calculate this oblique parameter. This can be problematic, as demonstrated in Ref.\cite{He:2022zjz}.
 \end{itemize}

\section{Gauge interactions of the vector-like quarks}\label{app:coupling}
The neutral and charged currents are modified by the presence of VLQs, the couplings between the exotic heavy quarks to the SM third generation quarks and the electroweak massive gauge bosons can be parametrized as follows:

\begin{eqnarray}
	Zq'q&=&  \frac{e}{2 s_w c_w}\gamma^\mu ( \lambda^{L}_{Zq'q} \mathbb{L}  + \lambda^{R}_{Zq'q} \mathbb{R} ),\nonumber \\
	Wq'q&=&  \frac{e}{\sqrt{2} s_w }\gamma^\mu ( \lambda^{L}_{Wq'q} \mathbb{L}  + \lambda^{R}_{Wq'q} \mathbb{R} )
\end{eqnarray}
Where $\lambda_{Vq'q}^{L,R}$ are the components for both left- and right-handed couplings of the Z and W bosons. Note that these couplings depend on the chosen representations of VLQs as follows:
\subsection{Vector-like singlet T}
In the $SU(2)_L$ vector-like singlet T scenario, the left and right handed charged current couplings $\lambda_{Wq'q}^{L,R}$  are given by:

\begin{eqnarray}
	\lambda_{Wtb}^L&=& \clx,\quad
	\lambda_{Wtb}^R= 0, \nonumber  \\
	\lambda_{WTb}^L&=& \slx , \quad
	\lambda_{WTb}^R= 0.  \label{eq:SMT_Wff}
\end{eqnarray}
the left and right handed neutral current couplings $\lambda^{L,R}_{Zq'q}$  are given by:
\begin{eqnarray}
	\lambda_{Ztt}^L&=& (\clx)^2 -\frac{4}{3} s_W^2 ,\quad
	\lambda_{Ztt}^R= -\frac{4}{3} s_W^2, \nonumber  \\
	\lambda_{Zbb}^L&=& -1+\frac{2}{3} s_W^2,\quad
	\lambda_{Zbb}^R= \frac{2}{3} s_W^2, \nonumber  \\
	\lambda_{ZTT}^L&=& (\slx)^2-\frac{4}{3} s_W^2 ,\quad
	\lambda_{ZTT}^R=-\frac{4}{3} s_W^2 , \nonumber  \\
	\lambda_{ZtT}^L&=& \slx \clx ,\quad
	\lambda_{ZtT}^R=  0 .
	\label{eq:SMT_Zff}
\end{eqnarray}

\subsection{Vector-like singlet B}
In the $SU(2)_L$ vector-like B singlet scenario, the left and right-handed quarks-W boson couplings $\lambda_{Wq'q}^{L,R}$  are given by:
\begin{eqnarray}
	\lambda_{Wtb}^L&=& \cld,\quad
	\lambda_{Wtb}^R= 0,  \nonumber  \\
	\lambda_{WtB}^L&=&  \sld\ ,\quad
	\lambda_{WtB}^R= 0. \label{eq:SMB_Wff}
\end{eqnarray}
While the neutral current couplings to the Z boson are given by:

\begin{eqnarray}
	\lambda_{Ztt}^L&=& 1-\frac{4}{3} s_W^2 ,\quad
	\lambda_{Ztt}^R= -\frac{4}{3} s_W^2,\nonumber  \\
	\lambda_{Zbb}^L&=& -(\cld)^2+\frac{2}{3} s_W^2,\quad
	\lambda_{Zbb}^R= \frac{2}{3} s_W^2, \nonumber  \\
	\lambda_{ZBB}^L&=& -(\sld)^2+\frac{2}{3} s_W^2 ,\quad
	\lambda_{ZBB}^R=\frac{2}{3} s_W^2 , \nonumber  \\
	\lambda_{ZbB}^L&=& -\sld \cld ,\quad
	\lambda_{ZbB}^R=  0.
	\label{eq:SMB_Zff}
\end{eqnarray}

\subsection{Vector-like doublet TB}
In the case of the $TB$ doublet model, the couplings between T , B and the SM third generation quarks  with the W-boson are given by:
\begin{eqnarray}
	\lambda_{Wtb}^L&=& \clu \cld + \slu \sld,\quad
	\lambda_{Wtb}^R= \sru \srd,  \nonumber  \\
	\lambda_{WTB}^L&=&  \clu \cld + \slu \sld ,\quad
	\lambda_{WTB}^R= \cru \crd, \nonumber  \\
	\lambda_{WTb}^L&=&  \slu \cld  - \clu \sld  ,\quad
	\lambda_{WTb}^R= -\cru \srd, \nonumber  \\
	\lambda_{WtB}^L&=& \clu \sld -\slu \cld ,\quad
	\lambda_{WtB}^R= -\sru \crd.  \label{eq:SMTB_Wff}
\end{eqnarray}
The left and right-handed neutral current couplings to the Z boson are given by:
\begin{eqnarray}
	\lambda_{Ztt}^L&=& 1-\frac{4}{3} s_W^2,\quad
	\lambda_{Ztt}^R= (\sru)^2-\frac{4}{3} s_W^2,\nonumber  \\
	\lambda_{Zbb}^L&=& -1+\frac{2}{3} s_W^2,\quad
	\lambda_{Zbb}^R= -(\srd)^2+\frac{2}{3} s_W^2, \nonumber  \\
	\lambda_{ZTT}^L&=& 1-\frac{4}{3} s_W^2  ,\quad
	\lambda_{ZTT}^R=  (\cru)^2-\frac{4}{3} s_W^2, \nonumber  \\
	\lambda_{ZBB}^L&=& -1+\frac{2}{3} s_W^2 ,\quad
	\lambda_{ZBB}^R= -(\crd)^2+\frac{2}{3} s_W^2, \nonumber  \\
	\lambda_{ZtT}^L&=& 0 ,\quad \quad
	\lambda_{ZtT}^R= -\sru \cru , \nonumber  \\
	\lambda_{ZbB}^L&=& 0  ,\quad \quad
	\lambda_{ZbB}^R= \srd \crd  .
	\label{eq:SMTB_Zff}
\end{eqnarray}

\subsection{Vector-like doublet XT}
For the XT doublet model, the couplings between the X and T heavy quarks and the third generation SM quarks with the W boson are defined as:
\begin{eqnarray}
	\lambda_{Wtb}^L&=& \clx,\quad
	\lambda_{Wtb}^R=   0   , \nonumber  \\
	\lambda_{WTb}^L&=&  \slx  ,\quad
	\lambda_{WTb}^R=  0, \nonumber  \\
	\lambda_{WtX}^L&=& -\slx  , \quad
	\lambda_{WtX}^R= -\srx , \nonumber  \\
	\lambda_{WXT}^L&=&  \clx ,\quad
	\lambda_{WXT}^R= \crx .  \label{eq:SMXT_Wff}
\end{eqnarray}
the left- and right-handed neutral current couplings to the Z boson $\lambda^{L,R}_{Zq'q}$  are defined as:
\begin{eqnarray}
	\lambda_{Ztt}^L&=& 1-2(\slx)^2 -\frac{4}{3} s_W^2 ,\quad
	\lambda_{Ztt}^R=  -(\srx)^2 -\frac{4}{3} s_W^2 ,\nonumber  \\
	\lambda_{Zbb}^L&=& -1+\frac{2}{3} s_W^2 ,\quad
	\lambda_{Zbb}^R=  \frac{2}{3} s_W^2 , \nonumber  \\
	\lambda_{ZTT}^L&=& 1-2(\clx)^2 -\frac{4}{3} s_W^2 ,\quad
	\lambda_{ZTT}^R=  -(\crx)^2 -\frac{4}{3} s_W^2 , \nonumber  \\
	\lambda_{ZXX}^L&=& 1- \frac{10}{3} s_W^2,\quad
	\lambda_{ZXX}^R=  1- \frac{10}{3} s_W^2 , \nonumber  \\
	\lambda_{ZtT}^L&=& 2 \slx \clx ,\quad
	\lambda_{ZtT}^R=  \crx \srx .
	\label{eq:SMXT_Zff}
\end{eqnarray}

\subsection{Vector-like doublet BY}
In the case of the BY doublet model, the interaction of vector-like B and Y quarks and the third generation SM quarks with the W boson are defined as:

\begin{eqnarray}
	\lambda_{Wtb}^L&=& c_L^d  ,\quad
	\lambda_{Wtb}^R=  0   , \nonumber  \\
	\lambda_{WtB}^L&=& s_L^d  ,\quad
	\lambda_{WtB}^R=   0  , \nonumber  \\
	\lambda_{WbY}^L&=& -s_L^d , \quad
	\lambda_{WbY}^R=   -s_R^d  , \nonumber  \\
	\lambda_{WBY}^L&=& c_L^d  ,\quad
	\lambda_{WBY}^R=   c_R^d  .  \label{eq:SMBY_Wff}
\end{eqnarray}
The neutral current couplings to the Z boson for both the left- and right-handed states are given by:
\begin{eqnarray}
	\lambda_{Ztt}^L&=& 1-\frac{4}{3} s_W^2 ,\quad
	\lambda_{Ztt}^R=  -\frac{4}{3} s_W^2 ,\nonumber  \\
	\lambda_{Zbb}^L&=& -1 + 2(s_L^d)^2 +\frac{2}{3} s_W^2,\quad
	\lambda_{Zbb}^R= (s_R^d)^2 +\frac{2}{3} s_W^2  , \nonumber  \\
	\lambda_{ZBB}^L&=& -1 + 2(c_L^d)^2 +\frac{2}{3} s_W^2,\quad
	\lambda_{ZBB}^R=  (c_R^d)^2 +\frac{2}{3} s_W^2 , \nonumber  \\
	\lambda_{ZYY}^L&=& - 1 +\frac{8}{3} s_W^2 ,\quad
	\lambda_{ZYY}^R=  - 1 +\frac{8}{3} s_W^2 , \nonumber  \\
	\lambda_{ZbB}^L&=& -2 s_L^d c_L^d ,\quad \quad
	\lambda_{ZbB}^R=  -s_R^d c_R^d  .
	\label{eq:SMBY_Zff}
\end{eqnarray}

\subsection{Vector-like triplet XTB}

In the framework of the XTB triplet model, the left- and right-handed charged current couplings $\lambda_{Wq'q}^{L,R}$  are expressed as follows:

\begin{eqnarray}
	\lambda_{Wtb}^L&=& \clu \cld + \sqt \slu \sld  ,\quad
	\lambda_{Wtb}^R=  \sqt \sru \srd   , \nonumber  \\
	\lambda_{WTB}^L&=& \slu \sld + \sqt \clu \cld  ,\quad
	\lambda_{WTB}^R=  \sqt \cru \crd   , \nonumber  \\
	\lambda_{WtB}^L&=&  \clu \sld - \sqt \slu \cld , \quad
	\lambda_{WtB}^R=  -\sqt \sru \crd   , \nonumber  \\
	\lambda_{WTb}^L&=& \slu \cld - \sqt \clu \sld  ,\quad
	\lambda_{WTb}^R=  -\sqt \cru \srd  , \nonumber  \\
	\lambda_{WtX}^L&=& -\sqt \slu  ,\quad
	\lambda_{WtX}^R=   -\sqt \sru , \nonumber  \\
	\lambda_{WXT}^L&=& \sqt \clu  ,\quad
	\lambda_{WXT}^R=   \sqt \cru  .  \label{eq:SMXTB_Wff}
\end{eqnarray}
While, the left- and right-handed neutral current couplings $\lambda^{L,R}_{Zf'f}$ are defined as
\begin{eqnarray}
	\lambda_{Ztt}^L&=& (\clu)^2-\frac{4}{3} s_W^2 ,\quad
	\lambda_{Ztt}^R=  -\frac{4}{3} s_W^2  ,\nonumber  \\
	\lambda_{Zbb}^L&=& -1-(\sld)^2+\frac{2}{3} s_W^2,\quad
	\lambda_{Zbb}^R=  -2(\srd)^2+\frac{2}{3} s_W^2 , \nonumber  \\
	\lambda_{ZTT}^L&=& (\slu)^2-\frac{4}{3} s_W^2 ,\quad
	\lambda_{ZTT}^R=  -\frac{4}{3} s_W^2 , \nonumber  \\
	\lambda_{ZBB}^L&=& -1-(\cld)^2+\frac{2}{3} s_W^2 ,\quad
	\lambda_{ZBB}^R=  -2 (\crd)^2+\frac{2}{3} s_W^2 , \nonumber  \\
	\lambda_{ZXX}^L&=& 2- +\frac{10}{3} s_W^2 ,\quad
	\lambda_{ZXX}^R=   2- +\frac{10}{3} s_W^2 , \nonumber  \\
	\lambda_{ZtT}^L&=& \slu \clu  ,\quad \quad
	\lambda_{ZtT}^R=  0 , \nonumber  \\
	\lambda_{ZbB}^L&=& \sld \cld ,\quad \quad
	\lambda_{ZbB}^R=  2\srd \crd.
	\label{eq:SMXTB_Zff}
\end{eqnarray}

\subsection{Vector-like triplet TBY}
In the TBY triplet model, the interactions between the vector-like T, B, and Y quarks, as well as the third generation SM quarks, with the W boson are defined as:
\begin{eqnarray}
	\lambda_{Wtb}^L&=& \clu \cld + \sqt \slu \sld  ,\quad
	\lambda_{Wtb}^R=  \sqt \sru \srd   , \nonumber  \\
	\lambda_{WTB}^L&=& \slu \sld + \sqt \clu \cld  ,\quad
	\lambda_{WTB}^R=  \sqt \cru \crd   , \nonumber  \\
	\lambda_{WtB}^L&=&  \clu \sld - \sqt \slu \cld , \quad
	\lambda_{WtB}^R=  -\sqt \sru \crd   , \nonumber  \\
	\lambda_{WTb}^L&=&  \slu \cld - \sqt \clu \sld  ,\quad
	\lambda_{WTb}^R=  -\sqt \cru \srd  , \nonumber  \\
	\lambda_{WbY}^L&=& -\sqt \sld  ,\quad
	\lambda_{WbY}^R=  -\sqt \srd  , \nonumber  \\
	\lambda_{WBY}^L&=& \sqt \cld  ,\quad
	\lambda_{WBY}^R=   \sqt \crd  .  \label{eq:SMTBY_Wff}
\end{eqnarray}
The left- and right-handed neutral current couplings $\lambda^{L,R}_{Zf'f}$ in this model are given by:
\begin{eqnarray}
	\lambda_{Ztt}^L&=& 1+(\slu)^2-\frac{4}{3} s_W^2  ,\quad
	\lambda_{Ztt}^R=  2(\sru)^2 -\frac{4}{3} s_W^2  ,\nonumber  \\
	\lambda_{Zbb}^L&=& -(\cld)^2+\frac{2}{3} s_W^2 ,\quad
	\lambda_{Zbb}^R=   \frac{2}{3} s_W^2 , \nonumber  \\
	\lambda_{ZTT}^L&=& 1+(\clu)^2-\frac{4}{3} s_W^2 ,\quad
	\lambda_{ZTT}^R=  2(\cru)^2 -\frac{4}{3} s_W^2  , \nonumber  \\
	\lambda_{ZBB}^L&=& -(\sld)^2+\frac{2}{3} s_W^2 ,\quad
	\lambda_{ZBB}^R=  \frac{2}{3} s_W^2  , \nonumber  \\
	\lambda_{ZYY}^L&=& -2+\frac{8}{3} s_W^2 ,\quad
	\lambda_{ZYY}^R=   -2+\frac{8}{3} s_W^2 , \nonumber  \\
	\lambda_{ZtT}^L&=& -\slu \clu ,\quad \quad
	\lambda_{ZtT}^R=  -2\sru\cru  , \nonumber  \\
	\lambda_{ZbB}^L&=& -\sld \cld ,\quad \quad
	\lambda_{ZbB}^R= 0 .
	\label{eq:SMXTY_Zff}
\end{eqnarray}

For simplification purposes and avoiding repetition of the global quantities of the $(Vf'f)$ couplings during the manipulation of the expressions for the oblique parameters. We introduce a new definition coupling given by:
\begin{equation}
	c^{L,R}_{Zf'f}=\frac{e}{2c_W s_W}\lambda^{L,R}_{Zf'f} , \text{ and } c^{L,R}_{Wf'f}=\frac{e}{\sqrt{2}s_W} \lambda^{L,R}_{Wf'f}.
\end{equation}

Additionally, the couplings in the SM framework are given as
\begin{eqnarray}
	c_{Wtb}^L&=&  \frac{e}{\sqrt{2} s_w },\quad
	c_{Wtb}^R= 0,\nonumber   \\
	c_{Ztt}^L&=&  \frac{e}{2 s_w c_w} (1 -\frac{4}{3} s_w^2) ,\quad
	c_{Ztt}^R= -\frac{2e s_w}{3 c_w}, \nonumber  \\
	c_{Zbb}^L&=& \frac{e}{2 s_w c_w}(-1+\frac{2}{3} s_w^2),\quad
	c_{Zbb}^R= \frac{e  s_w}{3 c_w}. \label{eq:SM_Vff}
\end{eqnarray}

\section{The PDG comparison}
This section aims to investigate the impact of newly introduced heavy quarks on the mass of the $W$-boson (PDG) within the context of the 2HDM+VLQs models. To facilitate a comparative analysis with the aforementioned study (CDF), we employ the $\chi^2_{M_W^\mathrm{PDG}} $ test instead of $\chi^2_{M_W^\mathrm{CDF}} $. The PDG values are as follows: $S = 0.05$ $\pm$ 0.08, $T = 0.09$ $\pm$ $0.07$, and the correlation factor between $S$ and $T$, denoted as $\rho _{ST}$, is 0.92. In line with the main text, we generate corresponding plots for visual representation.
	
	Clearly, we can observe from the plots that the degenerate case $m_{H^\pm}$ = $m_A$ = $m_H$ is highly favoured by the PDG data in all the scenarios.

\begin{figure}[H]
	\centering
	\includegraphics[height=4.5cm,width=5.cm]{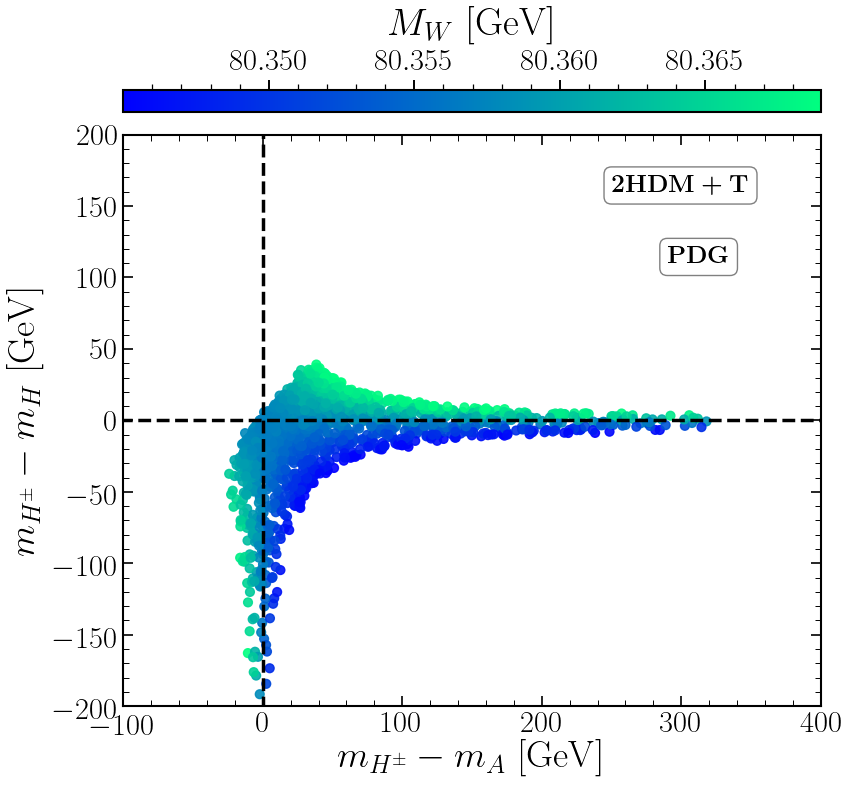}
	\includegraphics[height=4.5cm,width=5.cm]{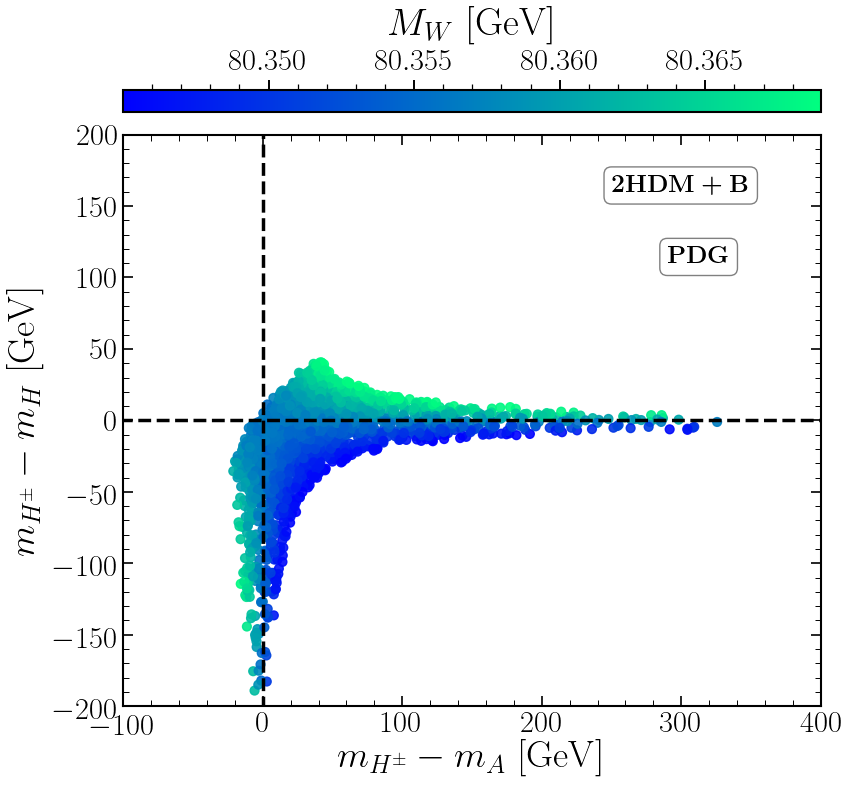}
	\includegraphics[height=4.5cm,width=5.cm]{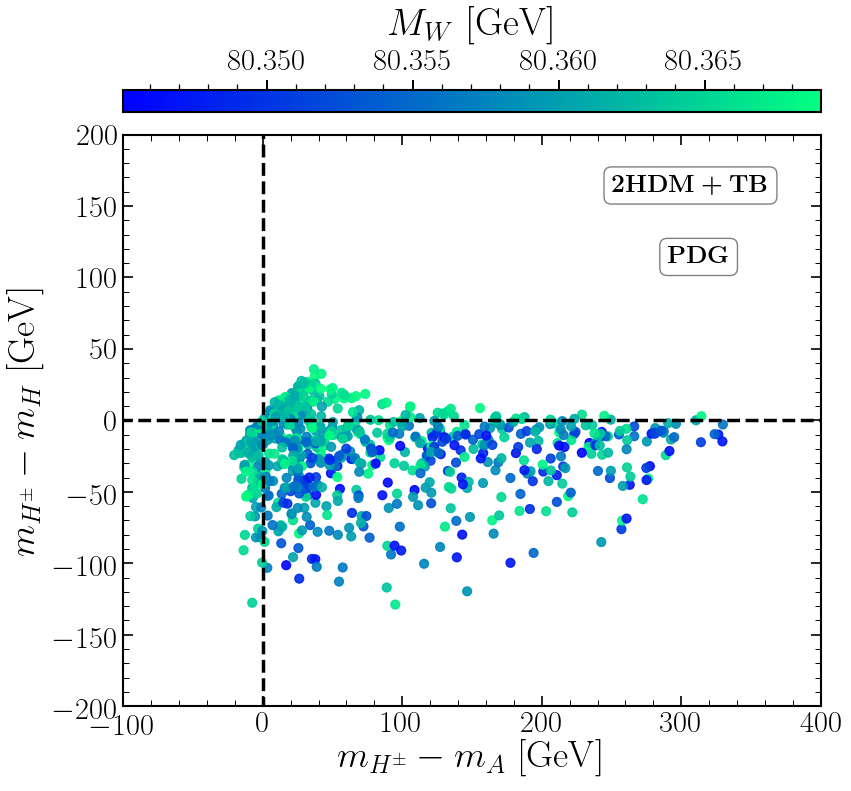}
	\includegraphics[height=4.5cm,width=5.cm]{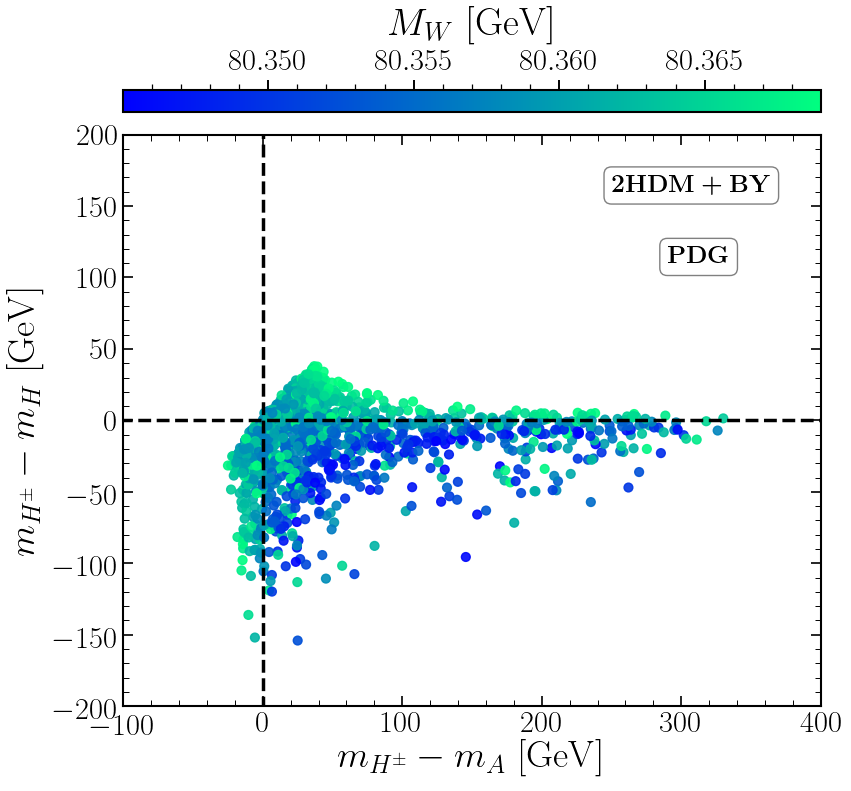}
	\includegraphics[height=4.5cm,width=5.cm]{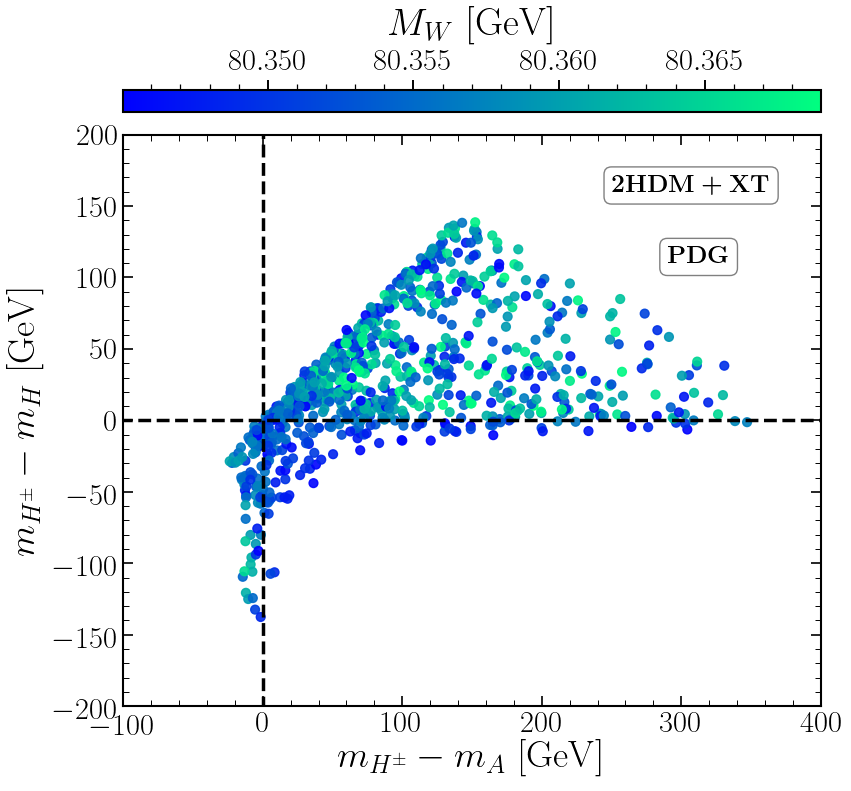}
	\includegraphics[height=4.5cm,width=5.cm]{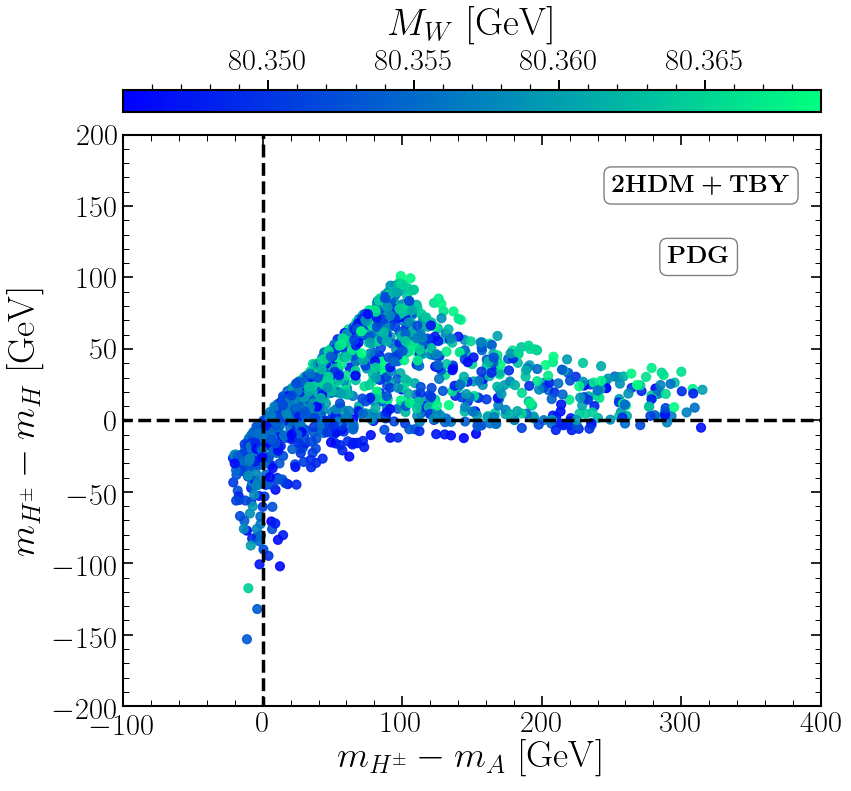}
	\includegraphics[height=4.5cm,width=5.cm]{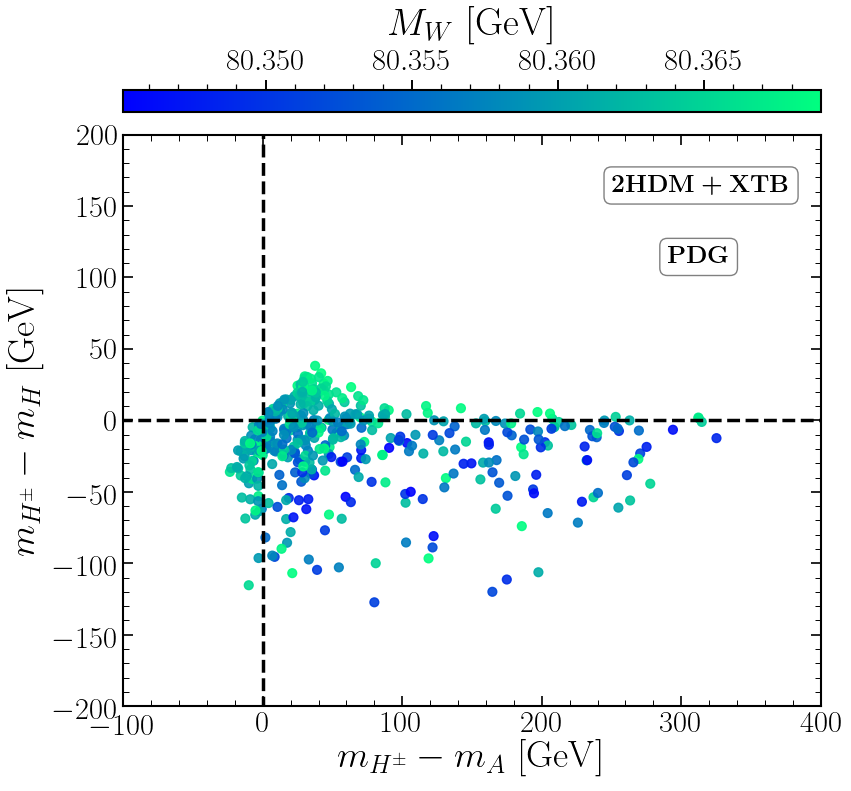}
	\caption{The points from the parameter scan of the  2HDM+VLQs models in the $m_{H^\pm}-m_A$  versus $m_{H^\pm}-m_H$ plane in the, with the colour code indicating the W-boson mass $M_W$. All points have $\chi^2_{M_W^\mathrm{PDG}}\le 4$. }\label{fig:fig11}
\end{figure}

\section{The CDF-II W-boson mass anomaly within the SM+VLQs}\label{SMVLQs_Results}
In the following, we will emphasize the impact of the new heavy quarks on the $W$-boson mass in the SM +VLQs models.




\begin{figure}[H]
	\centering
	\includegraphics[height=4.5cm,width=5.cm]{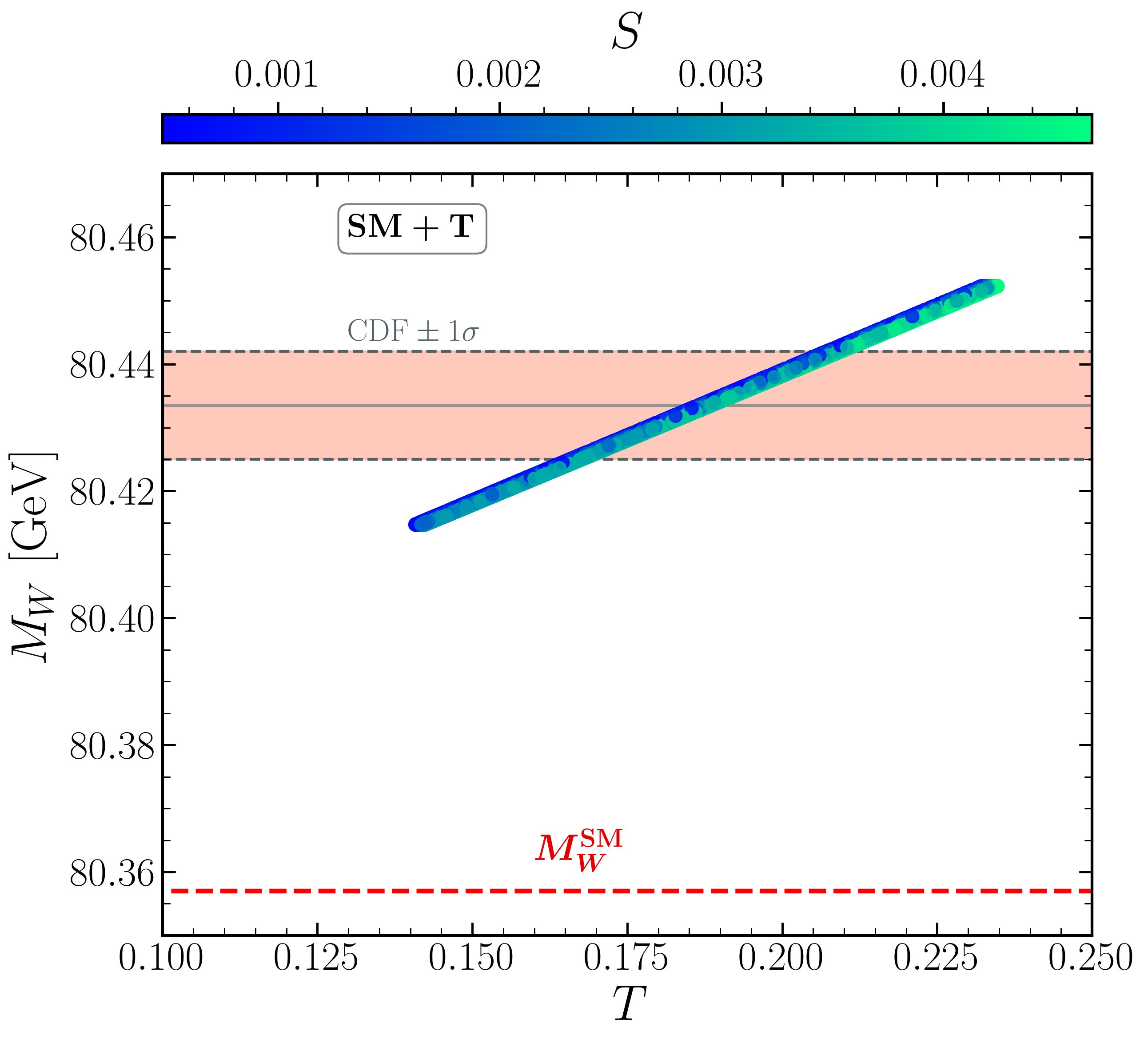}
	\includegraphics[height=4.5cm,width=5.cm]{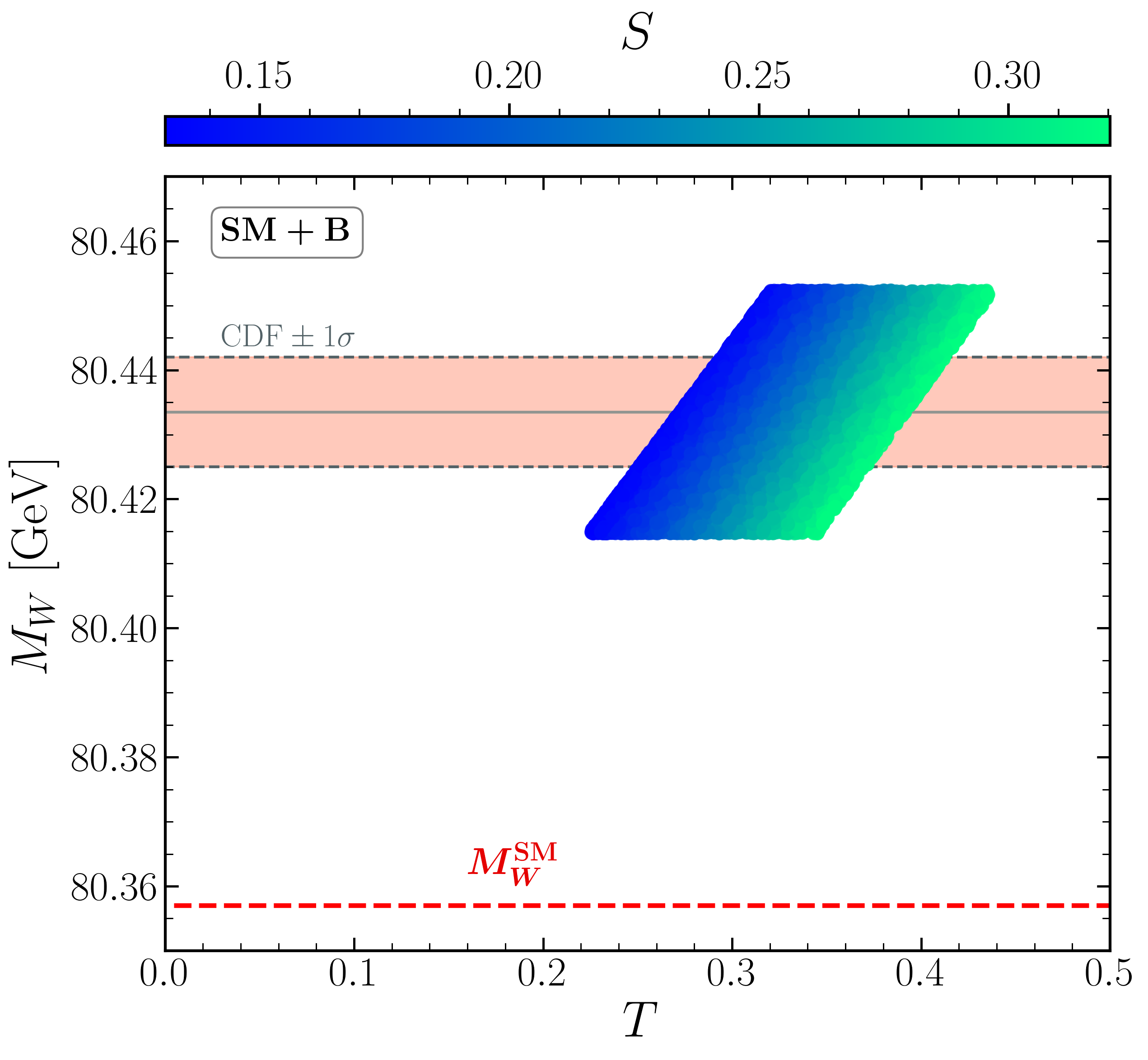}
	\includegraphics[height=4.5cm,width=5.cm]{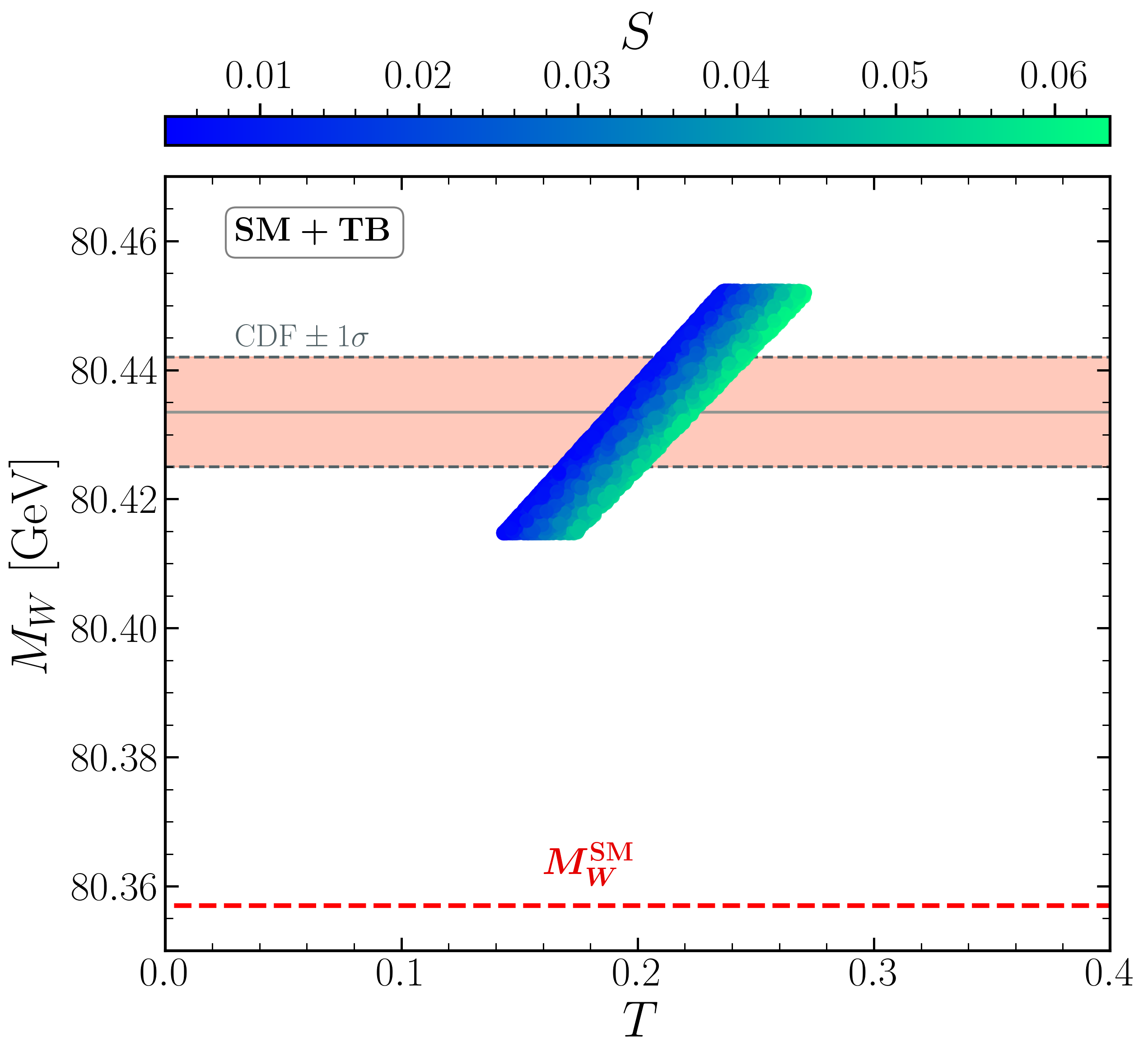}
	\includegraphics[height=4.5cm,width=5.cm]{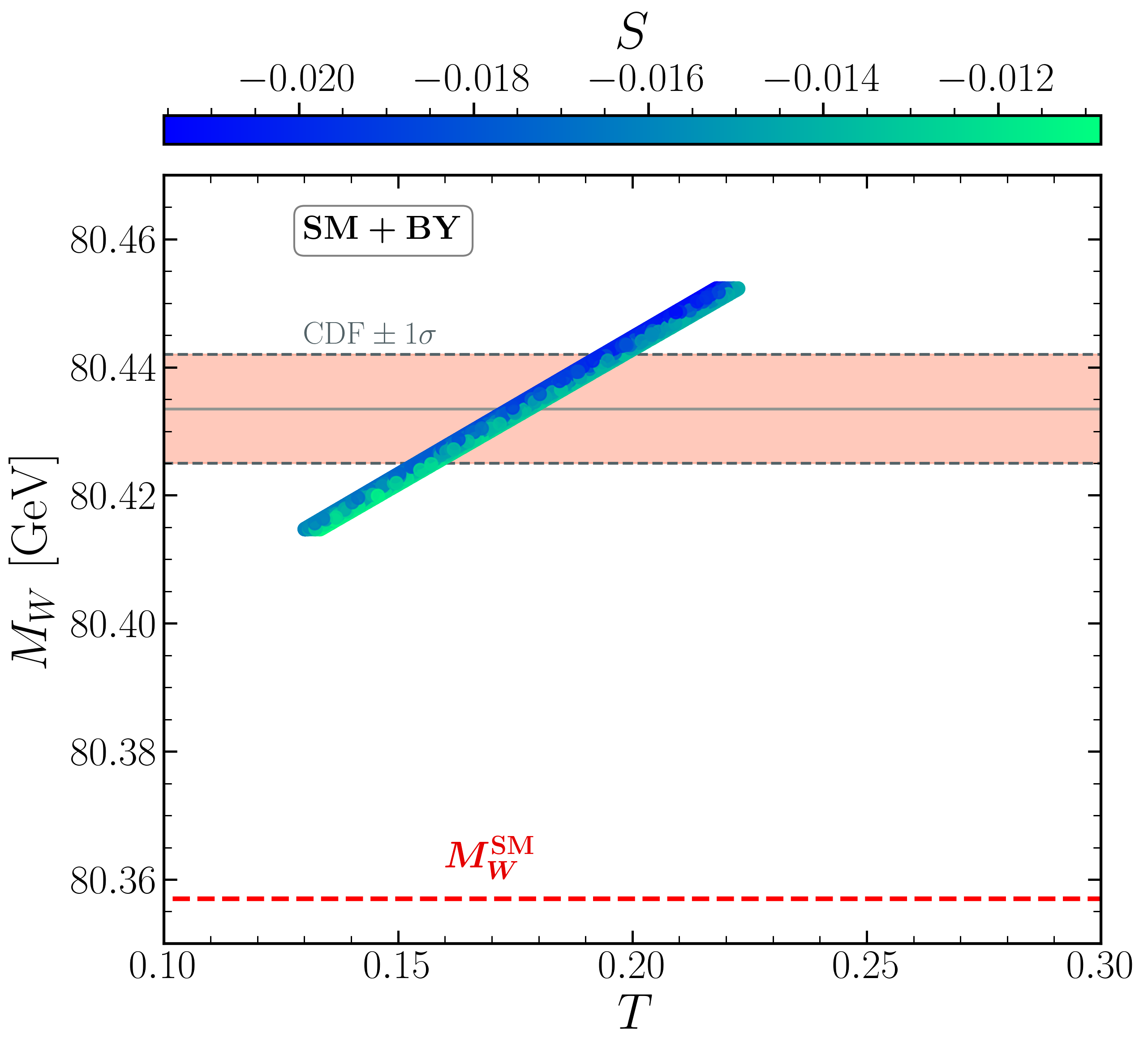}
	\includegraphics[height=4.5cm,width=5.cm]{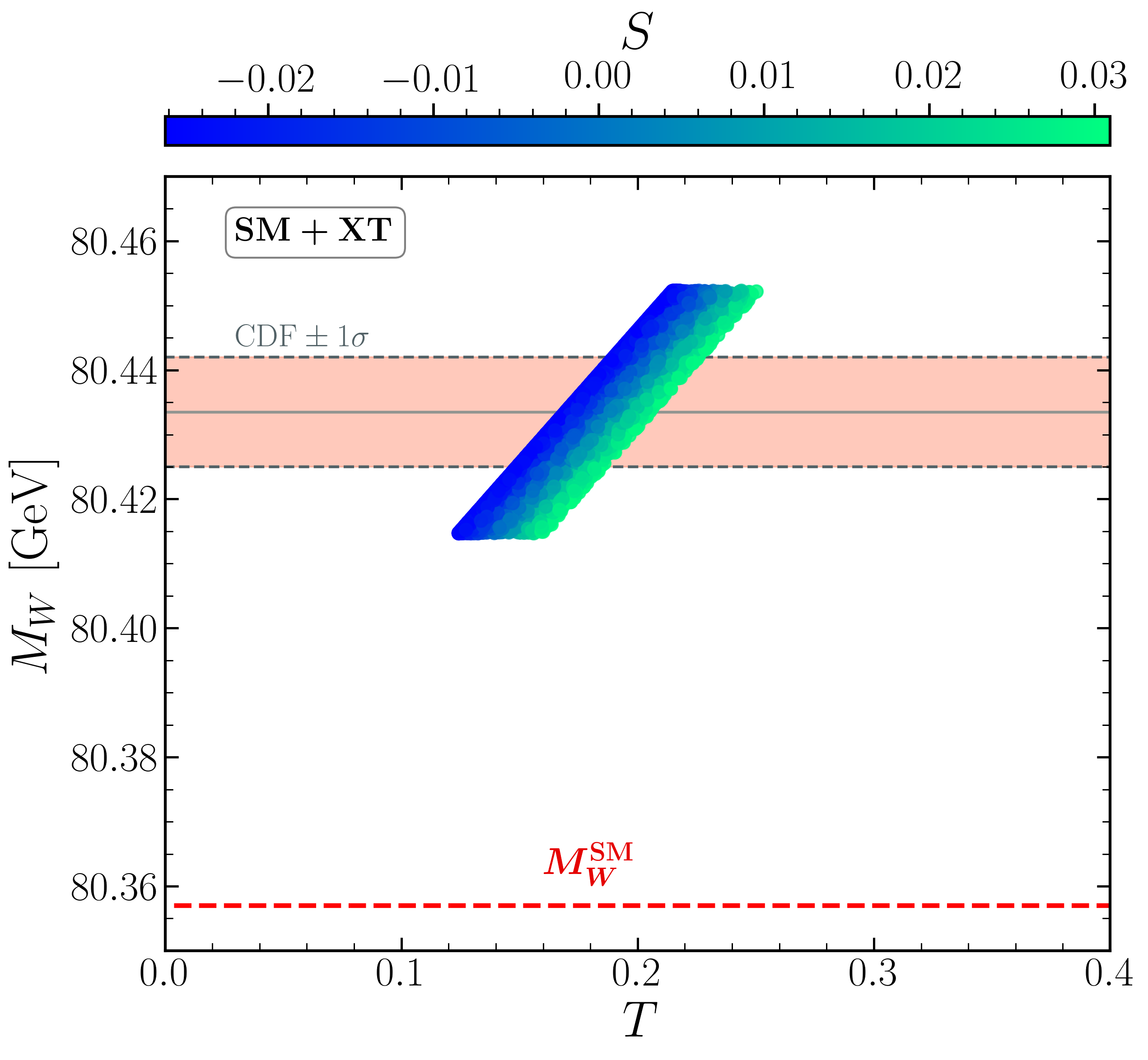}
	\includegraphics[height=4.5cm,width=5.cm]{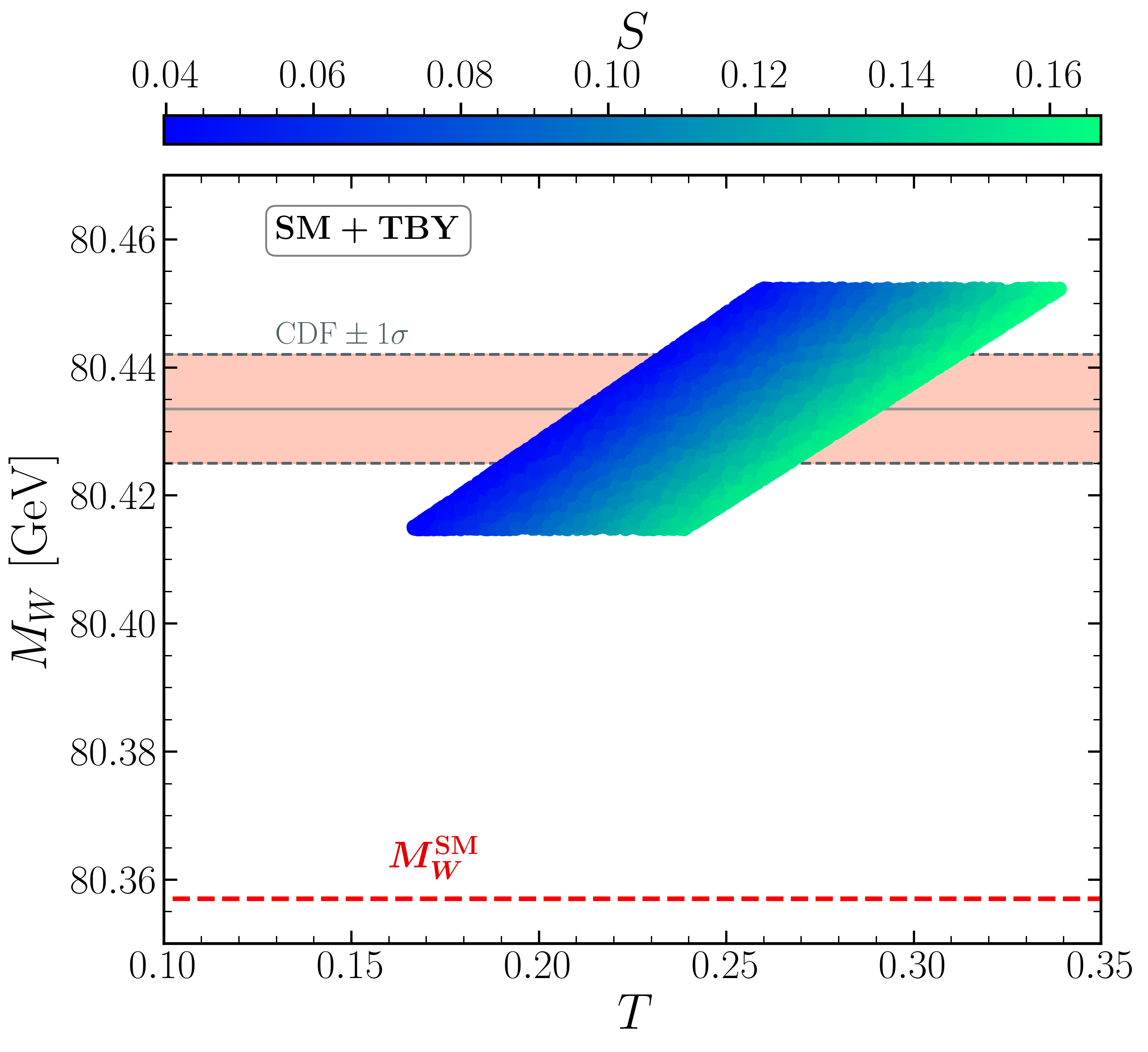}
	\includegraphics[height=4.5cm,width=5.cm]{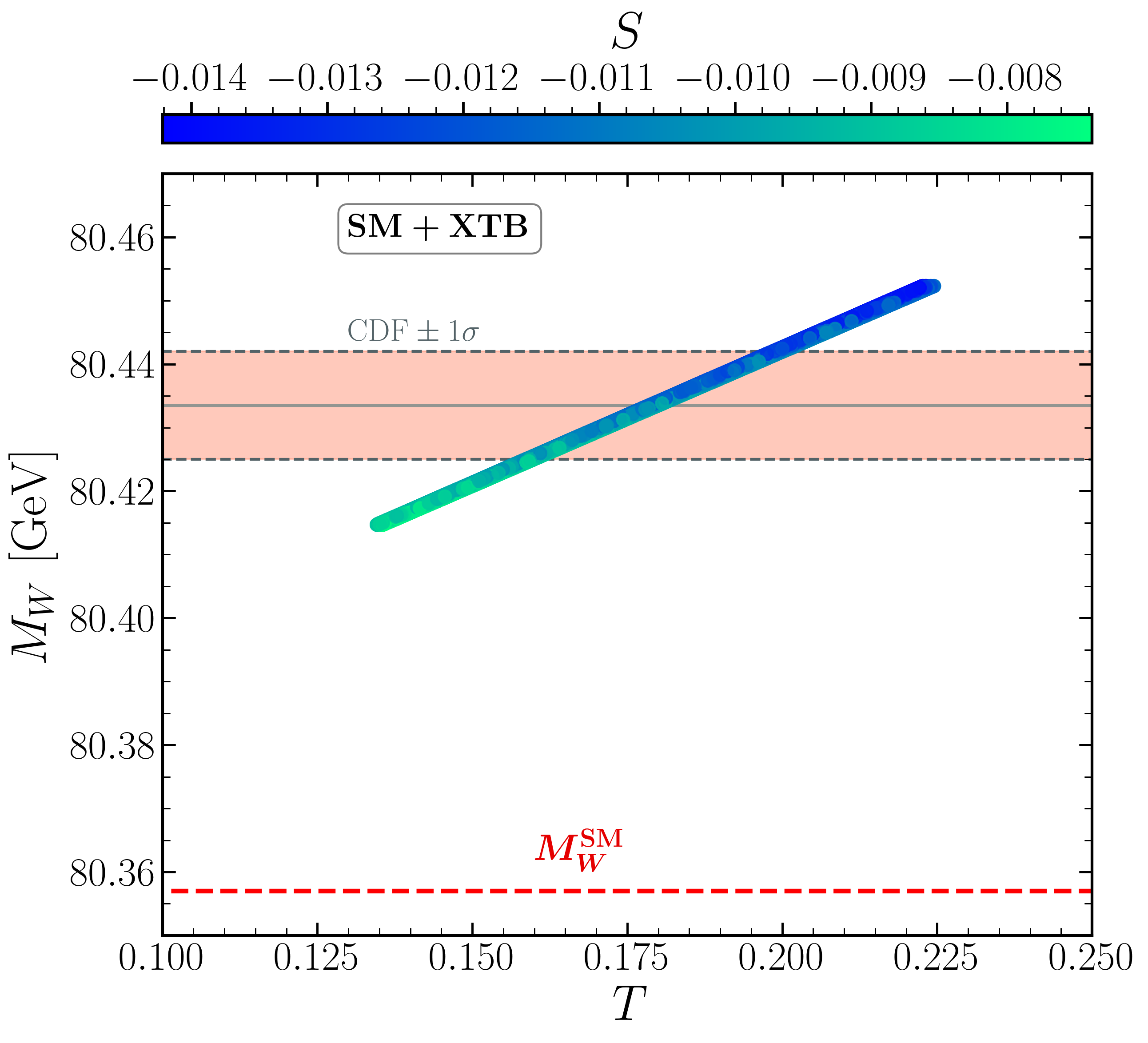}
	\caption{Scan result of the SM + (VLQs) in the  $(T,M_W)$. The light red band indicates the $M_W$ value with the associated 1$\sigma$ uncertainty measured recently by the CDF collaboration, while the light green band  shows the result of $\sin^2\theta_{\mathrm{eff}}$ and its associated 1$\sigma$ uncertainty measured  by the SLD collaboration.}\label{fig:SM1}
\end{figure}
In Fig.~\ref{fig:SM1}, we depict the $W$ mass calculated using Eq.\ref{eq4} as a function of the T parameter in the context of the SM + VLQs extensions. The oblique parameter $S$ is displayed on the color map. The orange band represents the new $M_W$ value measured by the CDF collaboration, along with its $1\sigma$ uncertainty. In all VLQs representations, the contribution of the T parameter required to explain the CDF's newly measured W-boson mass is positive. While the $S$ parameter is only positively restricted in the SM + T,B,TB and TBY extension, it is negative in the SM + BY  and XTB extension, and takes on both positive and negative values in the SM + XT extension.\\

\begin{figure}[H]
	\centering
	\includegraphics[height=4.75cm,width=5.25cm]{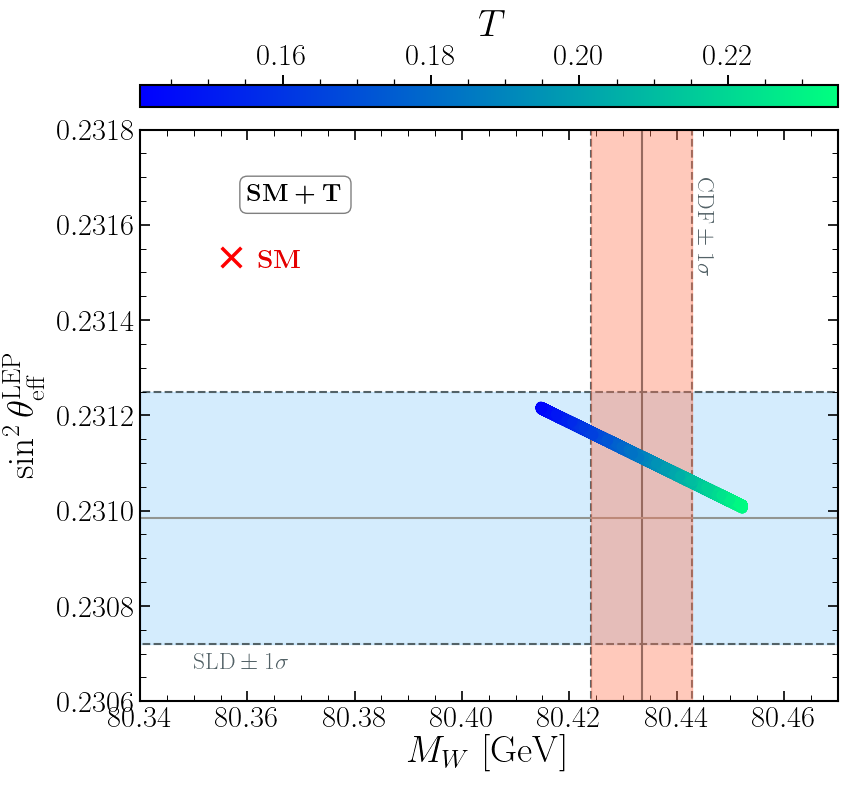}
	\includegraphics[height=4.75cm,width=5.25cm]{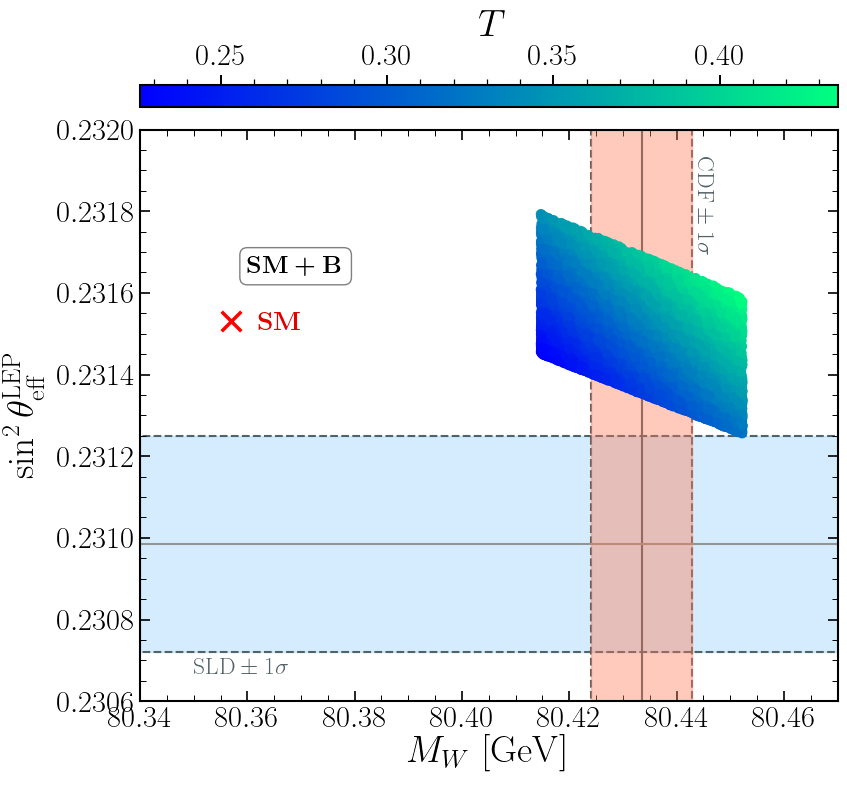}
	\includegraphics[height=4.75cm,width=5.25cm]{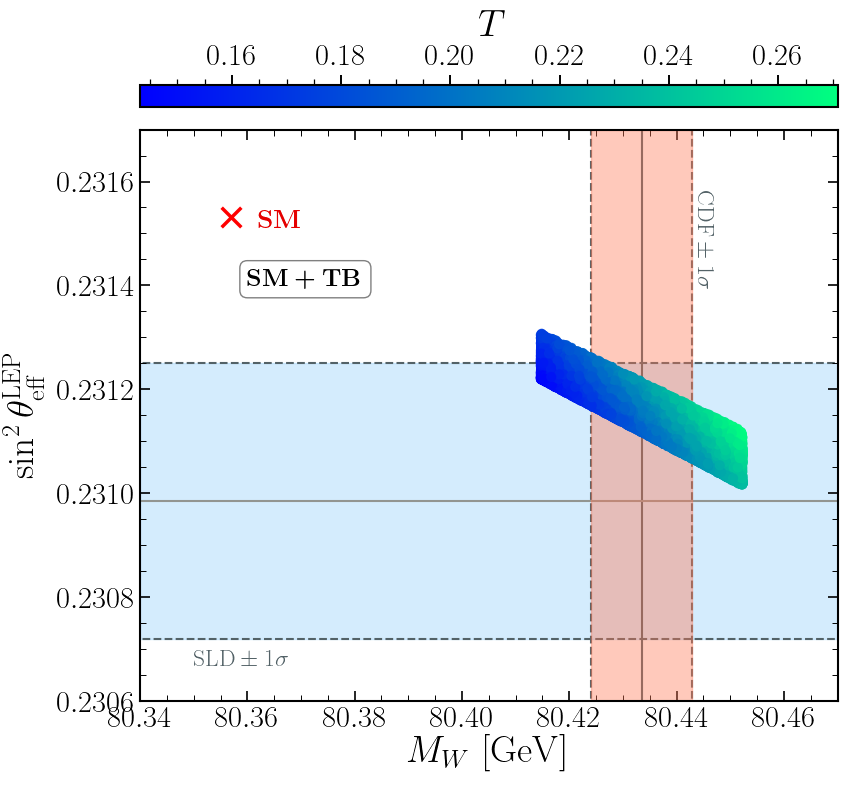}
	\includegraphics[height=4.75cm,width=5.25cm]{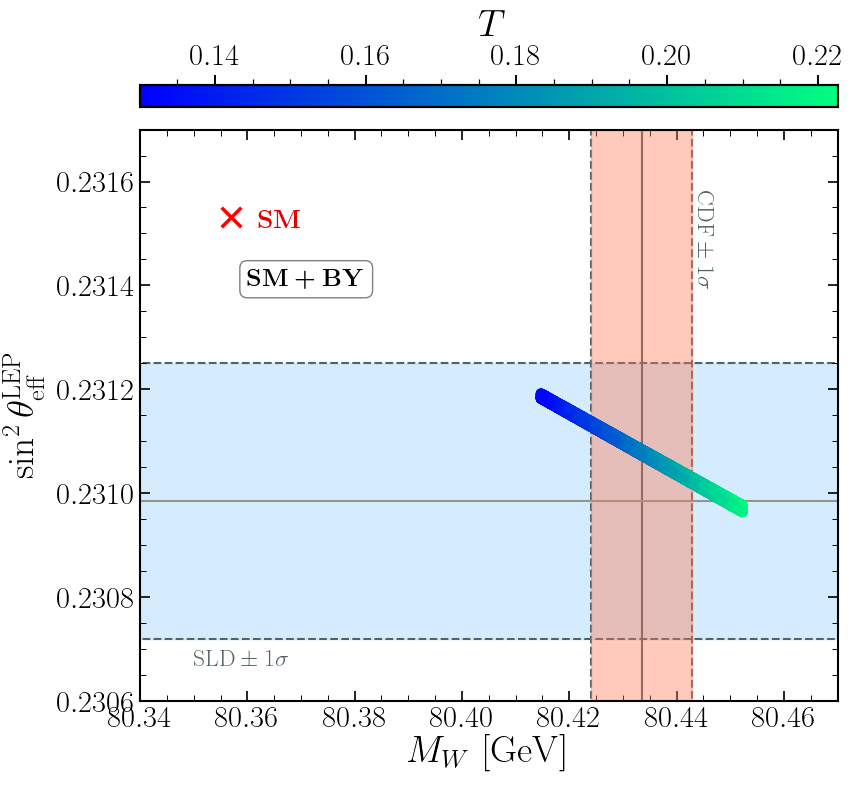}
	\includegraphics[height=4.75cm,width=5.25cm]{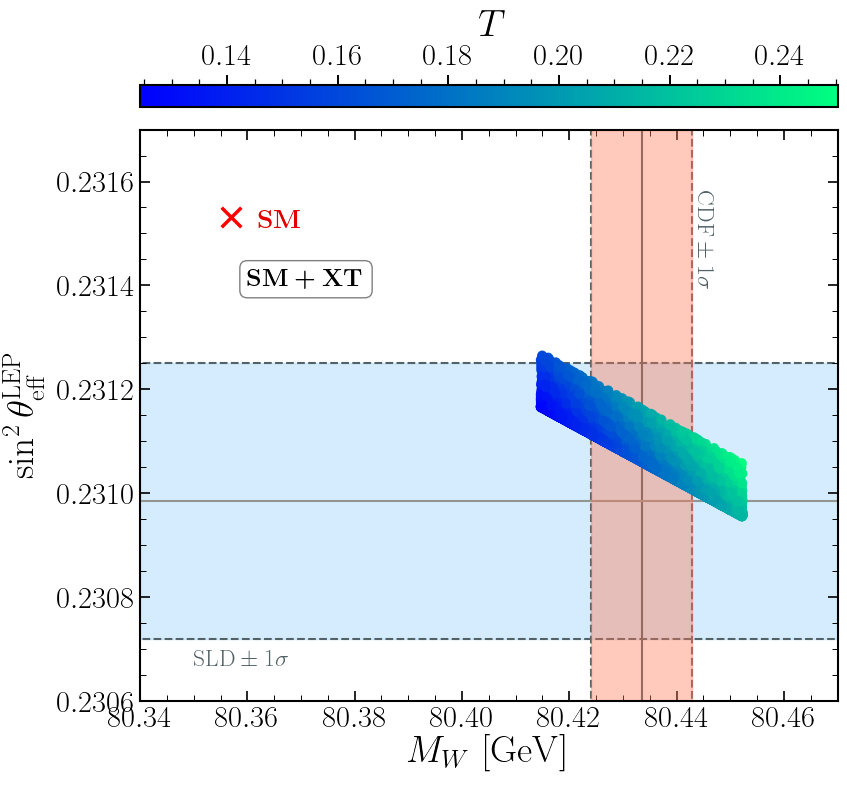}
	\includegraphics[height=4.75cm,width=5.25cm]{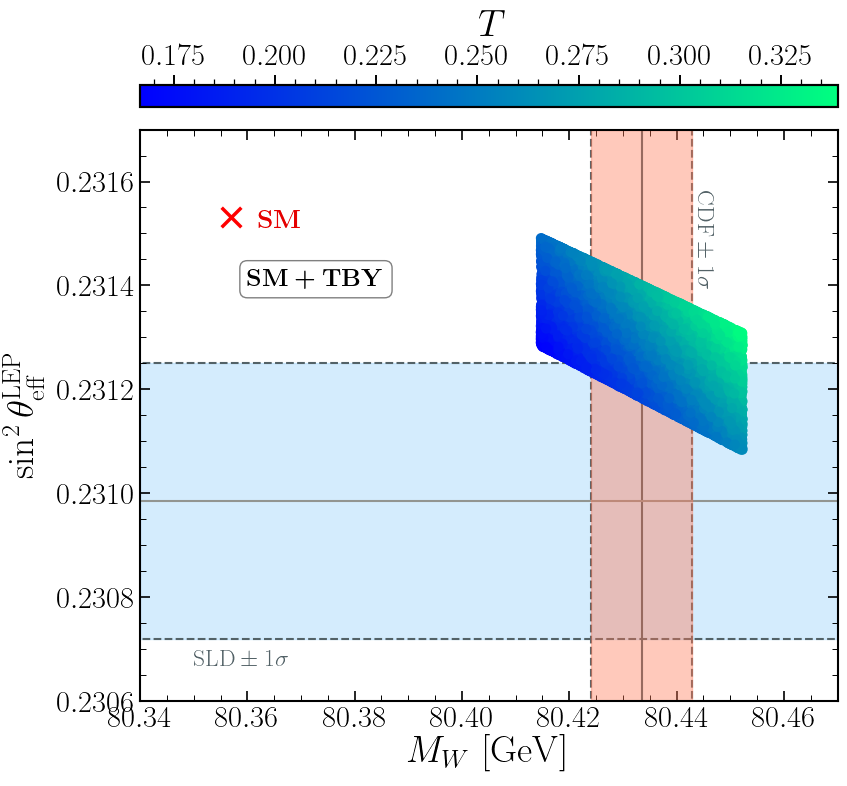}
	\includegraphics[height=4.75cm,width=5.25cm]{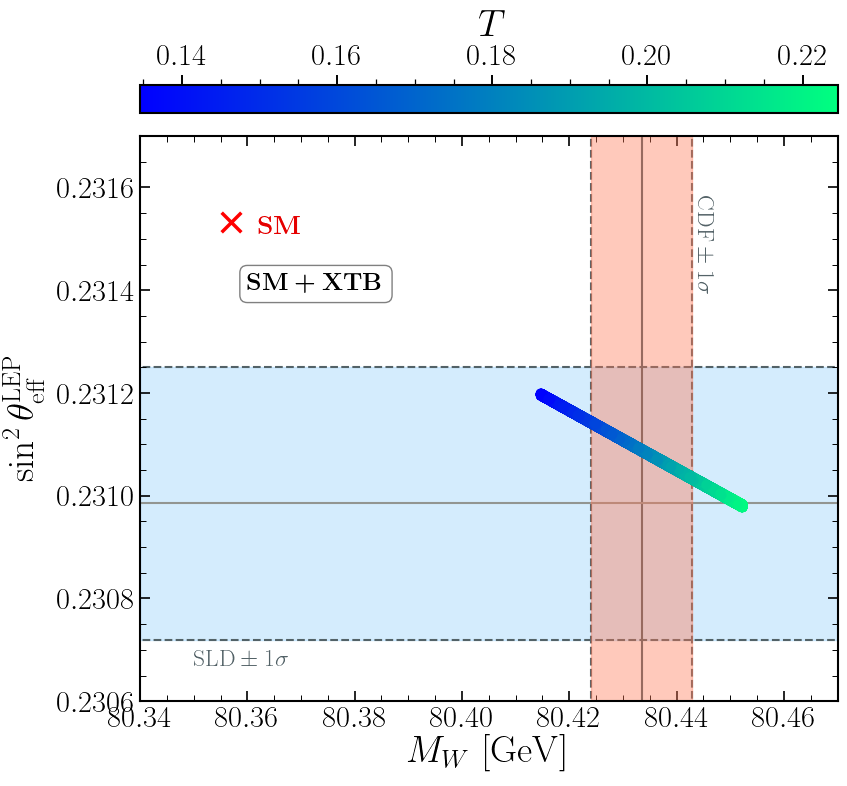}
	\caption{Scan result of the SM + (VLQs) in the $(M_W , \sin^2 \theta^{\mathrm{LEP}}_{\mathrm{eff}})$  plane. The light red band indicates the $M_W$ value with the associated 1$\sigma$ uncertainty measured recently by the CDF collaboration, while the light green band  shows the result of $\sin^2\theta_{\mathrm{eff}}$ and its associated 1$\sigma$ uncertainty measured  by the SLD collaboration.}\label{fig:SMT2}
\end{figure}

In Fig.~\ref{fig:SMT2} we present the same points in the $(M_W , \sin^2 \theta^{\mathrm{LEP}}_{\mathrm{eff}})$ plane with the T parameter in the color bar. The light blue band represents the measurement of the SLD collaboration for $\sin^2 \theta_{eff}$ within the $\pm 1 \sigma$. It is interesting to see that the points satisfying the CDF-II data are in good agreement with the data points preferred by the SLD measurement of the $\sin^2 \theta_{eff}$ within 1 $\sigma $ uncertainty in the case of SM+(T, BY and XTB) while the SM+(TB, XT and TBY) are only partially within the bands. Furthermore, it can be seen that within the SM+B singlet, there is a significant tension between the points preferred by the CDF-II data and the SLD measurement.

\newpage
	
\bibliography{ref} 
\bibliographystyle{JHEP}

\end{document}